\documentclass[aps,onecolumn,amssymb,superscriptaddress]{revtex4-1}

\usepackage{bm}
\usepackage{graphicx}
\usepackage{amsmath}
\usepackage{amsthm}
\usepackage{eucal}
\usepackage{amssymb}
\usepackage{mathrsfs}
\usepackage{enumerate}
\usepackage{xcolor}
\usepackage{array}
\usepackage{float}

\newcommand{\comments}[1]{}

\newcommand{\eref}[1]{Eq.~(\ref{#1})}

\newcommand{\fref}[1]{Figure \ref{#1}}

\newcommand{\sref}[1]{Section \ref{#1}}

\newcommand{\aref}[1]{Appendix \ref{#1}}

\begin{document}

\title{Exploiting fast-variables to understand population dynamics and evolution}
\author{George W.~A.~Constable}
\affiliation{Department of Evolutionary Biology and Environmental Studies,
University of Zurich, 8006 Zurich, Switzerland}
\author{Alan J.~McKane}
\affiliation{Theoretical Physics Division, School of Physics and Astronomy,
The University of Manchester, Manchester M13 9PL, United Kingdom}

\begin{abstract}
We describe a continuous-time modelling framework for biological population dynamics that accounts for demographic noise. In the spirit of the methodology used by statistical physicists, transitions between the states of the system are caused by individual events while the dynamics are described in terms of the time-evolution of a probability density function. In general, the application of the diffusion approximation still leaves a description that is quite complex. However, in many biological applications one or more of the processes happen slowly relative to the system's other processes, and the dynamics can be approximated as occurring within a slow low-dimensional subspace. We review these time-scale separation arguments and analyse the more simple stochastic dynamics that result in a number of cases. We stress that it is important to retain the demographic noise derived in this way, and emphasise this point by showing that it can alter the direction of selection compared to the prediction made from an analysis of the corresponding deterministic model.
\end{abstract}

\maketitle

\section{Introduction}
\label{sec:intro}

In a series of articles on biological evolution published in the Journal of Statistical Physics, it is natural to ask what expertise and insights statistical physicists can bring to the study of evolution, and in what way might their approach to the subject differ from biologists. If the subject is the one that will largely interest us in this paper --- the study of evolution within the framework of population genetics --- these questions are more easily answered. This is because in a system containing a large, but finite, number of individuals with given genetic characteristics, genetic drift leads to a stochastic dynamics which has many of the features which allow the application of the ideas and techniques of non-equilibrium statistical physics. We will use this formalism, but in addition our approach will have parallels with the traditional methodology of theoretical physicists. 

Firstly, we will stress the fundamental nature of the microscopic description. That is, we will start with genes as basic constituents, which can be in different states according to their type (allele), location (if the individual carrying the gene is located on an island), the sex of the individual carrying the gene, etc. 

Secondly, since the microscopic description contains too much detail which is irrelevant at the macroscale (or in our case, the mesoscale, where some stochastic element is retained) we will derive a \textit{reduced} or \textit{effective} model which contains parameters which depend on the parameters of the microscopic description and so encapsulate the relevant aspects of the microscopic description. 

Thirdly, we will be interested in \textit{generic} behaviour. By this we mean that we will attempt to formulate a microscopic description that does not have inbuilt assumptions that make the model more easily solvable. Instead we will try to formulate the model in such a way that it could be generalised to include many more effects, without changing its structure. In this philosophy, the simplifying assumptions are brought in during the process of obtaining the effective model, and should be clearly stated. 

Finally, although the whole basis of our work is mathematical, we will use intuition to explore the admissibility of the techniques we use outside of their regime of strict applicability, and check their correctness through the use of computer simulations. 

Although these ideas are familiar to the theoretical physics community, they tend to be utilised less in the biological sciences. For example, many biologists may use quite complex verbal arguments to gain insights. Conversely, our methodology may differ from that of many mathematical biologists, since rigorous justification will not be a central feature of our approach. In addition, many mathematical biologists are focussed on the deterministic dynamics found at the macroscale. Nevertheless, we view the approach we will discuss here as able to form a bridge between the intuition gleaned by biologists and the more analytic investigations of mathematicians. In this way we hope that our methods prove of interest to a wider audience outside of the theoretical physics community. 

In a previous paper~\cite{mckane_2014}, we have reviewed the process of setting-up a description of this class of biological systems in terms of its basic constituents, and from this deriving the mesoscopic equations governing the dynamics which generalise what might be the more familiar macroscopic equations. In particular, in Ref.~\cite{mckane_2014}, we give formulae for writing down the form of the mesoscopic dynamics in terms of quantities which appear in the microscopic formulation. This essentially is the first point of our methodology described above, and so while we will discuss it here, we will refer the reader to this earlier paper for more details. Instead we will focus on the second point above, namely obtaining an effective theory that is more amenable to analysis than the original.

\begin{figure}[th]
\begin{center}
\includegraphics[width=0.4\textwidth]{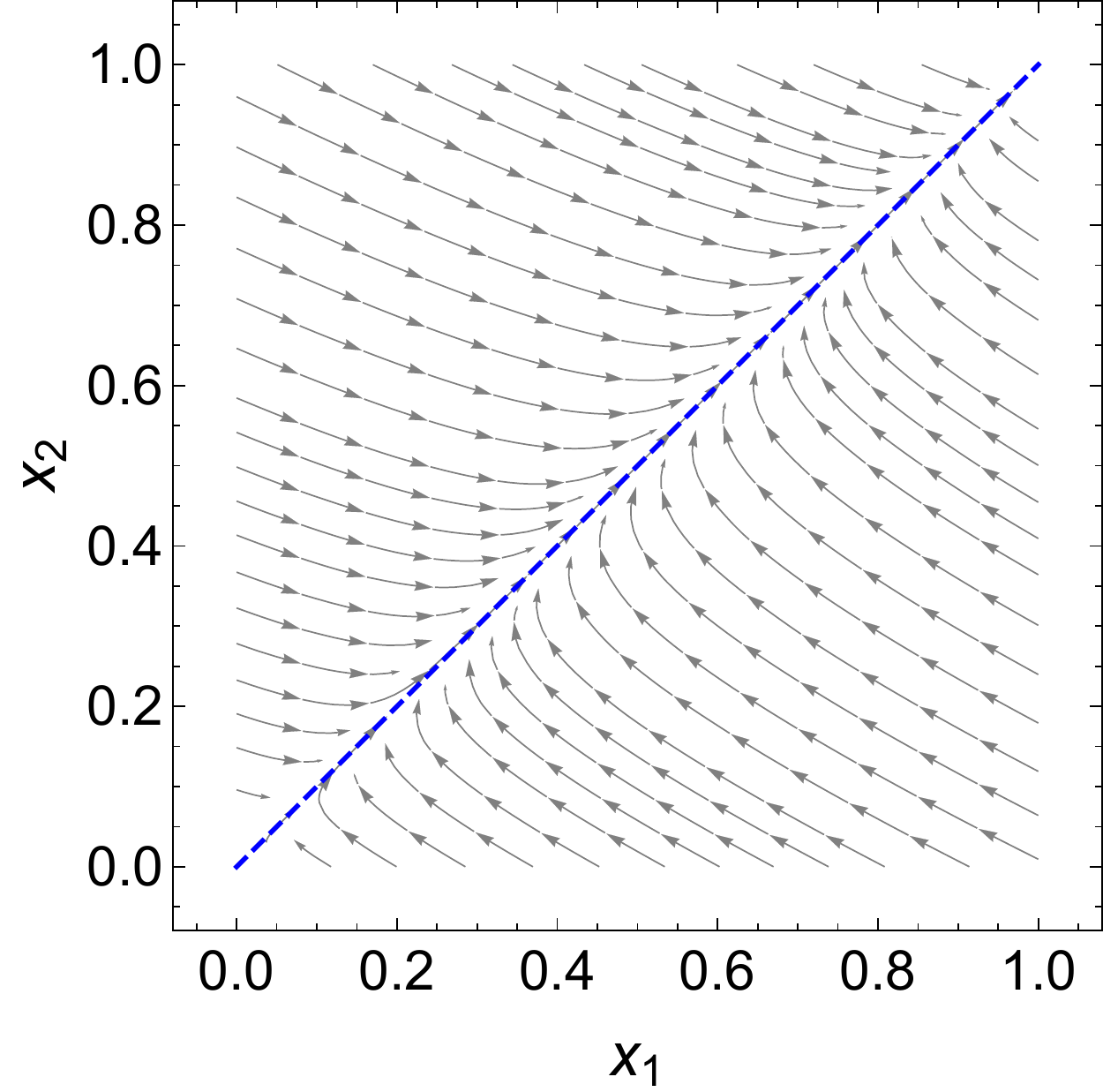}\hspace{0.08\textwidth}\includegraphics[width=0.4\textwidth]{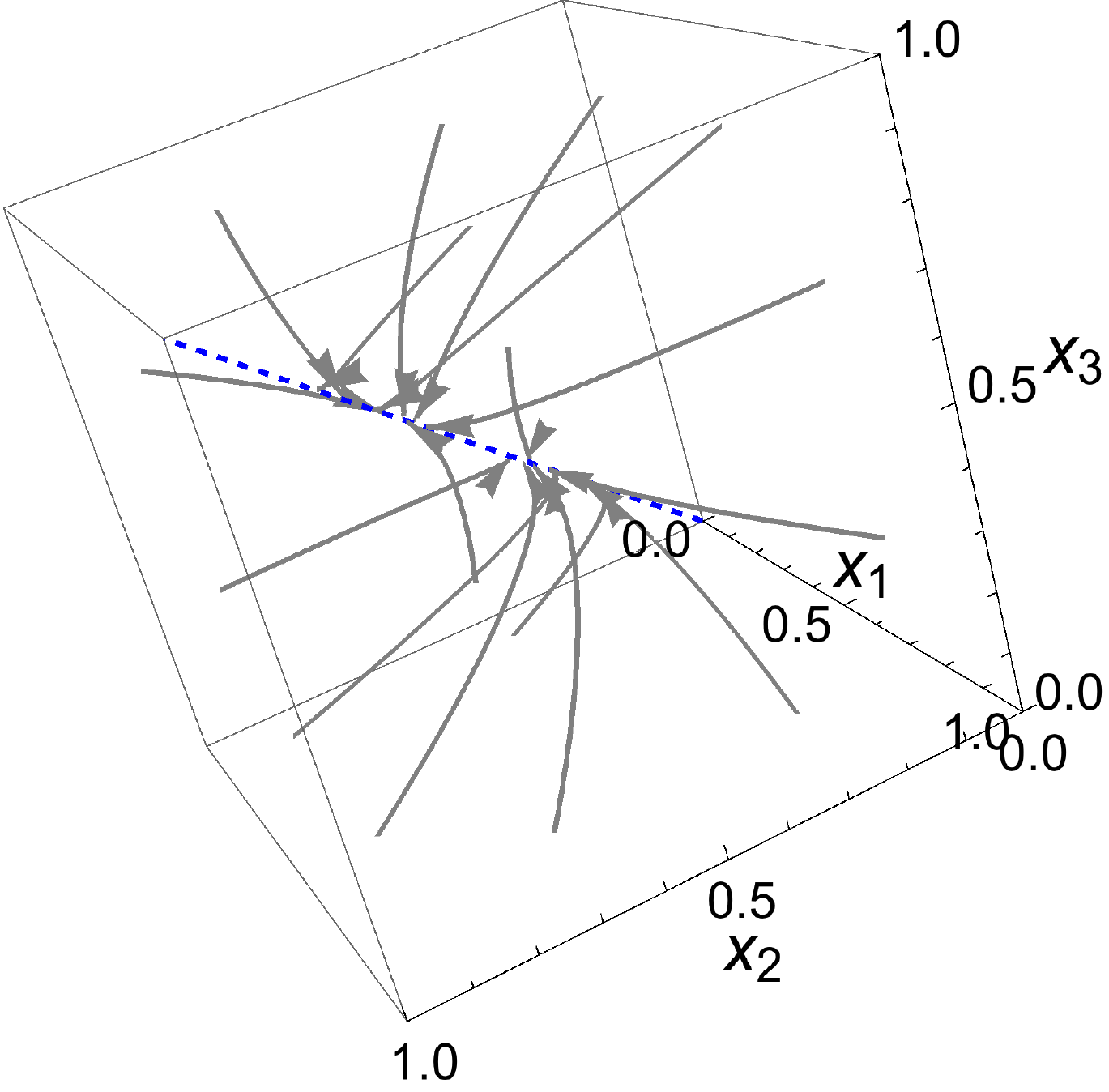}
\includegraphics[width=0.4\textwidth]{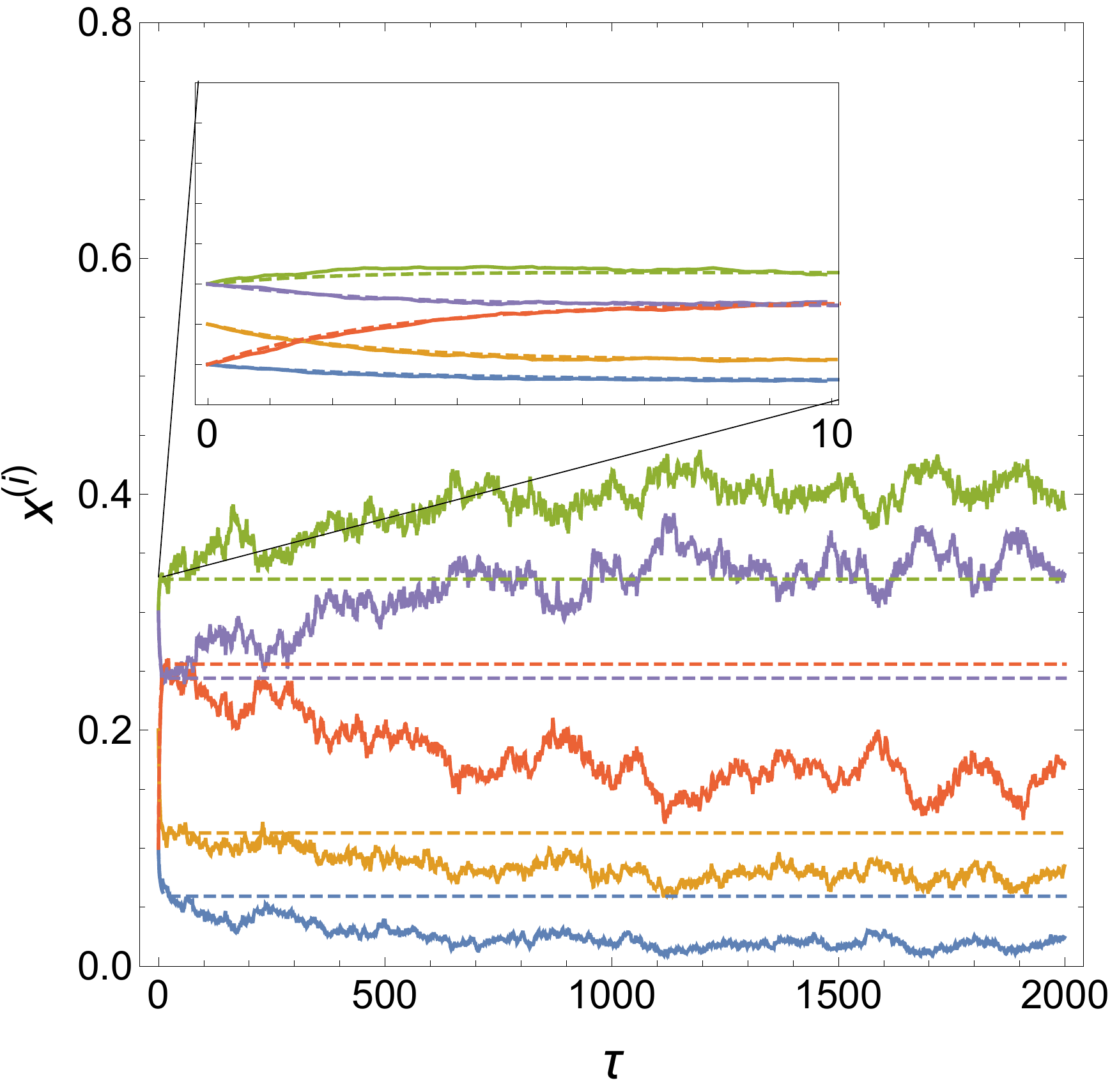}\hspace{0.08\textwidth}\includegraphics[width=0.4\textwidth]{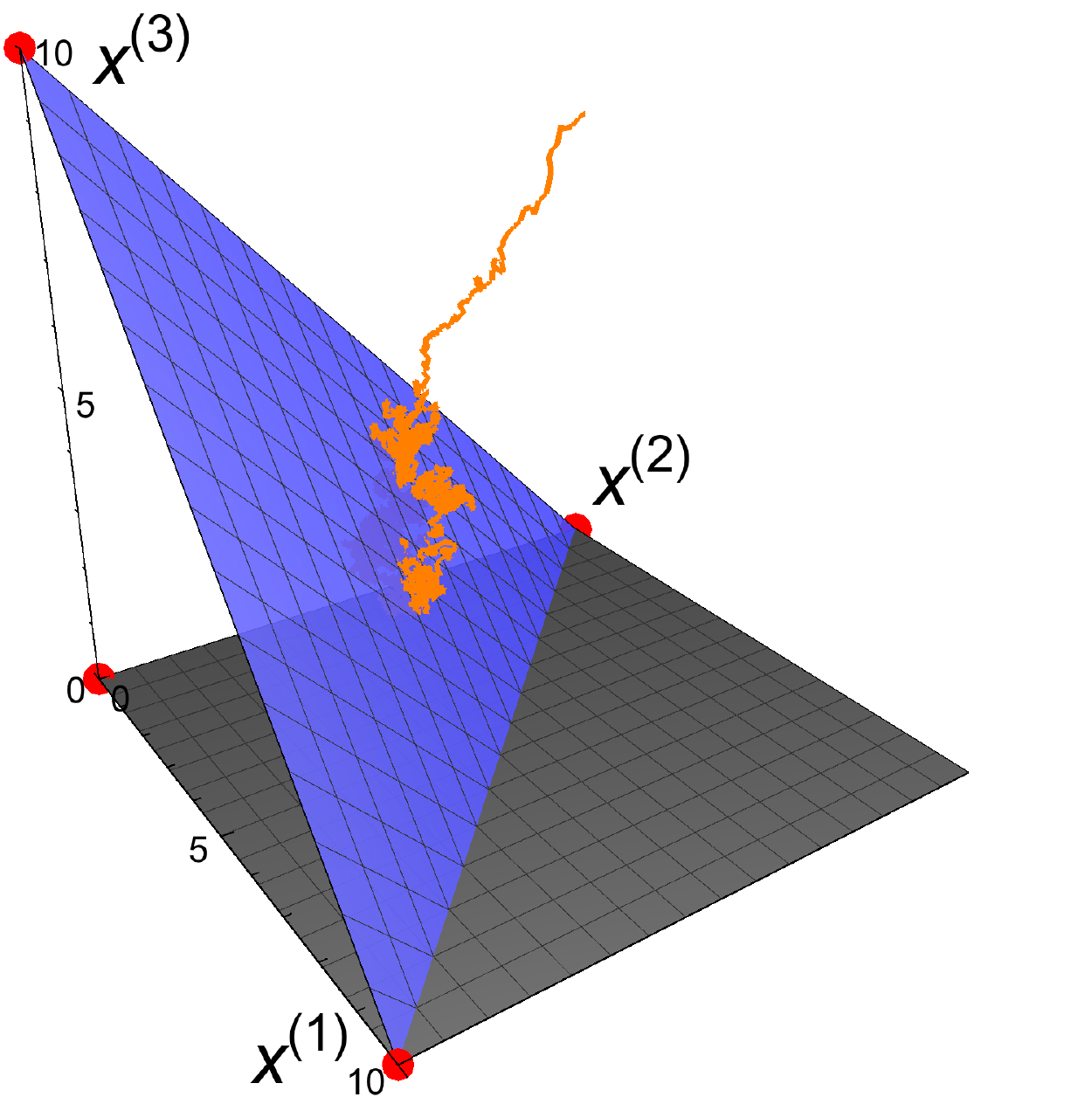}
\caption{Illustration of four systems featuring timescale separation that can be analysed with the methods reviewed in this paper. \textbf{Top left panel}: Phase plot for a haploid Moran model with two alleles on two islands with strong migration (described in \sref{sec:example}). The deterministic dynamics rapidly collapse to a slow subspace, indicated here by the blue dashed line. \textbf{Top right panel}: Deterministic trajectories (grey arrows) for a system similar to that in the top left panel but with three islands, addressed in \sref{sec:D_island_Moran}. Again, the deterministic dynamics rapidly collapse to a one-dimensional subspace indicated by the blue dashed line. \textbf{Bottom left panel}: Genotype frequencies as a function of time for a population genetic model, described in \sref{sec:heterogamety}. Stochastic trajectories (solid lines) initially rapidly relax on quasi-deterministic trajectories (inset) before reaching a one-dimensional slow subspace along which the system moves on a slower timescale. \textbf{Bottom Right panel}: A neutral three-species Lotka-Volterra model, addressed in \sref{sec:M_allele_SLVC}. Stochastic trajectories (orange) rapidly collapse along quasi-deterministic trajectories onto a two-dimensional slow subspace (blue surface), about which they are confined.}
\label{fig:figure_1}
\end{center}
\end{figure}

There may be several ways of reducing the complexity of the model, but here we will concentrate on one which is based on time-scale separation arguments. That is, we will seek to identify \textit{fast} modes which die away relatively quickly, and \textit{slow} modes which endure at long times. The dynamics of systems featuring such timescale separation are illustrated in Fig.~\ref{fig:figure_1}. This is, of course, a well-known procedure, perhaps the most famous example being in hydrodynamics, where the microscopic molecular dynamics can be replaced by a macroscopic dynamics with a few long-lasting variables. Although this dynamics is macroscopic, a mesoscopic extension can also be derived along the same lines~\cite{Fox1970}. In the theory of dynamical systems, the concept of a centre manifold (CM) is another manifestation of these ideas. During our discussion of this methodology, there will be several illustrations of the third and fourth points discussed above, namely the wish to use generic structures and the use of numerical simulations to check the precision of the approximations we utilise.

As we have already mentioned, one of our aims in writing this article is to make the ideas and techniques available to a larger audience. To help to achieve this we will present an application of the method in a pedagogical manner in Section 2. We have chosen one of the simplest possible systems: haploid individuals on one of two islands of equal size which can migrate from one island to the other. It will be assumed that are only two possible alleles which are modelled by a Moran process. After this informal, and hopefully easily accessible, introduction to the method, we will describe its application to a number of models in Section 3. These include: a haploid Moran model on an arbitrary number of islands with selection and mutation; a stochastic Lotka-Volterra competition model with an arbitrary number of islands and a stochastic Lotka-Volterra competition model with an arbitrary number of species; a derivation of the Hardy-Weinberg approximation from first principles; a model of epidemic spread on a network.  In Section 4, we will illustrate the method in the slightly more technical case where noise-induced dynamics are present. We will see that noise-induced selection can cause selection for genotypes that are neutral in a deterministic setting, and that further, this noise induced selection can, under certain conditions, be strong enough to reverse the direction of deterministic selection. We will illustrate this behaviour with reference to Lotka-Volterra competition models, where we will see that this effect can help alleviate the dilemma of cooperation, and a model of transitions between sex-chromosome systems. Finally, in Section 5 we conclude with a discussion.


\section{Pedagogical Example}
\label{sec:example}
In this section we will explain as simply as possible how to apply the ideas discussed in the Introduction to a concrete example. The example we choose is a Moran model with migration. We ask how this can be reduced to an effective one-island model.

\subsection{Setting-up the model}
\label{sec:setup}

We set up the model at the microscale, that is, at the level of haploid individuals who each carry an allele of type $1$ or of type $2$. The individuals can reside on one of two islands, both of which can only carry a fixed number of individuals, which we denote by $N$. We therefore denote by $n_1$ the number of individuals carrying allele $1$ on island $1$, by $N-n_1$ the number of individuals carrying allele $2$ on island $1$, by $n_2$ the number of individuals carrying allele $1$ on island $2$, and by $N-n_2$ the number of individuals carrying allele $2$ on island $2$. So the state of the whole system is given by only two variables, which we can form into the two-dimensional vector $\bm{n}=(n_1,n_2)$. We would like to reduce this description to one involving only one variable, which gives the fraction of individuals in the system carrying allele $1$. This would allow us to calculate, for example, the probability that allele $1$ or allele $2$ fix, and the mean time to fixation.

There is not enough information about the birth, death and migration of individuals to model them in any other way than as random processes, so the model is specified by giving the probability per unit time that the system transitions from its current state, given by the vector $\bm{n}$, to a new state $\bm{n}'$. We write these transition rates as $T(\bm{n}'|\bm{n})$, with the initial state on the right and the final state on the left (some authors use the reverse convention). The probability distribution function (pdf) of the system, $p(\bm{n},t)$, is then given by the master equation
\begin{equation}
\frac{\mathrm{d}p(\bm{n},t)}{\mathrm{d}t} = \sum_{\bm{n}' \neq \bm{n}}\left[ T(\bm{n}|\bm{n}')p(\bm{n}',t) - T(\bm{n}'|\bm{n})p(\bm{n},t)\right].
\label{master_generic}
\end{equation}
This is relatively easy to understand. The first term on the right-hand side is made up of the probability of being in state $\bm{n}'$ multiplied by the probability of being in that state and making a transition to state $\bm{n}$. It therefore represents the probability of starting in state $\bm{n}'$ and making a transition to state $\bm{n}$. In the same way the second term on the right-hand side represents the probability of starting in state $\bm{n}$ and making a transition to state $\bm{n}'$. Their difference, summed over all states $\bm{n}'$, different to $\bm{n}$, gives the rate of increase of $p(\bm{n},t)$ with time.

The form we take for the $T(\bm{n}'|\bm{n})$ depends on the model choice. Here we choose the Moran model, because it is simple: it amalgamates births and deaths and asks that birth, death and migration events happen in such a way that the population size of each island, $N$, is kept fixed. These are not the most realistic assumptions, and we discuss ways to relax them later in the paper, but they have the merit that the number of model parameters is kept to a minimum. The method of constructing $T(\bm{n}'|\bm{n})$ also seems a little more complicated, due to the requirement of keeping a fixed number of individuals on each island. This is done as follows: 
\begin{itemize}
\item[(i)] Pick an island (with probability $1/2$, since the islands are identical) and then pick an individual randomly from that island. Allow the individual to reproduce by duplicating the individual.
\item[(ii)] With probability $m$, the progeny migrates to the other island. In this case choose an individual on the other island at random to die.
\item[(iii)] With probability $(1-m)$, the progeny remains on the same island. In this case choose an individual on the same island at random to die.
\end{itemize}
It should be noticed that the processes of birth and death are inextricably linked and that they are assumed to happen at rate $1$, this choice being possible through a choice of time units. On top of this process, migration is imposed with a probability of occurrence equal to $m$ ($0 \leq m \leq 1$). 

Following these rules, if the model is neutral the transition rate from the state $(n_1,n_2)$ to the state $(n_1 + 1,n_2)$ is 
\begin{equation}
T(n_1 + 1,n_2|n_1,n_2) = \frac{1}{2}\,(1-m)\,\frac{n_1}{N}\frac{(N-n_1)}{N} + \frac{1}{2}\,m\,\frac{n_2}{N}\frac{(N-n_1)}{N}.
\label{neut_transition_rate_simple}
\end{equation}
Similar expressions can be be found for $T(n_1 - 1,n_2|n_1,n_2)$ and for $T(n_1, n_2 \pm 1|n_1,n_2)$. However, we would like to include selection in the model. In this case we have to weight the choice of picking an allele by the relative fitness of that allele on a particular island. Since we are aiming to be as simple as possible to illustrate the basic ideas, we will assume that this fitness weighting is the same on both islands, though it is simple enough to relax this condition. Therefore we will denote the fitness weighting of allele $1$ to be $W^{(1)}(\bm{n})$ and of allele $2$ to be $W^{(2)}(\bm{n})$. Then the four transition rates from state $(n_1,n_2)$ to the new state are
\begin{eqnarray}
T(n_1 + 1,n_2|n_1,n_2) &=& \frac{1}{2}\,\left( 1 - m \right)\,\frac{W^{(1)}(\bm{n})n_1}{\mathcal{W}_1(\bm{n})}\,\frac{(N-n_1)}{N} \nonumber \\
&+& \frac{1}{2}\,m\,\frac{W^{(1)}(\bm{n})n_2}{\mathcal{W}_2(\bm{n})}\,\frac{(N-n_1)}{N}, 
\nonumber \\
T(n_1 - 1,n_2|n_1,n_2) &=& \frac{1}{2}\,\left( 1 - m \right)\,\frac{W^{(2)}(\bm{n})(N-n_1)}{\mathcal{W}_1(\bm{n})}\,\frac{n_1}{N} \nonumber \\
&+& \frac{1}{2}\,m\,\frac{W^{(2)}(\bm{n})(N-n_2)}{\mathcal{W}_2(\bm{n})}\,\frac{n_1}{N}, 
\nonumber \\
T(n_1,n_2 + 1|n_1,n_2) &=& \frac{1}{2}\,\left( 1 - m \right)\,\frac{W^{(1)}(\bm{n})n_2}{\mathcal{W}_2(\bm{n})}\,\frac{(N-n_2)}{N} \nonumber \\
&+& \frac{1}{2}\,m\,\frac{W^{(1)}(\bm{n})n_1}{\mathcal{W}_1(\bm{n})}\,\frac{(N-n_2)}{N}, 
\nonumber \\
T(n_1,n_2 - 1|n_1,n_2) &=& \frac{1}{2}\,\left( 1 - m \right)\,\frac{W^{(2)}(\bm{n})(N-n_2)}{\mathcal{W}_2(\bm{n})}\,\frac{n_2}{N} \nonumber \\
&+& \frac{1}{2}\,m\,\frac{W^{(2)}(\bm{n})(N-n_1)}{\mathcal{W}_1(\bm{n})}\,\frac{n_2}{N}, 
\label{transition_rate_simple}
\end{eqnarray}
where $\mathcal{W}_i(\bm{n})= W^{(1)}(\bm{n})n_i + W^{(2)}(\bm{n})(N-n_i)$, $i=1,2$, is the fitness of the individuals on island $i$. Here superscripts are a label for the two different alleles, whereas subscripts are a label for the two different islands. For further background on how to arrive precisely at the forms given by Eq.~(\ref{transition_rate_simple}) the reader is referred to previous discussions in the literature~\cite{blythe_mckane_models_2007,mckane_2014,constable_phys}. 

If $W^{(1)}$ and $W^{(2)}$ are independent of $\bm{n}$, then the selection is known as frequency independent selection. This will be assumed in this pedagogical treatment, and we write $W^{(1)} = 1 + \alpha^{(1)}s + \mathcal{O}(s^2)$ and $W^{(2)} = 1 + \alpha^{(2)}s + \mathcal{O}(s^2)$, where $s$ is a selection coefficient and $\alpha^{(1)},\alpha^{(2)}$ are constants. Since $s$ is typically very small, we do not expect it will be necessary to include the order $s^2$ terms in the expressions for $W^{(1)}$ and $W^{(2)}$.

Equations (\ref{master_generic}) and (\ref{transition_rate_simple}) define the microscopic model, and once an initial condition $p(\bm{n},0)$ for the pdf has been given, specify the dynamics for all $t$. All other systems discussed in this paper will have a similar structure; the form of the transition rates will differ depending on the system, but in all cases the substitution of these rates into the master equation (\ref{master_generic}) will give the dynamics. As we discussed in the Introduction, the validity of our methods and approximations are checked via computer simulations, and these use the microscopic model defined by Eqs.~(\ref{master_generic}) and (\ref{transition_rate_simple}). The simulations use the Gillepsie algorithm~\cite{gillespie_1976,gillespie_1977} which is developed within the same formalism discussed in this section. However, the master equation is difficult to study analytically. It is for this reason that we make the diffusion approximation, replacing the microscopic model with a mesoscopic version.

The diffusion approximation was applied very early on in the development of population genetics~\cite{fisher_1922} and is widely used~\cite{crow_kimura_into}. We do insist, however, that it should be derived from an underlying microscopic model, since there are potentially many microscopic models that give the same mesoscopic model, and so simply defining the model at the mesoscale can lead to ambiguities. The idea itself is very simple: if $N$ is large enough, the ratios $n_i/N$, which are formally fractions, are assumed to be continuous, and denoted by $x_i$. At the same time the master equation is expanded in powers of $N^{-1}$, and terms in $N^{-3}$ and higher are neglected. 

This process can be carried out directly, but formulae exist for the analogues of the transition rates which appear in the mesoscopic equations~\cite{mckane_2014}. To use these we need to introduce what are in effect stoichiometric vectors corresponding to the four ``reactions'' in Eq.~(\ref{transition_rate_simple}). In other words we write the final state, $\bm{n}'$, in terms of the initial state, $\bm{n}$, as $\bm{n}' = \bm{n} + \bm{\nu}_\mu$, where $\mu=1,\ldots,4$ specify each of the four reactions. So for example, in the first reaction of Eq.~(\ref{transition_rate_simple}), $n_1$ increases by $1$ and $n_2$ does not change, so $\bm{\nu}_1 = (1,0)$. Similarly, $\bm{\nu}_2 = (-1,0), \bm{\nu}_3 = (0,1)$ and $\bm{\nu}_4 = (0,-1)$. These identifications allow us to use Eqs.~(18) and (21) of Ref.~\cite{mckane_2014} to show that the $\bm{A}$ and $\bm{B}$ functions which appear in the mesoscopic equations are
\begin{eqnarray}
A_1(\bm{x}) &=& -\frac{1}{2}m\left(x_1 - x_2\right) \nonumber \\
&+& \frac{1}{2}\left(\alpha^{(1)} - \alpha^{(2)}\right)s\left[ (1-m)x_1\left( 1 - x_1 \right) + m x_2\left( 1 - x_2 \right) \right] + \mathcal{O}\left( s^2 \right), \nonumber \\
A_2(\bm{x}) &=& -\frac{1}{2}m\left(x_2 - x_1\right) \nonumber \\
&+& \frac{1}{2}\left(\alpha^{(1)} - \alpha^{(2)}\right)s\left[ (1-m)x_2\left( 1 - x_2 \right) + m x_1\left( 1 - x_1 \right) \right] + \mathcal{O}\left( s^2 \right), \nonumber \\
\label{A_simple}
\end{eqnarray}
and 
\begin{eqnarray}
B_{11}(\bm{x}) &=& x_1\left(1-x_1\right) + \frac{1}{2}m\left(x_1 - x_2\right)\left(2x_1 - 1 \right) + \mathcal{O}(s), \nonumber \\
B_{22}(\bm{x}) &=& x_2\left(1-x_2\right) + \frac{1}{2}m\left(x_2 - x_1\right)\left(2x_2 - 1 \right) + \mathcal{O}(s),
\label{B_simple}
\end{eqnarray}
with $B_{12}=B_{21}=0$. These then specify the model, and the general dynamical equations which allow us to find the dynamics are either the Fokker-Planck equation (FPE)
\begin{equation}
\frac{\partial P(\bm{x},t)}{\partial t} =
- \frac{1}{N}\,\sum_{i=1}^2 \frac{\partial }{\partial x_i} 
\left[ A_i(\bm{x}) P(\bm{x},t) \right] + 
\frac{1}{2N^2} \sum_{i,j=1}^2 \frac{\partial^2 }{\partial x_i \partial x_j} \left[ B_{ij}(\bm{x}) P(\bm{x},t) \right], 
\label{FPE_simple}
\end{equation}
or the equivalent It\={o} stochastic differential equation (SDE) 
\begin{equation}
\frac{\mathrm{d}x_i}{\mathrm{d}\tau} = A_i(\bm{x}) + \frac{1}{\sqrt{N}} \eta_i(\tau), \ \ \ \ i=1,2,
\label{SDE_simple}
\end{equation}
where $\tau = t/N$ is a rescaled time and $\eta_i(\tau)$ is a Gaussian white noise with zero mean and with a correlator 
\begin{equation}
\left\langle \eta_i(\tau) \eta_j(\tau') \right\rangle = B_{i j}(\bm{x}) \delta\left( \tau - \tau' \right), \ \ \ i,j=1,2.
\label{correlator_simple}
\end{equation}
As for the microscopic model, substitution of the specific forms given by Eqs.~(\ref{A_simple}) and (\ref{B_simple}) into the generic forms of the FPE or SDE gives the behaviour of the mesoscopic model for all time, provided an initial condition $P(\bm{x},0)$ is given. For more details on derivation and meaning of these equations standard texts on the theory of stochastic processes~\cite{risken_1989,gardiner_2009} may be consulted, or previous articles on the application of these ideas in a biological context~\cite{blythe_mckane_models_2007,mckane_2014,constable_phys}. 

We end this section with two general points. First, if we take the limit $N \to \infty$ of Eq.~(\ref{SDE_simple}) we obtain the \textit{macroscopic} model, which is deterministic, since the noise is eliminated by taking the limit. The dynamics of the deterministic model is then given by $\mathrm{d}x_i/\mathrm{d}\tau = A_i(\bm{x})$, $i=1,2$. Second, we keep terms of order $s$ in $\bm{A}$, but neglect them in $\bm{B}$, since we are envisaging keeping terms of order $s/N$ or $1/N^2$ in the FPE, but discarding terms of order $s^2/N, s/N^2$ and $1/N^3$. This essentially assumes that $s \sim N^{-1}$, although we will keep $s$ and $N$ to be independent variables throughout the paper.

\subsection{First stage of the reduction process: identifying the fast and slow variables}
\label{sec:first_stage}

Although the mesoscopic equations (\ref{FPE_simple}) and (\ref{SDE_simple}) are potentially more manageable than the differential-difference equation (\ref{master_generic}), they are still formidably difficult to analyse --- equation (\ref{FPE_simple}) is a partial differential equation in three variables, and in most of the other systems discussed later in this paper, the corresponding equation may have tens or even hundreds of variables. To find a simpler, or reduced, form we want to eliminate the variables which decay away quickly, since they are not relevant to making predictions about the medium- to long-term behaviour. This subset of slow variables will form a slow-subspace (SS), so that instead of allowing the system to explore the whole space of variables, we only allow it to move within this subspace.

In practice, instead of searching for a SS directly, we frequently search for a CM which is composed not of slow-variables, which hardly change with time, but of conserved variables that do not change with time at all. In population genetics, for example, neutral theories may contain conserved quantities, due to symmetries in the system (the different alleles behave in the same way), and the effects of selection can be added as perturbative corrections, given the extremely small size of selection coefficients. The CM is found by looking for fixed points of the macroscopic equation (the macroscopic limit of the SDE with no noise term present).

This first stage of the reduction therefore consists of the following steps:
\begin{itemize}
\item[1.] Identify a CM, perhaps by setting some parameters to zero in order to increase the symmetry of the deterministic equations (this could be the neutral limit of the deterministic dynamics, for example).
\item[2.] Find the Jacobian at the fixed points that constitute the CM, and so find the eigenvalues and eigenvectors of the Jacobian evaluated on the CM. 
\item[3.] Form a projection operator from the eigenvectors found in 2, which is used to operate on quantities in the full mesoscopic system in order to eliminate the fast variables.
\item[4.] Use this projection operator, or use conservation laws, to find the point where the system first reaches the CM. This will be the new initial condition for the reduced system.
\end{itemize}

We will now illustrate these four steps on the pedagogical example.
\begin{itemize}
\item[(i)] Setting $s=0$ in Eq.~(\ref{A_simple}), we see that there is a line of fixed points $x_1 = x_2$. This is the CM. 
\item[(ii)] The Jacobian of the deterministic system $\mathrm{d}x_i/\mathrm{d}\tau = A_i(\bm{x})$, $i=1,2$, with $s=0$, is
\begin{equation}
J = \left( \begin{array}{cc} 
-m/2 & \ m/2 \\ 
\ m/2 & -m/2
\end{array}
\right)\,.
\label{Jacobian_simple}
\end{equation}
This matrix has zero determinant and a trace equal to $-m$, which immediately gives its two eigenvalues as $\lambda^{\{ 1\} }=0$ and $\lambda^{\{ 2\} }=-m$. Two typical features we would expect are illustrated here: the number of zero eigenvalues equals the number of dimensions of the CM (since there is no dynamics at all on the CM --- it is comprised only of fixed points) and the non-zero eigenvalue has a real part which is negative (so that it can be identified as the fast mode which dies away quickly). In this case the non-zero eigenvalue is real, which is a reflection of the fact that the Jacobian, $J$, is symmetric. Another consequence of $J$ being symmetric is that we would normally be required to find right- and left-eigenvectors of $J$, but these coincide for symmetric matrices and are given by
\begin{align}\label{eigenvectors_simple}
\bm{v}^{\{ 1\} } = \frac{1}{\sqrt{2}}\,\left( \begin{array}{c} 
1 \\
1 
\end{array} \right), 
\ \ \bm{v}^{\{ 2\} } = \frac{1}{\sqrt{2}}\,\left( \begin{array}{c}
1 \\
-1 
\end{array} \right).
\end{align}
We would expect that the eigenvector corresponding to the zero eigenvalue would lie in the CM, and indeed $\bm{v}^{\{ 1\} }$ lies on the line $x_1 = x_2$. The normalisation of the eigenvectors has been chosen so that they are orthonormal: $\sum^2_{i=1} v^{\{ \mu\} }_iv^{\{ \nu\} }_i = \delta_{\mu \nu}$, where $\mu,\nu=1,2$ and $\delta_{\mu \nu}$ is the Kronecker delta.
\item[(iii)] The projection operator is defined by $P_{i j} = v^{\{ 1\} }_i v^{\{ 1 \} }_j$, constructed only from the eigenvectors of the zero mode(s). To illustrate its use, we operate with it on a general vector of the full system given by $\phi_i = C_1 v^{\{ 1\} }_i + C_2 v^{\{ 2\} }_i$, where $C_1$ and $C_2$ are constants. Then $\sum^2_{i = 1} P_{i j} \phi_j = C_1 v^{\{ 1\} }_i$, that is, the term involving the fast mode(s) in $\phi_i$, $C_2 v^{\{ 2\} }_i$, has been wiped out using the orthogonality of the eigenvectors.
\item[(iv)] If the point at which the system begins is $x^{\rm IC}_i$, then we would expect it to reach the CM at $x^{\rm CMIC}_i = \sum^{2}_{j=1} P_{i j} x^{\rm IC}_j$, since only the slow (zero) modes will have survived by this time. Here $i=1,2$ and `IC' and `CMIC' stand for `initial condition' and `centre-manifold initial condition' respectively. Applying the projection operator one finds that $x^{\rm CMIC}_1 = x^{\rm CMIC}_2 = (x^{\rm IC}_1 + x^{\rm IC}_2)/2$. Another way to obtain this result is to use the conserved quantity which exists in this degenerate system. From Eq.~(\ref{A_simple}), one sees that $\mathrm{d}(x_1 + x_2)/\mathrm{d}\tau = 0$ when $s=0$. Therefore, $x_1 + x_2$ is unchanged in time, and so $x^{\rm IC}_1 + x^{\rm IC}_2 = x^{\rm CMIC}_1 + x^{\rm CMIC}_2 = 2 x^{\rm CMIC}_1$ or $2 x^{\rm CMIC}_2$. 
\end{itemize}

These results will be used to construct the reduced model. As an illustration of the fourth general point made in the Introduction relating to intuition and the checking of approximations through simulations, we note that the trajectory from $\bm{x}^{\rm IC}$ to $\bm{x}^{\rm CMIC}$ is stochastic, not deterministic. Nevertheless, we will use $\bm{x}^{\rm CMIC}$ as the initial condition for the reduced system on the CM, even though it has been deduced through a deterministic argument. We expect the deterministic dynamics to dominate the collapse from $\bm{x}^{\rm IC}$ to the CM under the condition that the rate of migration (which controls the collapse of the system to the CM) is much stronger than the rate of genetic drift (which causes deviations from the deterministic collapse to the CM). Since the rate of genetic drift grows linearly with the population size (see \eref{FPE_simple}), this condition can be expressed $m \gg N^{-1}$.

In addition, the projection of the IC onto the CM as described is only strictly true when trajectories to the CM are linear (as in this case) or when the initial condition is close to the CM (in which case the linear approximation is applicable). If the deterministic trajectories to the CM are non-linear, more sophisticated mathematical techniques may be necessary to calculate $\bm{x}^{\rm CMIC}$  when $\bm{x}^{\rm IC}$ lies far from the CM~\cite{roberts_1989}. We will also continue to use the eigenvectors of the neutral model when constructing the $s \neq 0$ reduced model. It would be possible to find perturbative corrections to these in $s$, but we expect the effects to be sufficiently small as to be completely negligible. These are judgements made on the basis of intuition. Their validity will be examined through numerical simulations, and a comparison made of analytic results found on the basis of these assumptions with results obtained by simulations of the original model.

Finally it is worth noting that, as addressed in point (iv) above, an alternative line of attack is possible in which one transforms into the fast-slow basis of the problem, removes the fast-variables and then transforms back into the original, biologically relevant, variables of the problem (for an illustration of such an approach, see \cite{constable_2013}). For the system at hand, the fast-slow basis is straightforward to obtain and is given by $(x_1 - x_2)$ and $(x_1 + x_2)$. However in more general problems such a basis may not be straightforward to obtain analytically (we will explore such a scenario in \sref{sec:heterogamety}) while the projection method that we develop here will continue to yield insight. We therefore continue to explore the current pedagological example using the projection formalism.

\subsection{Second stage of the reduction process: construction of the reduced model}
\label{sec:second_stage}

In the first stage we worked with a version of the model which had a CM, but our real interest is in the actual, realistic, model which will typically only have a SS. This will not change materially the process of collapse described in Sec.~\ref{sec:first_stage}: we still expect what is in effect a deterministic collapse onto the SS. However, we would like to be able to assume that the system would then stay in the subspace which is defined by the slow variables. This is true (at least deterministically) when there is a CM, since the system ceases to move once it reaches the CM (because $\bm{A}=\bm{0}$). But this is not true when there is a SS. We therefore demand that $\bm{A}$ has no component in the fast directions. These conditions give the equation for the SS. There will still be a (weak) dynamics in the direction of the slow variables. We will also ask that there is no noise in the fast directions, only in the slow directions. In this way, the system is effectively constrained to evolve in the SS. 

This second stage of the reduction therefore consists of the following steps:
\begin{itemize}
\item[1.] Ask that $\bm{A}$ has no components in the fast directions. This gives the equation that the SS must have for this to be true.
\item[2.] Apply the projection operator to the SDE of the full system, to obtain the SDE of the reduced system.
\end{itemize}

We can now illustrate these steps on the pedagogical example.
\begin{itemize}
\item[(i)] The condition that $\bm{A}$ has no component in the fast direction is $\bm{v}^{\{ 2 \} }\cdot \bm{A} = 0$, where $\bm{A}$ is the full ($s \neq 0$) form given by Eq.~(\ref{A_simple}). This condition is simply $A_1(\bm{x}) - A_2(\bm{x})=0$. Substituting $x_i = X_i^{(0)} + sX_i^{(1)} + \mathcal{O}(s^2)$ into this equation we can determine the SS. We know that when $s=0$ the SS is the CM and that $X_1^{(0)} = X_2^{(0)}$, however using $A_1(\bm{x}) - A_2(\bm{x})$ we find that it is also true that $X_1^{(1)} = X_2^{(1)}$, so the SS is defined by $x_1 = x_2$ to order $s$. It therefore has exactly the same form as the CM, and is linear. This is not true in general, and is only a feature of this simple pedagogical example. The variable $x_1$, or $x_2$, will be denoted by $z$ on the SS.
\item[(ii)] Applying the projection operator to the SDE (\ref{SDE_simple}) gives $( 2/\sqrt{2}) (\mathrm{d}z/\mathrm{d}\tau)$ for the left-hand side, since $x_1 = x_2\equiv z$ on the SS. The first term on the right-hand side gives $(2/\sqrt{2}) \bar{A}(z)$ for a similar reason: $A_1(\bm{x})=A_2(\bm{x})$ on the SS $x_1=x_2$, and we write $A_1$ or $A_2$ on the SS in terms of $z$ as $\bar{A}(z)$. Therefore the reduced SDE is given by
\begin{equation}
\frac{\mathrm{d}z}{\mathrm{d}\tau} = \bar{A}(z) + \frac{1}{\sqrt{N}} \zeta(\tau),
\label{reduced_SDE_simple}
\end{equation}
where 
\begin{equation}
\bar{A}(z) = \frac{1}{2}\left(\alpha^{(1)} - \alpha^{(2)}\right)s\,z\left( 1 - z \right) + \mathcal{O}\left( s^2 \right), 
\label{A_bar_simple}
\end{equation}
and where
\begin{equation}
\zeta(\tau) = \frac{1}{\sqrt{2}}\,\frac{1}{\sqrt{2}}\left[ v^{\{ 1\} }_1 \eta_1(\tau) +  v^{\{ 1\} }_2 \eta_2(\tau) \right] = \frac{1}{2\sqrt{2}} \left[ \eta_1(\tau) + \eta_2(\tau) \right].
\label{zeta_defn}
\end{equation}
From the properties of the noise $\bm{\eta}(\tau)$, including the correlation function in Eq.~(\ref{correlator_simple}), it follows that $\zeta(\tau)$ is a Gaussian white noise with zero mean and with a correlator 
\begin{equation}
\left\langle \zeta(\tau) \zeta(\tau') \right\rangle = \frac{1}{8} \left [B_{1 1} + B_{2 2} \right] \delta\left( \tau - \tau' \right) \equiv \bar{B}(z) \delta\left( \tau - \tau' \right).
\label{reduced_correlator_simple}
\end{equation}
where
\begin{equation}
\bar{B}(z) = \frac{1}{4}\,z\left( 1 - z \right) + \mathcal{O}\left( s \right). 
\label{B_bar_simple}
\end{equation}
\end{itemize}

The reduction we have described here has produced an effective mesoscopic model defined by Eqs.~(\ref{reduced_SDE_simple}), (\ref{A_bar_simple}), (\ref{reduced_correlator_simple}) and (\ref{B_bar_simple}), which has only one variable, and which is sufficiently simple that it can be analysed mathematically. The pedagogical example was chosen on the grounds that its simplicity allowed the method to be clearly explained, and unfortunately this means that the reduction does not give that much useful information. For instance, the final results do not depend on $m$, and all we have done is to verify a known result~\cite{maruyama1969} that with this type of selection and with the same selection pressure on all islands, the population behaves in a similar way to a well-mixed population of size equal to that of the two islands added together. Actually, if one takes the calculation to higher order in $s$, then one can find an exception to this: migration-selection balance can occur if selection acts in opposing directions on the different islands.

In later sections we will describe applications of the method which will yield informative reduced models. Although the essentials of the reduction method will be the same as those that we have described in this section, a few aspects will make it seem slightly more complicated. Apart from the number of variables being greater, requiring additional indices, we mention
\begin{itemize}
\item[(a)] The generalisation of the Jacobian (\ref{Jacobian_simple}) will typically not be a symmetric matrix. This means that the non-zero eigenvalues will be complex in general, and the left- and right-eigenvectors will not coincide. We will use the notation $\bm{u}^{\{ \mu\} }$ and $\bm{v}^{\{ \mu\} }$ respectively for the left- and right-eigenvectors corresponding to the eigenvalue $\lambda^{\{ \mu\} }$, where $\mu=1,2,\ldots$. They will be chosen to be orthonormal, that is, $\bm{u}^{\{ \mu\} }\cdot\bm{v}^{\{ \nu\} } = \delta_{\mu \nu}$.
\item[(b)] As a consequence of this a typical term in the projection operator will involve left- and right-eigenvectors and the condition for there to be no deterministic dynamics in the $\mu$-direction will be $\bm{u}^{\{ \mu \} }\cdot\bm{A}(\bm{x}) = 0$, that is, will involve $\bm{u}^{\{ \mu \} }$ rather than $\bm{v}^{\{ \mu \} }$. 
\end{itemize}
These are simply small technical details, and the method as discussed here is in essence that used in the other applications which we will now go on to discuss. However, having accounted for these points, we can define more general forms of $\bar{A}(z)$ and $\bar{B}(z)$ that will be relevant for later problems. In particular, for a system with initially $M$ variables we find an effective one-dimensional approximation of form \eref{reduced_SDE_simple} with;
 \begin{eqnarray}
\bar{A}(z) = \sum_{i=1}^M u^{ \{ 1\} }_i  \left. A_i(\bm{x}) \right|_{\mathrm{CM}}
\label{eq_general_ABar}
\end{eqnarray}and
\begin{eqnarray}
\bar{B}(z) &=& \sum_{i,j=1}^M u^{ \{ 1\} }_i u^{ \{ 1\} }_j B_{ij}(\bm{x}) |_{\mathrm{CM} } \,,
\label{eq_general_BBar}
\end{eqnarray}
and where we recall that  $\bm{u}^{ \{ 1\} }$ is the left-eigenvector corresponding to the zero eigenvalue. Since $\bm{u}^{ \{ 1\} }$ is perpendicular to all of the fast directions~\cite{constable_phys}, these two terms can be viewed as the deterministic and noisy components of the full problem respectively projected onto the SS; that is using $P_{ij} = v^{ \{ 1 \} }_i u^{ \{ 1 \} }_j $.

Finally, since one of the themes of this paper relates to the utilisation of techniques from theoretical physics to population genetics, and other areas of theoretical biology, we could import the use of the bra-ket notation from quantum mechanics~\cite{dirac_1958}. In this notation the right-eigenvector $\bm{v}^{\{ \nu \} }$ is written as the ket $| \nu \rangle$ and the left-eigenvector $\bm{u}^{\{ \mu \} }$ as the bra $\langle \mu |$. Then the orthogonality relation becomes $\langle \mu | \nu \rangle = \delta_{\mu \nu}$ and the projection operator $P_{ij} = v^{ \{ 1 \} }_i u^{ \{ 1 \} }_j $ may be written as $P = | 1 \rangle \langle 1 |$. Though undoubtedly more elegant, for consistency with earlier work we shall not use the bra-ket notation here.

\section{Applications of the fast-mode elimination procedure}
\label{sec:applications}
In this section we will discuss some applications of the formalism which we have described in Sec.~\ref{sec:example}, making reference to previous work, notable features of the various models and possible future work.

\subsection{The $\mathcal{D}$-island Moran model}
\label{sec:D_island_Moran}
The modelling and analysis of migration effects in population genetics has always been challenging since it involves a spatial aspect in an essential way. Historically, it was Wright~\cite{wright_1931} who first studied migration models in population genetics, however he in fact did not assume a spatial structure, since migratory individuals were chosen from a global, well-mixed, population. The stepping stone model~\cite{kimura_1964_SSM} was one of the first which did have real spatial structure. It consisted of a line of islands, but where migration could only take place between an island and its neighbours on either side. If we view migration as an interaction event between two islands, this is a one-dimensional model with nearest-neighbour interactions.

Expressed in this way, an obvious generalisation is to a network of islands with interaction strengths between the islands proportional to the probability of migration between the islands. Such a model was investigated by Nagylaki~\cite{nagylaki1980SM}, although with discrete generations and strong migration, where the probability of a migration event is of the same order as a birth or a death. Assumptions either made the analysis difficult to follow or were not thought to be widely applicable, and many further studies of this kind followed which made a different set of assumptions. These, and further discussions and analysis, can be found in the book by Rousset~\cite{rousset_2004}, while more recent results that utilise probability generating functions~\cite{reichl_1998} are developed in \cite{houch_2011}.

In Sec.~\ref{sec:example} a simple two island model was introduced. The $\mathcal{D}$-island model is a generalisation of this but with added features. Details are given in earlier papers of ours~\cite{constable_phys,constable_bio,constable2015c}, but examples of more general features are the migration probability to island $i$ from island $j$, $m_{ij}$, which will in general not be symmetric ($m_{ij} \neq m_{ji}$), and the fact that islands will be allowed to contain different number of individuals ($\beta_i N$ on island $i$). The migration matrix is not so far defined for $i = j$, however if one sets the probability of the chosen individual not migrating equal to $m_{ii} = 1 - \sum^{\mathcal{D}}_{j \neq i} m_{ji}$, then this defines $m_{ij}$ for all $i,j$.

The factor $N$ which appears in the popoulation size is the only large parameter in  the model, and the approximations made in Sec.~\ref{sec:example} are reliant on this. So, for example, $\beta_i$ should not be so small or large that we still cannot treat the factor $\beta_i N$ as being of a similar magnitude to $N$. Similarly, the number of islands $\mathcal{D}$ should not be so large that it can be thought of as being of order $N$. It is likely that in some cases the approximations will continue to be good outside of their strict range of validity, but at present this can only be tested by comparing the analytic results with simulations.

We will discuss the model with and without mutation separately, since typical questions which we are interested in answering differ. We begin with the model with no mutation.

\subsubsection{The $\mathcal{D}$-island model without mutation}
\label{sec:D_island_Moran_without}
This model has been analysed in detail in Refs.~\cite{constable_phys,constable_bio}, where further details may be found. Here we only note that, in addition to the points already made, that the probability of choosing an island on which an individual is then chosen to die or migrate, $f_i$, has to be done with a probability proportional to $\beta_i$, if we are to get sensible results. When the approximation described in Sec.~\ref{sec:example} are made, one again finds that the reduced model has only one degree of freedom, and is given by Eqs.~(\ref{reduced_SDE_simple}) and (\ref{reduced_correlator_simple}). Even the functional forms of $\bar{A}(z)$ and $\bar{B}(z)$ are unchanged, although now they contain parameters which are functions of most of the parameters of the starting model:
\begin{equation}
\bar{A}(z) = a_1 s\,z\left( 1 - z \right) + \mathcal{O}\left( s^2 \right), \ \ \ \bar{B}(z) = b_1 z\left( 1 - z \right) + \mathcal{O}\left( s \right), 
\label{A_and_B_bar_Dislands}
\end{equation}
where
\begin{equation}
a_1 = \sum^\mathcal{D}_{i,j=1} u^{\{ 1\} }_i \frac{m_{ij}f_j}{\beta_i} \alpha_j, \ \ \ 
b_1 = \sum^\mathcal{D}_{i,j=1} \left[ u^{\{ 1\} }_i \right]^2\frac{m_{ij}f_j}{\beta^2_i}.
\label{aone_bone}
\end{equation}
Here $\alpha_i$ is the relative fitness of the first allele over the second on island $i$ and $\bm{u}^{\{ 1 \} }$ is the left-eigenvector corresponding to the zero eigenvalue. So even with the added complexity of each island differing in size, arbitrary migration probabilities, selective advantage of allele $1$ over allele $2$ varying from island to island, and the number of islands itself arbitrary, the reduction gives a standard (i.e. non-spatial model) Moran model with selection with effective parameters which can be seen from Eq.~(\ref{aone_bone}) to contain information from virtually all the parameters of the original model. It should be noted that in order to prove results pertaining to the spectrum of the eigenvalues of the Jacobian, we assumed that the migration matrix had a structure which in effect meant that no subgroup of islands were isolated from any other~\cite{constable_phys}, and so the results displayed in Eq.~(\ref{aone_bone}) are subject to this restriction. This rules out, for example, the case where the islands may be divided into two subgroups, with no migration between the subgroups.


\begin{figure}[t]
 \includegraphics[width=0.45\textwidth]{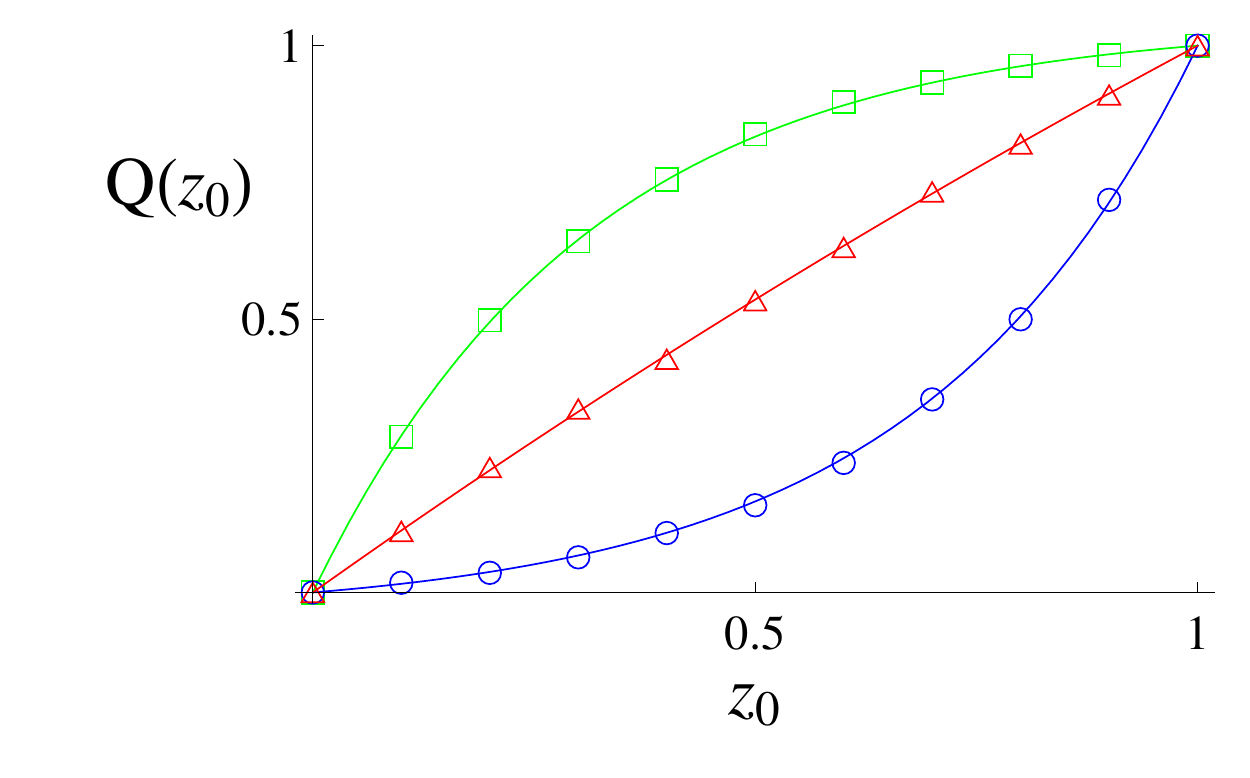}
 \includegraphics[width=0.45\textwidth]{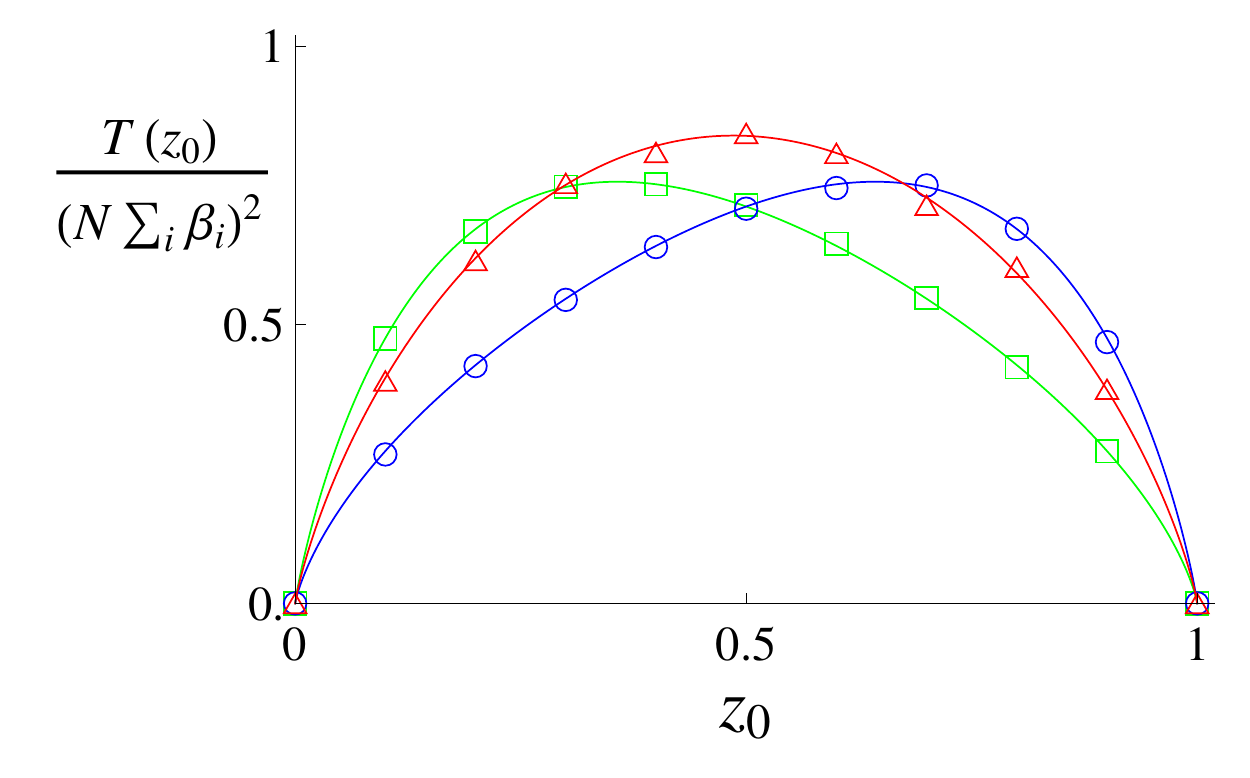}
\caption{Plots for the probability of fixation (left panel) and mean time to fixation (right panel) as a function of the projected initial conditions for the $\mathcal{D}=3$ island Moran model. The solid lines are calculated from the reduced model and the various symbols indicate the results obtained from simulations. For each different colour/symbol different $\bm{\alpha}$ vectors are used; green squares, $\bm{\alpha}=(1,1,-1)$, red triangles, $\bm{\alpha}=(-1,-1,1)$, and blue circles, $\bm{\alpha}=(1,-2,-1)$. All other parameters are kept constant; $s=0.005$, $N=200$, $\bm{\beta}=(3,2,1)$ and the migration matrix $m$ is fixed, though not given here. Simulation results are the average of $5000$ runs.}
\label{fig:Moran_Dislands}
\end{figure}


As mentioned the reduced model is the mesoscopic version of the Moran (or, in fact, the Wright-Fisher model) with selection. The probability of either allele $1$ or allele $2$ fixing for a given initial state and also the mean time to fixation may be straightforwardly found~\cite{crow_kimura_into}, given as they are by the solution to second-order ordinary differential equations~\cite{constable_bio}. These are denoted by $Q(z_0)$ and $T(z_0)$ respectively. Here $z_0$ is the initial condition for the reduced system, which was referred to as $\bm{x}^{\rm CMIC}$ in the final paragraph of Sec.~\ref{sec:first_stage} on the first stage of the reduction process. We have changed notation to the new coordinate $z$ and also indicated that it is an initial condition through use of the subscript $0$, rather than the superscript CMIC. It is clear that both $Q$ and $T$ have to depend on precisely where the reduced system is initialised.

In Fig.~\ref{fig:Moran_Dislands}, fixation times calculated from the reduced model and simulations of the underlying microscopic model, from which it was derived, are shown. The very good agreement between the two gives us confidence in the method, with more comparisons~\cite{constable_phys,constable_bio} also giving further support. The calculation can also be taken to next order in $s$, and an effective term of order $s^2$ found in the expression for $\bar{A}(z)$ given to first order in $s$ in Eq.~(\ref{A_and_B_bar_Dislands}). In this case $s$ is tentaively assumed to be of order $N^{-1/2}$ (although a direct indentification is not made), so that terms of order higher than $N^{-2}$ have been ignored in the expansion of the master equation. Novel effects can be investigated through use of the $s^2$ term, and a greater range of parameters explored. We refer the reader to the original literature for a discussion of these~\cite{constable_phys,constable_bio}.

\subsubsection{The $\mathcal{D}$-island model with mutation}
\label{sec:D_island_Moran_with}
So far in this article we have examined the effects of migration, selection and genetic drift, but not that other important process of population genetics: mutation. The inclusion of mutation has a drastic effect on the long term behaviour of the system, since it is now possible in principle for one allele to mutate to another at any time, and so concepts such as fixation probabilities and mean time to fixation do not apply. On the other hand, the pdf of the allele frequencies is now non-trivial and becomes stationary at long times. This is an interesting quantity which characterises the model and we use it to test the accuracy of the effective model found through the reduction procedure.

The microscopic model is constructed in a similar way to that described in Sec.~\ref{sec:example}. There is more than one way that mutation can be included; here we follow the procedure described in Ref.~\cite{constable2015c}. In this case we allow birth/death/migration events to happen a fraction $\xi$ of the time, and mutation events a fraction $(1-\xi)$ of the time. The mutation rate from the first allele to the second on island $i$ is denoted by $\kappa^{(1)}_i$ and from the second to the first on island $i$ by $\kappa^{(2)}_i$. It may be a little unusual to allow rates to vary from one island to the other, but allowing for this does not markedly increase the complexity of the calculation and so we include it.

If we denote the rates without mutation (such as are shown in Eq.~(\ref{transition_rate_simple}) for the case of two islands) as $T_{\rm S}(\bm{n}'|\bm{n})$, where the subscript S indicates that selection has been included, but not mutation, then the corresponding rates with mutation and selection included are
\begin{eqnarray}
T_{\rm MS}(n_i + 1|n_i) &=& \xi T_{\rm S}(n_i + 1|n_i) + \left( 1 - \xi \right) \kappa^{(1)}_i \frac{(\beta_i N - n_i)}{\beta_i N}, \nonumber \\
T_{\rm MS}(n_i - 1|n_i) &=& \xi T_{\rm S}(n_i - 1|n_i) + \left( 1 - \xi \right) \kappa^{(2)}_i \frac{n_i}{\beta_i N}.
\label{rates_with_mutation}
\end{eqnarray}
Here the dependence of the transition rates, $T_{\rm S}(\bm{n}'|\bm{n})$, on elements of $\bm{n}$ that do not change in the transition has been suppressed. We may rescale time in the master equation by a factor of $\xi$ and absorb a factor of $(1-\xi)/\xi$ in the mutation rates. This effectively means that we can drop the factors $\xi$ and $(1-\xi)$ from Eq.~(\ref{rates_with_mutation}). 

Making the diffusion approximation, the mesoscopic model is given by Eq.~(\ref{SDE_simple}) with the noise correlator given by Eq.~(\ref{correlator_simple}). In this case 
\begin{eqnarray}
\left. A_i(\bm{x}) \right|_{\rm MS} &=&  \left. A_i(\bm{x}) \right|_{\rm S} + \frac{1}{\beta_i} \left[ \kappa^{(1)}_i - \left( \kappa^{(1)}_i + \kappa^{(2)}_i \right) x_i \right], \nonumber \\
\left. B_{i j}(\bm{x}) \right|_{\rm MS} &=&  \left. B_{i j }(\bm{x}) \right|_{\rm S} + \mathcal{O}\left( \bm{\kappa}^{(1)}, \bm{\kappa}^{(2)} \right),
\label{A_B_mut}
\end{eqnarray}
where $\bm{\kappa}^{(1)} = (\kappa^{(1)}_1,\ldots,\kappa^{(1)}_\mathcal{D})$ and $\bm{\kappa}^{(2)} = (\kappa^{(2)}_1,\ldots,\kappa^{(2)}_\mathcal{D})$. Since mutation has been modelled as a linear process, the $\kappa$ dependence in $A_i(\bm{x})$ in Eq.~(\ref{A_B_mut}) is exact. We will neglect the $\kappa$ dependence in $B_{ij}(\bm{x})$ for precisely the same reasons that we neglected the dependence on the selection coefficients: we are assuming that the elements of $\bm{\kappa}^{(1)}$ and $\bm{\kappa}^{(2)}$ are so small that they can be thought to be of the same order as $N^{-1}$. We therefore only keep terms of order $\bm{\kappa}^{(1)}/N$, $\bm{\kappa}^{(2)}/N$ and $1/N^2$ in the FPE.

The reduction process itself is similar to that discussed previously since, as just mentioned,  mutation rates are generally very small, and so they can be treated as perturbations of the neutral model in exactly the same way as was done for selection strengths. Therefore the reduced model is given by Eqs.~(\ref{SDE_simple}) and (\ref{correlator_simple}) with $\bar{B}(z)$ unchanged from the form given in Eq.~(\ref{A_and_B_bar_Dislands}). Perhaps not surprisingly $\bar{A}(z)$ is modified by the addition of an extra term depending on the mutation rates~\cite{constable2015c}:
\begin{equation}
\bar{A}_{\rm MS}(z) = \bar{A}_{\rm S}(z) + \hat{\kappa}^{(1)} - \left(\hat{\kappa}^{(1)} + \hat{\kappa}^{(2)}\right) z =  a_1 s\,z\left( 1 - z \right) + \hat{\kappa}^{(1)} - \left(\hat{\kappa}^{(1)} + \hat{\kappa}^{(2)}\right) z,
\label{A_bar_with_mut}
\end{equation}
where 
\begin{equation}
\hat{\kappa}^{(1)} = \sum^{\mathcal{D}}_{i=1} \frac{u^{\{ 1 \} }_i\kappa^{(1)}_i}{\beta_i}, \quad \hat{\kappa}^{(2)} = \sum^{\mathcal{D}}_{i=1} \frac{u^{\{ 1 \} }_i\kappa^{(2)}_i}{\beta_i}.
\label{kappa_hats}
\end{equation}
and where $a_1$ is defined in Eq.~(\ref{aone_bone}). Here, as in Sec.~\ref{sec:D_island_Moran_without}, we retain terms only to order $s$.

\begin{figure}[t]
\includegraphics[width=0.70\textwidth]{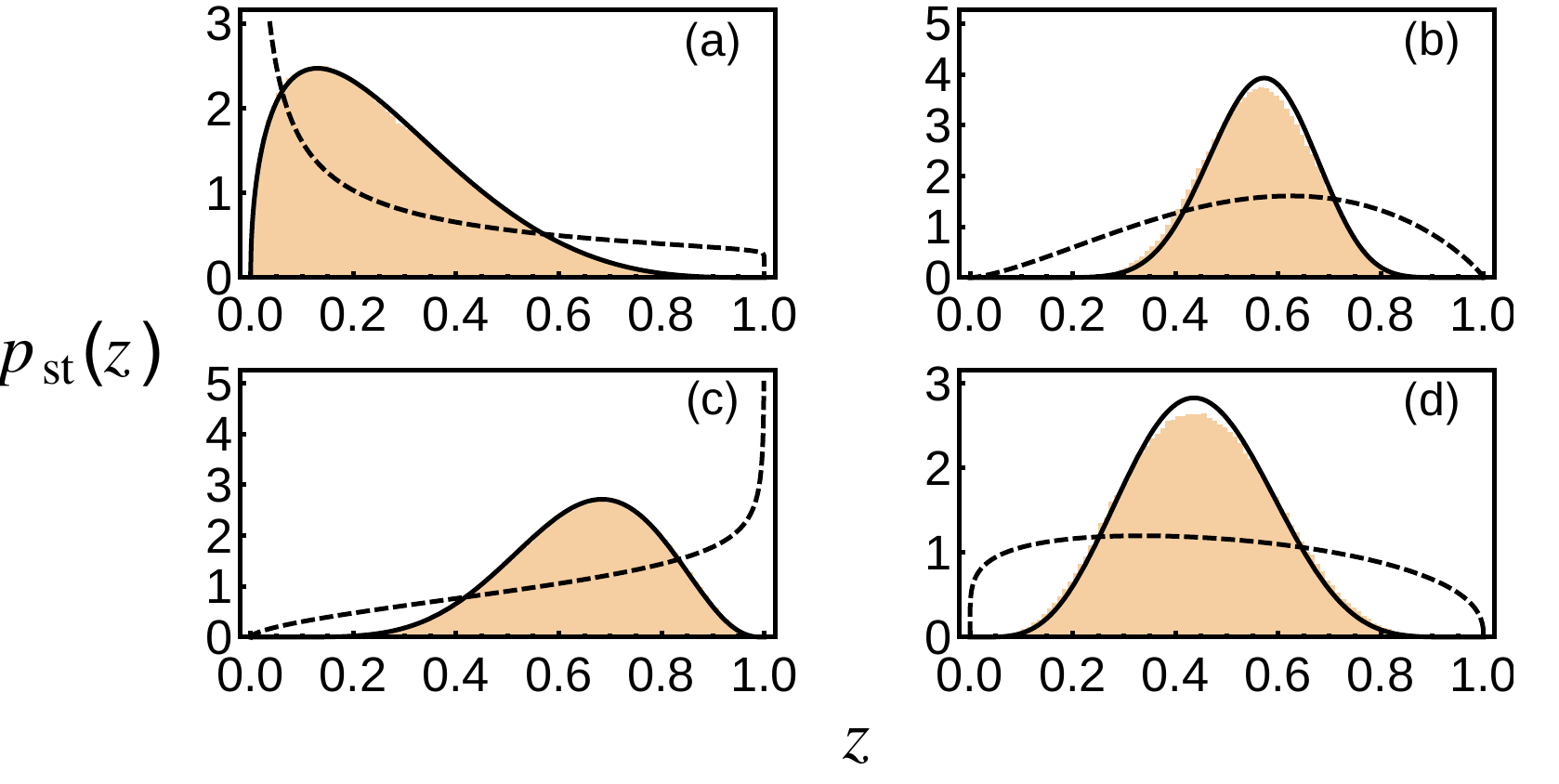}
\caption{The stationary pdf for the $\mathcal{D}$-island Moran model on the slow subspace for a range of systems with various parameters which are omitted here for brevity but which can be found in Ref.~\cite{constable2015c}. The solid black line is obtained from an analysis of the reduced model, the orange histogram from simulations of the original microscopic model, and the dashed line from a well mixed model with the same total system size and average mutation rates (weighted by island size).}
\label{fig:Moran_mutation}
\end{figure}

The stationary pdf of the effective theory can be straightforwardly found from the FPE corresponding to the SDE (\ref{SDE_simple}). Using the explicit forms for $\bar{A}(z)$ and $\bar{B}(z)$ one finds that~\cite{constable2015c}
\begin{equation}
p_{\rm st}(z) = \mathcal{N} z^{c_1} \left( 1 - z \right)^{c_2}\exp{ (c_3 z)},
\label{P_stat_mut}
\end{equation}
where $\mathcal{N}$ is a normalisation constant and 
\begin{equation}
c_1 = \frac{N}{b_1}\hat{\kappa}^{(1)} - 1, \quad c_2 = \frac{N}{b_1}\hat{\kappa}^{(2)} - 1, \quad c_3 = a_1 s \frac{N}{b_1}.
\label{c_1_c_2_c_3}
\end{equation}
A comparison between simulations of the original model and calculations from the reduced model in Fig.~\ref{fig:Moran_mutation} shows that the reduced model captures well the features of the full model.


There are several ways in which the work discussed here and in the previous section could be taken forward. There is scope for biologists to tailor the technique to their own interests, perhaps including additional processes with their own set of parameters and dropping others. Mathematicians may be able to provide conditions under which the approximations made would be expected to be valid, perhaps giving upper bounds for the negative real parts of the non-zero eigenvalues. Another extension is to perform a similar analysis, but starting not from the Moran model, but from one which is closer to those used in ecological modelling. We now go on to discuss this.

\subsection{The stochastic Lotka-Volterra competition model}
\label{sec:SLVC}
The theory of evolution has had a very convoluted history, and a reflection of this was the significant contribution made to the subject by theoretical studies, at least compared to other areas of the biological sciences. This had repercussions on the nature of the mathematical models used in these studies: they tended to be unrelated to broader questions relating to the organism, and more focussed on the combinatorics of allele selection. This was a good strategy when trying to test the ideas of Darwinian evolution, but it tended to isolate the theoretical development of the subject from developments elsewhere. 

An example is the Wright-Fisher model~\cite{wright_1931,fisher_1930}, one of the first models of genetic drift, as well as the precursor of the Moran model~\cite{moran_1957}. In this model there is no competition between individuals --- a trait which is obviously a feature of Darwinian evolution. The population therefore grows very quickly, but is kept under control by sampling from the very large pool of individuals that come into existence, in order to form the next generation. Only a fixed number, $N$, of individuals are retained to form each new generation (Wright-Fisher) or births and deaths are coupled so the population at any given time is always equal to $N$ (Moran). This leads to an artificiality in the way that the models are set-up.

In this section we will use as a starting point models in which the population is regulated by competition, rather than by a fictitious constraint which fixes the population size. This is a closer reflection of reality, and indeed the formulation of aspects of the models seem less contrived. It is a constant theme in the biological modelling literature that models of evolution should have a more ecological flavour, and this approach conforms to these views. An apparent disadvantage is that the number of variables is increased. To see that, recall that the Moran model of Sec.~\ref{sec:D_island_Moran} had only $\mathcal{D}$ variables, since the number of individuals carrying the second allele on island $i$ could be expressed in terms of the number of individuals carrying the first allele on the same island \textit{i.e.}~$N-n_i$. If the population of an island is not fixed, the number of individuals carrying the first and second allele can independently vary, and so the number of variables will double. Below we will describe how application of the reduction methods to models with competition show that they reduce to Moran-like models in the medium-to-long term~\cite{constable_2015,constable_2017}. The competition will be chosen to be of the simple Lotka-Volterra type~\cite{roughgarden_1979}, but in principle more complex competitive processes could be utilised. Since these are stochastic models, we will refer to them as stochastic Lotka-Volterra competition (SLVC) models.

\subsubsection{The $\mathcal{D}$-island SLVC model}
\label{sec:D_island_SLVC}
As indicated above this system has $2\mathcal{D}$ variables which in the microscopic model are $n^{(\alpha)}_i$, where $i=1,\ldots,\mathcal{D}$ labels the islands and $\alpha=1,2$ labels the allele. The comment made above concerning SLVC models being less contrived can be illustrated here in the way that migration is modelled. The procedure for doing this in the Moran model involves considerable care in making sure that there are no biases built into the way the transition rates are constructed (see Eq.~(\ref{transition_rate_simple}) in the simple two-island case) while keeping the population of both islands fixed. For example, one has to ensure that a death on island $j$ occurs before a migration from island $i$ to this island is allowed. In the SLVC model, one simply specifies birth, death and competition rates, respectively $b^{(\alpha)}_i, d^{(\alpha)}_i$ and $c^{(\alpha \beta)}_i$, which are all independent of each other. For the neutral version of the model, the birth, death and competition rates are the same for all alleles (and are denoted by a superscript $0$); selection is introduced through a small perturbation in $\epsilon$, where $\epsilon$ is the selection strength: 
\begin{equation}
b^{(\alpha)}_i = b^{(0)}_i \left( 1 + \epsilon \hat{b}^{(\alpha)}_{i} \right),
\ \ \ d^{(\alpha)}_i = d^{(0)}_i \left( 1 + \epsilon \hat{d}^{(\alpha)}_{i} \right), \ \ \ c^{(\alpha \beta)}_{i} = c^{(0)}_i \left( 1 + \epsilon \hat{c}^{(\alpha \beta)}_i \right).
\label{parameters_non_neut}
\end{equation}

The diffusion approximation is applied in the same way as in the Moran model, although now the large parameter is not $N$, which is no longer present in the definition of the model, but $V$, which is some measure of the size of the system, such as the volume. Although there are $2\mathcal{D}$ variables initially, $2\mathcal{D}-1$ of these are fast, and so the reduced model has again only one variable. It may be possible in some parameter regimes to see a clear cut decay first to the $\mathcal{D}$ variables of a Moran-type model with fixed populations on each island, and then a slower decay to an effective one-island model, which parallels the discussion in Sec.~\ref{sec:D_island_Moran}, but in many cases these time-scales will be similar or will overlap. The time-scales are related to the inverse of the eigenvalues of the Jacobian and, in general, these are complicated functions of all the model parameters.

While it is true that the reduced SLVC $\mathcal{D}$-island model does reduce to a system which has a mesoscopic description given by Eqs.~(\ref{reduced_SDE_simple}) and (\ref{reduced_correlator_simple}) --- although with $N$ replaced by $V$ --- and with $\bar{B}(z)=b_1 z(1-z)$ to leading order in $\epsilon$, the form of $\bar{A}(z)$ is a little different. It is found to be given by~\cite{parrarojas_2017}
\begin{equation}
\bar{A}(z) = \epsilon\,z\left( 1 - z \right)\left( a_1 - a_2 z \right) + \mathcal{O}\left( \epsilon^2 \right).
\label{A_bar_Dislands}
\end{equation}
In this case $a_1, a_2$ and $b_1$ are functions of the rates given which appear in Eq.~(\ref{parameters_non_neut}), $\beta_i$, and the left-eigenvector of the Jacobian corresponding to the zero eigenvalue. This change in the form of $\bar{A}(z)$ may be slight, but it could give significantly different fixation probabilities and mean time to fixation. One reason for this is that it is now possible to have a fixed point in the deterministic dynamics. This dynamics will be given by Eq.~(\ref{reduced_SDE_simple}) without the noise term, and so fixed points are solutions of $\bar{A}(z)=0$. When $\bar{A}(z)$ has the structure shown in Eq.~(\ref{A_bar_Dislands}), an internal fixed point (\textit{i.e.}~one not at the boundaries $z=0$ or $z=1$) is possible if $a_2 \neq 0$: $z^* = a_1/a_2$, where the asterisk denotes a fixed point. It will only exist if $0 < a_1/a_2 < 1$, but if it is stable, it may prolong the time taken for the system to fix (reach the points $z=0$ or $z=1$). Similarly if it is unstable, it may lead to a shorter mean time to fixation.

\begin{figure}[t]
\centering
\includegraphics[width=0.45\textwidth]{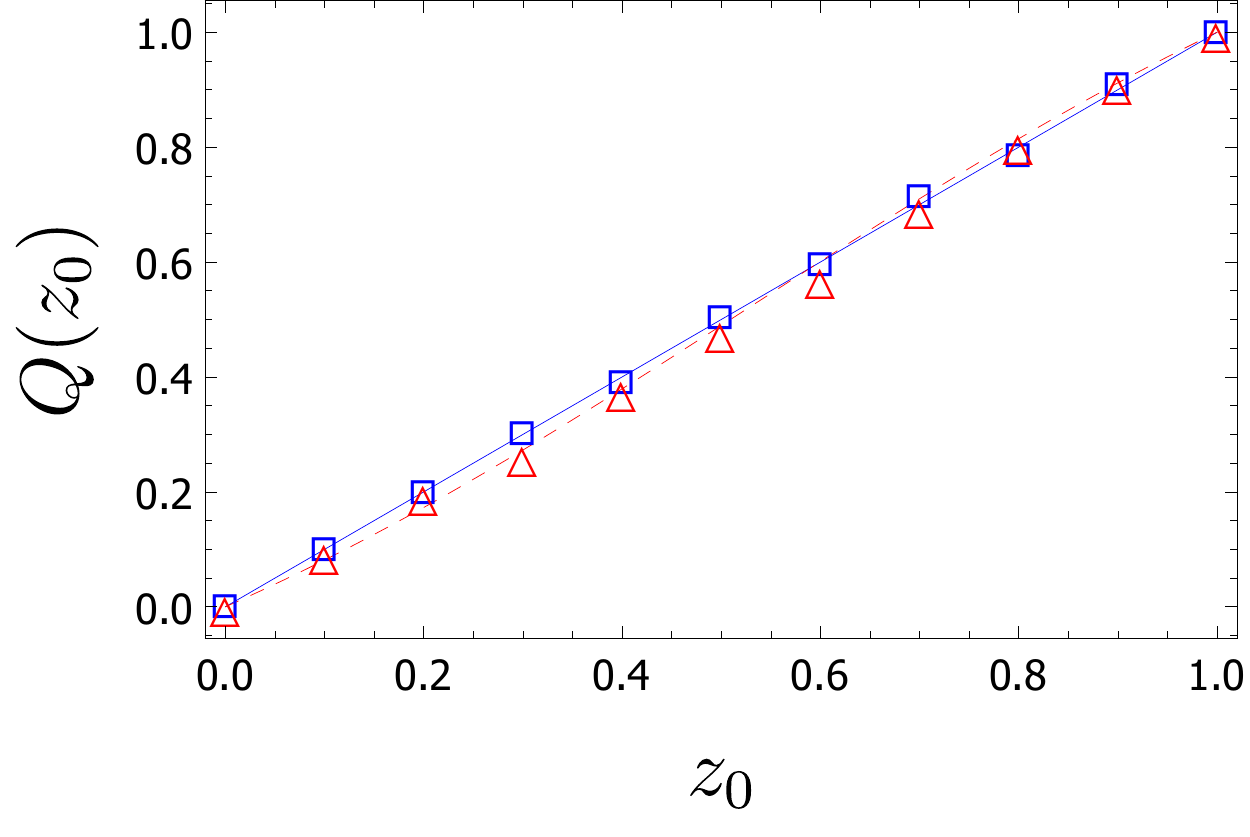}
\includegraphics[width=0.45\textwidth]{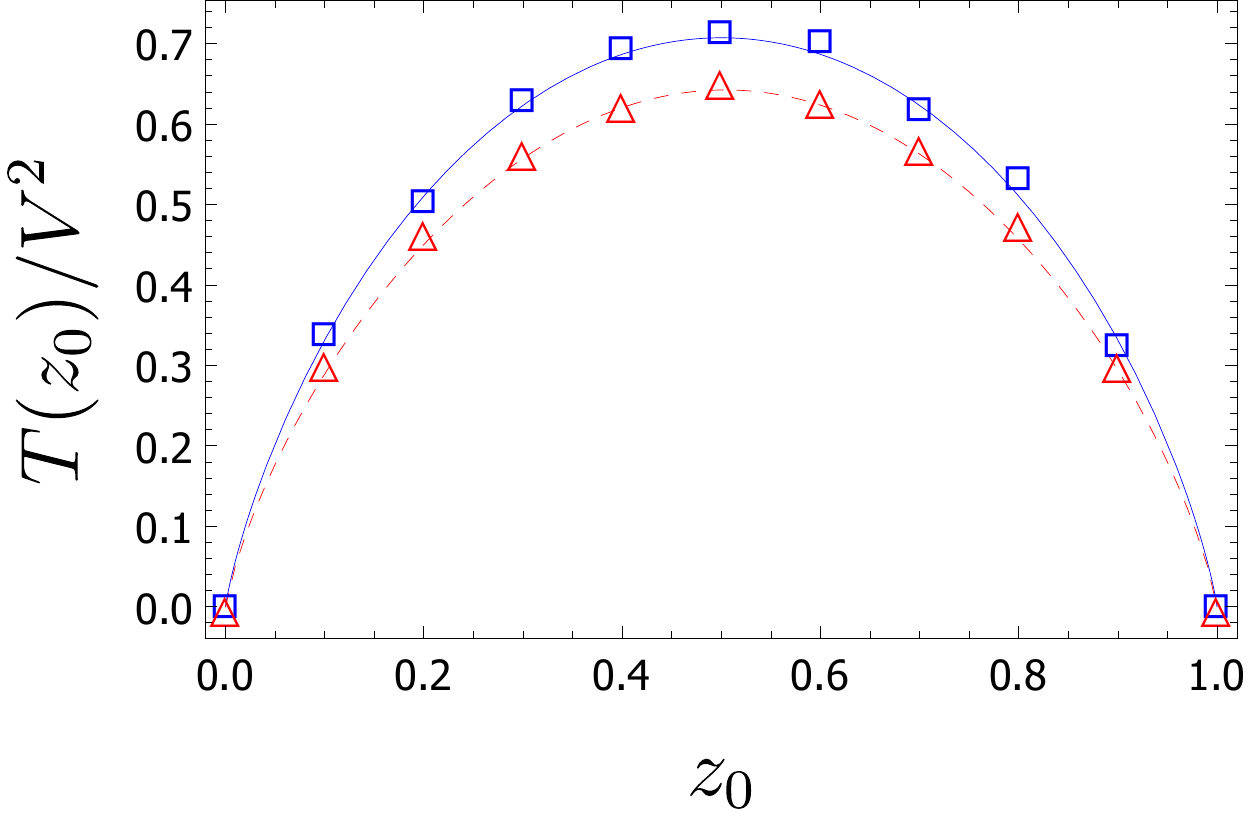}
\caption{Fixation probability of allele $1$ (left panel) and mean unconditional time to fixation (right panel) as a function of the projected initial condition $z_0$ for an SLVC model with $\mathcal{D}=4$ islands in the neutral case (blue) and in the case with selection (red, $\epsilon=0.05$). Symbols: mean obtained from 3000 stochastic simulations of the microscopic system; lines: theoretical predictions for the fixation probability and mean time to fixation obtained from the reduced model. Here the parameter $V$ is equal to $150$.}
\label{fig:SLVC_Dislands}
\end{figure}

To test the approximation we again compare the fixation probabilities and mean fixation time derived from the reduced model and those found from simulations of the original model. Although the form of $\bar{A}(z)$ is slightly more complicated than before, it is nevertheless still straightforward to work with the ordinary differential equations for $Q(z_0)$ and $T(z_0)$. The results shown in Fig.~\ref{fig:SLVC_Dislands} indicate that the reduction method is again working well in this case.


\subsubsection{The $M$-allele SLVC model}
\label{sec:M_allele_SLVC}
So far we have only discussed individuals which are haploid and carry one of two possible alleles. Here we discuss the generalisation to $M$ alleles. This is interesting for a number of reasons, not least because we are able to recognise patterns that are not apparent in the two allele case. As in Sec.~\ref{sec:D_island_SLVC}, the variables in the microscopic model are $n^{(\alpha)}$, where now $\alpha=1,\ldots,M$ and there is no island label, because we will assume that the population is well-mixed. In the corresponding haploid multiallelic Moran model there are only $M-1$ variables $n^{(a)}$, $a=1,\ldots,M-1$, since the fixed population constraint $\sum^M_{\alpha=1} n^{(\alpha)} = N$, means that $n^{(M)}$ can be expressed in terms of the other $M-1$ variables: $n^{(M)}= N - \sum^{M-1}_{a=1} n^{(a)}$. Here the Greek indices $\alpha, \beta,\ldots$ will always run from $1$ to $M$ and the Roman indices $a, b,\ldots$ will always run from $1$ to $M-1$.
 
Previously we used the reduction method to obtain an effective model which was amenable to analysis. Here will have a different perspective: we will ask if we can perform a reduction on the multiallelic SLVC model with $M$ variables to the multiallelic Moran model with $M-1$ variables. If this is so, then the more natural SLVC model will give the same results as the Moran model at medium and long times. From a mathematical point of view, the difference between the reduction described in Sec.~\ref{sec:D_island_Moran} and in Sec.~\ref{sec:D_island_SLVC} is that previously there were $\mathcal{D}-1$ or $2\mathcal{D}-1$ fast modes, and a single slow mode, whereas here there is one fast mode and $M-1$ slow modes, thus giving an effective model which is $(M-1)$ dimensional~\cite{constable_2017}. 

The reduction procedure has been carried out in Ref.~\cite{constable_2017}. An equation analogous to Eq.~(\ref{reduced_SDE_simple}) is found, but now in $(M-1)$ variables:
\begin{equation}
\frac{\mathrm{d}z^{(a)}}{\mathrm{d}\tau} = A^{(a)}(\underline{z}) + \frac{1}{\sqrt{V}}\,\zeta^{(a)}(\tau), \ \ a=1,\ldots,M-1,
\label{SDE_reduced_Mallele}
\end{equation}
where $\zeta^{(a)}(\tau)$ is a Gaussian noise with zero mean and with a correlator
\begin{equation}
\left\langle \zeta^{(a)}(\tau) \zeta^{(b)}(\tau') \right\rangle 
= \frac{2 b^{(0)} c^{(0)}}{\left( b^{(0)} - d^{(0)} \right)^2}\,\left[ z^{(a)} \delta_{a b} - z^{(a)} z^{(b)} \right]\,\delta \left( \tau - \tau' \right).
\label{reduced_correlator_Mallele}
\end{equation}
Here, the notation of underlining a vector is used for $(M-1)$-dimensional vectors and bold for $M$-dimensional vectors. The correlation function is given to zeroth order in $\epsilon$, which is the neutral model result. It is exactly the form found in the $M$-allele Moran model, up to some rescalings which are absorbed into the new time $\tau$~\cite{constable_2017}. In the neutral case $\underline{A}=\underline{0}$, and the SLVC model reduces exactly to the Moran model at medium to long times, after rescaling the time.

If selection is included, $\underline{A}$ is no longer equal to zero. In order to aid the comparison with the Moran model, it is useful to introduce two quantities:
\begin{equation}
\hat{C}^{(a b)} \equiv \hat{c}^{(a b)} - \hat{c}^{(a M)} - \hat{c}^{(M b)} + \hat{c}^{(M M)},
\label{C_hat_defn}
\end{equation}  
and
\begin{equation}
\Phi^{(\alpha)} \equiv \frac{b^{(0)} \hat{b}^{(\alpha)} - d^{(0)} \hat{d}^{(\alpha)}}{b^{(0)} - d^{(0)}}.
\label{Phi_defn}
\end{equation}
Then $A_a(\underline{z})$ takes the form 
\begin{eqnarray}
A^{(a)}(\underline{z}) &=& \epsilon z^{(a)} \left\{ \left[ \left( \Phi^{(a)} - \hat{c}^{(a M)} \right) - \left( \Phi^{(M)} - \hat{c}^{(M M)} \right) \right] \right. \nonumber \\
&-& \sum^{M-1}_{b=1} \left[ \left( \Phi^{(b)} - \hat{c}^{(b M)} \right) - \left( \Phi^{(M)} - \hat{c}^{(M M)} \right) \right] z^{(b)} \nonumber \\
&-& \left. \sum^{M-1}_{b = 1} \hat{C}^{(a b)} z^{(b)} + \sum^{M-1}_{b,c = 1} \hat{C}^{(b c)} z^{(b)} z^{(c)} \right\} + \mathcal{O}(\epsilon^2).
\label{A_tilde_Mallele}
\end{eqnarray}
For now, we note that, just as in Eq.~(\ref{A_bar_Dislands}), $\underline{A}$ is cubic in the components of $\underline{z}$. This is an important point in mappings between reduced SLVC and Moran models, as we will now see.

We begin our discussion with the $M$-allele Moran model in the case where the selection is frequency independent, that is, when the weight functions $W^{(\alpha)}$, analogous to those introduced in Eq.~(\ref{transition_rate_simple}), are independent of $\underline{n}$. Specifically we assume that $W^{(\alpha)}=1 + \rho^{(\alpha)}s$, where the $\rho^{(\alpha)}$ are constants. This is the case that we have examined so far in this paper. Then we find, after making the diffusion approximation, a model given by Eqs.~(\ref{SDE_reduced_Mallele}), (\ref{reduced_correlator_Mallele}) and (\ref{A_tilde_Mallele}), but only if $\hat{C}^{(a b)}=0$ for all $a$ and $b$~\cite{constable_2017}. If this condition holds, the reduced SLVC model and the Moran model with frequrncy independent selection match, provided that we make the identification $\rho^{(\alpha)} = \Phi^{(\alpha)} - \hat{c}^{(\alpha M)}$. We also need to match up the selection strength used in the SLVC model ($\epsilon$) to the one used in the Moran model ($s$). The relation between them is: $s=\epsilon (b^{(0)}-d^{(0)})/b^{(0)}$. Although some care has to be taken with making the identification between the two models~\cite{constable_2017}, one can note that the function $\underline{A}$ in the Moran model with frequency dependent selection is quadratic, and Eq.~(\ref{A_tilde_Mallele}) is cubic in general, so the condition $\hat{C}^{(a b)}=0$ gives the possibility of a direct mapping between the two models.

We have assumed that selection is frequency independent so far, since this is the usual supposition made by many population geneticists and historically was the standard assumption used. However this may simply be a theoretical prejudice, since if one wishes to allow the fitness weightings $W^{(\alpha)}$ to depend on the composition of the population, one has to devise a model for this dependence, and so frequency independence is the simplest and most convenient choice. In addition, there are hints from experimental investigations that even if there are attempts to suppress factors that might lead to frequency dependent selection, it still seems to emerge~\cite{maddamsetti_2015}. Therefore it seems important to devise a natural way of including frequency dependence in modelling selection. Fortunately, there does exist a methodology to do this. It is based on ideas from game theory, where each allele ``plays'' a game with every other allele in the population~\cite{nowak_2006}. In the way we choose to implement this~\cite{constable_2017}, the fitness weightings are taken to have the form
\begin{equation}
W^{(\alpha)}(\underline{n}) =  1 + s \left[ \sum_{b=1}^{M-1} g^{(\alpha b)} \frac{n^{(b)}}{N} + g^{(\alpha M)}\left( 1 - \sum_{b=1}^{M-1}\frac{n^{(b)}}{N} \right) \right] \,,
\label{W_freq_dep}
\end{equation}
where $g^{(\alpha \beta)}$ is the payoff to allele $\alpha$ from interacting with type $\beta$. 

We can now make the diffusion approximation, just as in the frequency independent case, but now using $W^{(\alpha)}(\underline{n})$ given by Eq.~(\ref{W_freq_dep}), rather than the $\underline{n}$-independent form $W^{(\alpha)}=1 + \rho^{(\alpha)}s$. Clearly, the structure of $W^{(\alpha)}(\underline{n})$ in Eq.~(\ref{W_freq_dep}) can potentially lead to more complicated $\underline{z}$ dependence in $\underline{A}$, and indeed $A^{(a)}(\underline{z})$ is now found to be cubic and given by~\cite{constable_2017}
\begin{equation}
sz^{(a)} \left[ \mathcal{G}^{(aM)} + \sum^{M-1}_{b=1} G^{(ab)}z^{(b)} - \sum^{M-1}_{b=1} \mathcal{G}^{(bM)} z^{(b)} - \sum^{M-1}_{b,c=1} G^{(bc)} z^{(b)} z^{(c)} \right],
\label{Replicator_A}
\end{equation}
which is of exactly the same form as that given by Eq.~(\ref{A_tilde_Mallele}). To get the precise correspondence between the two models one must take $G^{(a b)} = - \hat{C}^{(a b)}$ for all $a$ and $b$, where $G^{(a b)}$ has the same structure as is displayed for $\hat{C}^{(a b)}$ in Eq.~(\ref{C_hat_defn}), namely
\begin{equation}
\mathcal{G}^{(a \beta)} \equiv g^{(a \beta)} - g^{(M \beta)}; \quad G^{(a b)} \equiv \mathcal{G}^{(a b)} - \mathcal{G}^{(a M)}.
\label{relative_fitnesses}
\end{equation}
In addition the identification $g^{(\alpha M)} = \Phi^{(\alpha)} - \hat{c}^{(\alpha M)}$ has to be made~\cite{constable_2017}. The fact that it is only $\mathcal{G}^{(a M)}$ and $G^{(a b)}$, and not $g^{(\alpha \beta)}$ alone, that appear in the expression for $\underline{A}$ is interesting, since the quantity $\mathcal{G}^{(a \beta)}$ can be interpreted as a relative fitness, namely the payoff to allele $a$ against an opponent $\beta$ relative to the payoff to allele $M$ against the same opponent. Similarly, $G^{(a b)}$ is a relative relative fitness. Therefore, as one would expect, it is not the actual payoffs which are important, but their values relative to the other payoffs.

In Sec.~\ref{sec:D_island_SLVC} we discussed how existence of an interior fixed point, that is, one not on the boundaries, could lead to different fixation probabilities and mean times to fixation. To investigate the possible existence of such fixed points in the frequency dependent $M$-allele case, we set $A^{(a)}(\underline{z})$, given by Eq.~(\ref{Replicator_A}), to zero. Now summing this expression over $a$ gives
\begin{equation}
0 = \left[ 1 - \sum^{M-1}_{a=1} z^{(a)} \right] \left\{ \sum^{M-1}_{b=1} \mathcal{G}^{(bM)} z^{(b)} + \sum^{M-1}_{b,c=1} G^{(bc)} z^{(b)} z^{(c)} \right\}.
\label{fixed_point_proof}
\end{equation}
If the fixed point is not to be on the boundary, then $\sum^{M-1}_{a=1} z^{(a)} \neq 1$ and so the second bracket in Eq.~(\ref{fixed_point_proof}) must vanish. Substituting this condition into the expression (\ref{Replicator_A}), which is itself taken to be zero, gives the fixed point equation to be $\mathcal{G}^{(aM)} + \sum^{M-1}_{b=1} G^{(ab)}z^{(b)} = 0$, since $z^{(a)} \neq 0$ for internal fixed points. Since this non-boundary fixed point equation is linear, there can generically be at most one fixed point. The position of this fixed point can therefore easily be found, and a determination made as to whether it lies in the SS and is therefore admissible. A similar analysis for the frequency independent case yields the condition $\rho^{(a)}=\rho^{(M)}$ for all $a$. However, if all the $\rho^{(\alpha)}$ are equal there is no selection, so in the case of frequency independent selection there are no interior fixed points.

\begin{figure}[t]
\setlength{\abovecaptionskip}{-2pt plus 3pt minus 2pt}
\begin{center}
\includegraphics[height=0.25\textwidth]{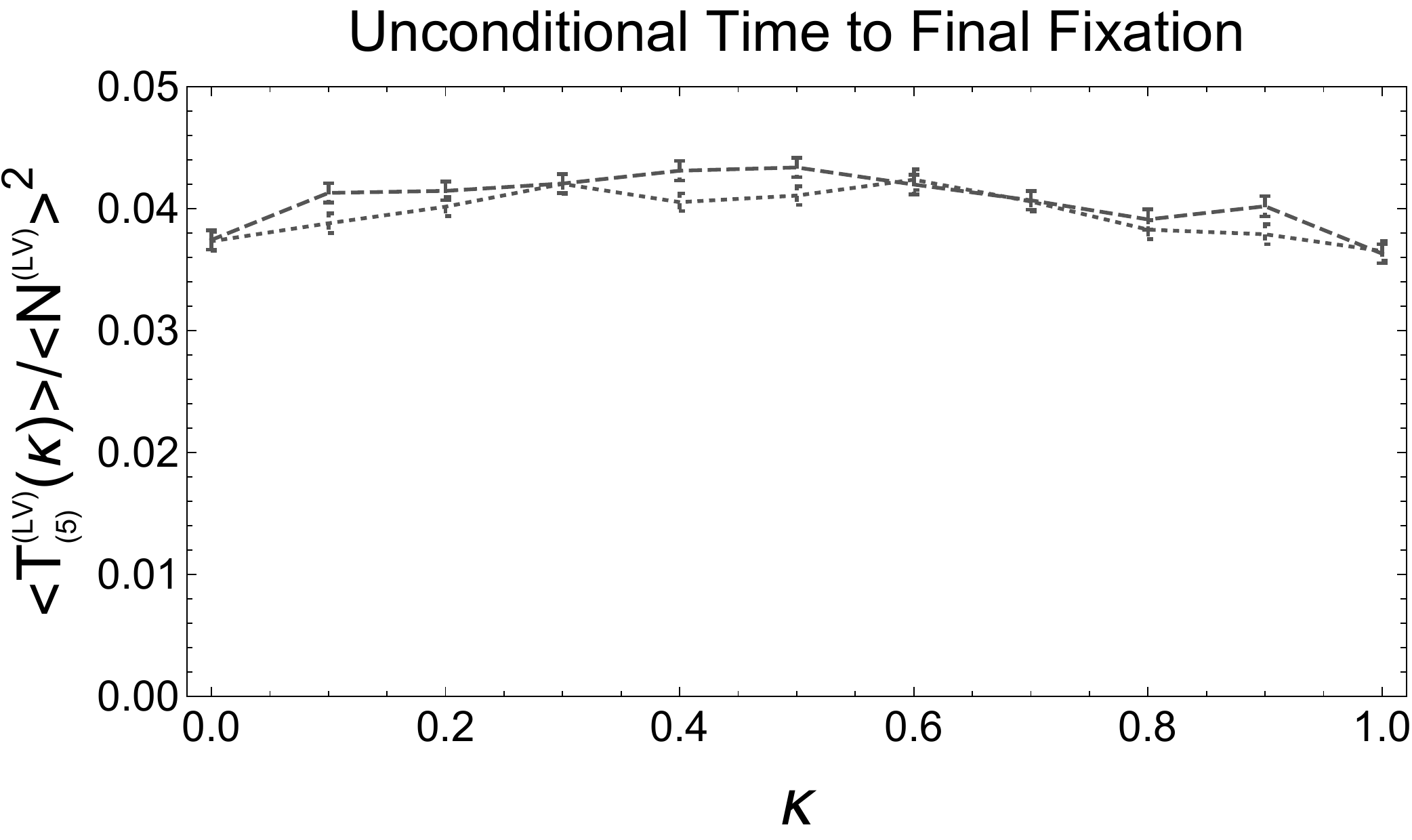}
\includegraphics[height=0.25\textwidth]{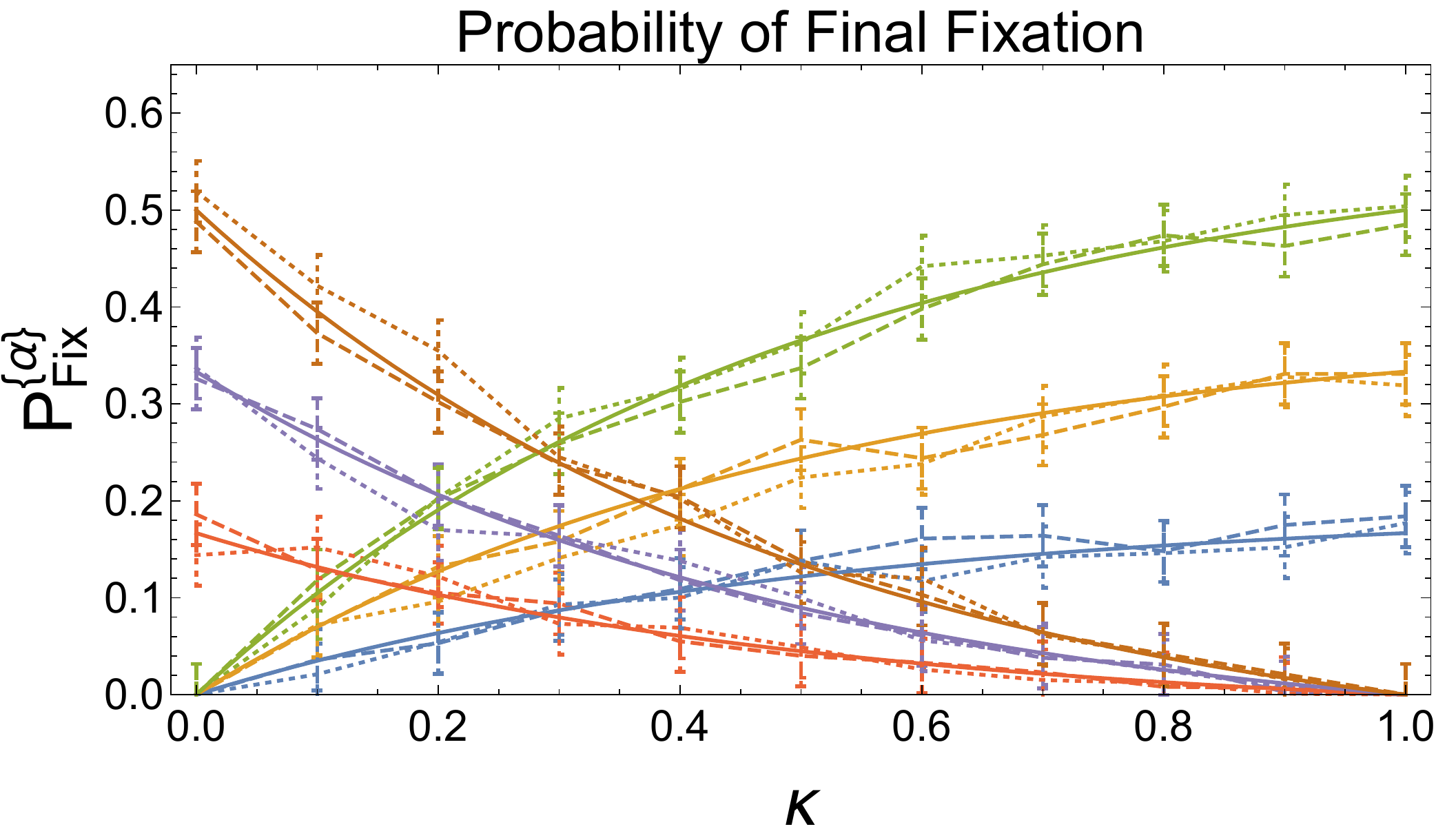}
\includegraphics[height=0.025\textwidth]{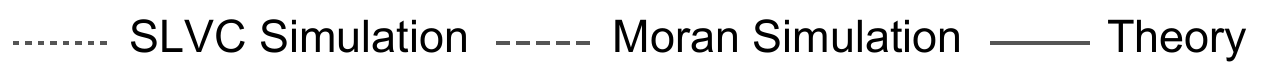}
\includegraphics[height=0.025\textwidth]{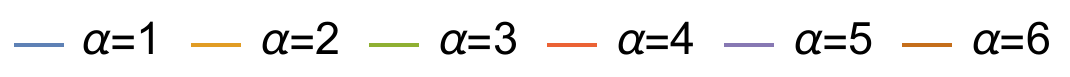}
\end{center}
\caption{Plots of the unconditional mean time until the fixation of a single allele/species and the probability of the fixation of an allele/species for the Moran and SLVC models with frequency-independent selection in the case $M=6$ alleles/species. In these plots all alleles in the Moran model are under one of two selection pressures, while in the SLVC model all species have differing parameters that combine to give two selection pressures, making the system mappable to the Moran model presented. Analytical results are only available for the probability of fixation. Simulations results are the mean of $10^{3}$ stochastic simulations of the Moran and SLVC models. Parameters used are given Ref.~\cite{constable_2017}, where the parameterization of $\bm{x}^{(0)}$ in terms of $\kappa$ is also described.
}\label{fig:SLVC_Malleles}
\end{figure}

The finding that the more realistic SLVC model reduces to the Moran model with frequency dependent selection is another reason to use frequency dependence in the modelling of selection in the Moran model. Although, as we have already remarked, the resulting Moran model is still $(M-1)$-dimensional, and so difficult to analyse, some progress can still be made in some cases~\cite{constable_2017}. In this way the SLVC model may, in effect, be analysed. An example of such a situation is shown in Fig.~\ref{fig:SLVC_Malleles}.


\subsection{Diploid Moran model with sexual reproduction: The Hardy-Weinberg assumption from first principles}\label{sec:HW}

In the Moran model discussed in Sec.~\ref{sec:example} and Sec.~\ref{sec:D_island_Moran}, individuals are haploid (they carry only a single allele) and reproduction occurs asexually (individuals simply duplicate themselves). While this is a relevant case for certain simple organisms, it is less so for many complex organisms which are diploid and reproduce sexually, such as animals. 

Suppose we want to model such a system, with diploid individuals and two possible alleles at a single locus, reproducing sexually with any other individual in the population (here, there are no sexes). A mechanistic approach might be to attempt to model the three possible genotypes in the population; here we will denote the homozygotes by A$^{(1)}$A$^{(1)}$ and A$^{(2)}$A$^{(2)}$, while the heterozygotes will be denoted by A$^{(1)}$A$^{(2)}$. If we fix the population size to be $N$, this leaves two free variables. As we have previously discussed, this makes obtaining analytic quantities in the model far more difficult than in the asexual haploid case, where the system was described by a single variable.

\hspace{0.2cm}

\noindent \textbf{Classic Approach.} Classic studies in population genetics circumvented this complexity by building single variable models that implicitly exploited a separation of timescales. It was noticed early on in theoretical population genetics that if there were no fitness differences between the genotypes (i.e. the system was neutral) the frequency of genotypes in such a diploid system would quickly relax to Hardy-Weinberg frequencies~\cite{hardy_1908,weinberg_1908}, where the number of each genotype could be described in terms of a single variable, the frequency of one of the alleles. In the terminology of our present paper, the system would quickly relax to a CM. Denoting the allele frequencies by $x^{(1)}=n^{\rm (A^{(1)})}/(2N)$ and $x^{(2)}=n^{\rm (A^{(2)}) }/2N=(1-x^{(1)})$, and the genotype frequencies by $y^{(1)}=n^{\rm (A^{(1)}A^{(1)}) }/N$, $y^{(2)}=n^{\rm (A^{(1)}A^{(2)}) }/N$, $y^{(3)}=n^{\rm (A^{(2)}A^{(2)})}/N=(1-y^{(1)}-y^{(2)})$, this is given by~\cite{ewens_2004}
\begin{eqnarray}
y^{(1)} = (x^{(1)})^{2} \,, \qquad y^{(2)} = 2 x^{(1)} (1-x^{(1)}) \,, \qquad y^{(3)} = (1-x^{(1)})^{2} \,. \label{eq_HW_freq}
\end{eqnarray}

Rather than model the dynamics of the diploid population, the dynamics of the \emph{alleles} were modelled with the assumption that they existed at Hardy-Weinberg frequencies. This was assumed to also hold when selection was sufficiently weak that the deviations from these ``equilibrium'' frequencies were not too great~\cite{moran_1957} (note the conceptual similarities with our approach). Genotypes AA are assumed to be under selective pressure A$^{(1)}$A$^{(1)}$, genotypes A$^{(1)}$A$^{(2)}$ under $(1 + s h)$ and genotypes A$^{(2)}$A$^{(2)}$ under $1$~\cite{ewens_2004}. Note then that choosing $h>1$ corresponds to overdominance, while $h \leq 1$ corresponds to underdominance~\cite{ewens_2004}. The details are given in \aref{sec:app_HW}, however upon applying the diffusion approximation one obtains an FPE (\ref{FPE_simple}) or SDE (\ref{SDE_simple}), with~\cite{ewens_2004}
\begin{eqnarray}
A( x^{(1)} ) &=& s x^{(1)} (1-x^{(1)} ) \left[ x^{(1)} + h ( 1 - 2 x^{(1)} ) \right] \,, \nonumber \\
 B( x^{(1)} ) &=& 2 x^{(1)} ( 1 - x^{(1)} ) \,.\label{eq_HW_classic}
\end{eqnarray}

\noindent \textbf{Mechanistic Approach.} Whereas \eref{eq_HW_classic} was developed using an \textit{a priori} assumption that the system lay at Hardy-Weinberg equilibrium, we may now use the methods detailed in \sref{sec:example} to formally obtain an approximation for the dynamics on the CM. We note that a similar approach was taken recently in~\cite{hossjer_2016} and that this separation of timescales has been long noted and exploited~\cite{watterson_1964,ethier_1980,ethier_1988}. We begin by modelling the genotypes themselves; genotypes encounter each other at a rate proportional to their frequency in the population, weighted by a joint probability $W^{( \alpha \beta )}$ that genotype $\alpha$ successfully mates with genotype $\beta$. In this way we account for selection. In a similar fashion to \sref{sec:setup}, we assume selection is small and formalize this by setting 
\begin{eqnarray}
W^{( \alpha \beta )} = 1 + s \alpha^{( \alpha \beta )} \,.
\label{W_alpha_beta}
\end{eqnarray}
Expanding the master equation for small $s$ as in \sref{sec:setup}, we obtain a two dimensional description of the dynamics. We now seek to understand how this two-dimensional description is related to the classic description.


\begin{figure}[t]
\begin{center}
 \includegraphics[width=0.33\textwidth]{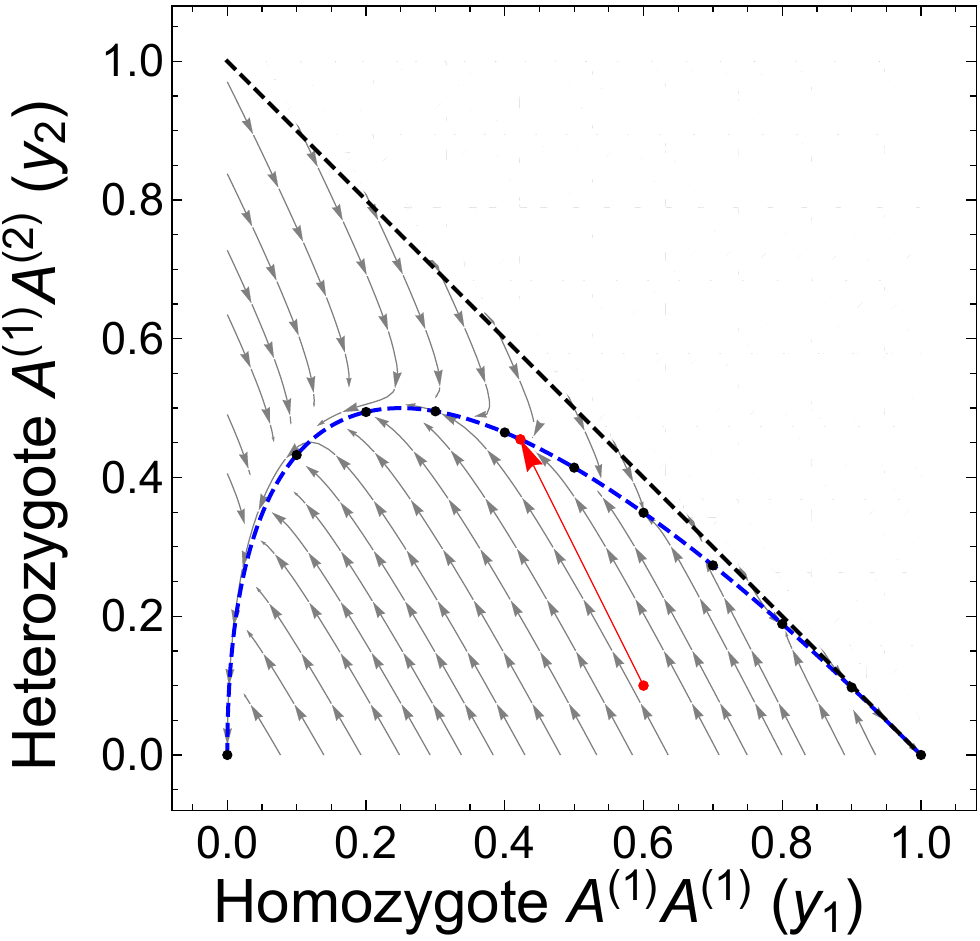}  \includegraphics[width=0.57\textwidth]{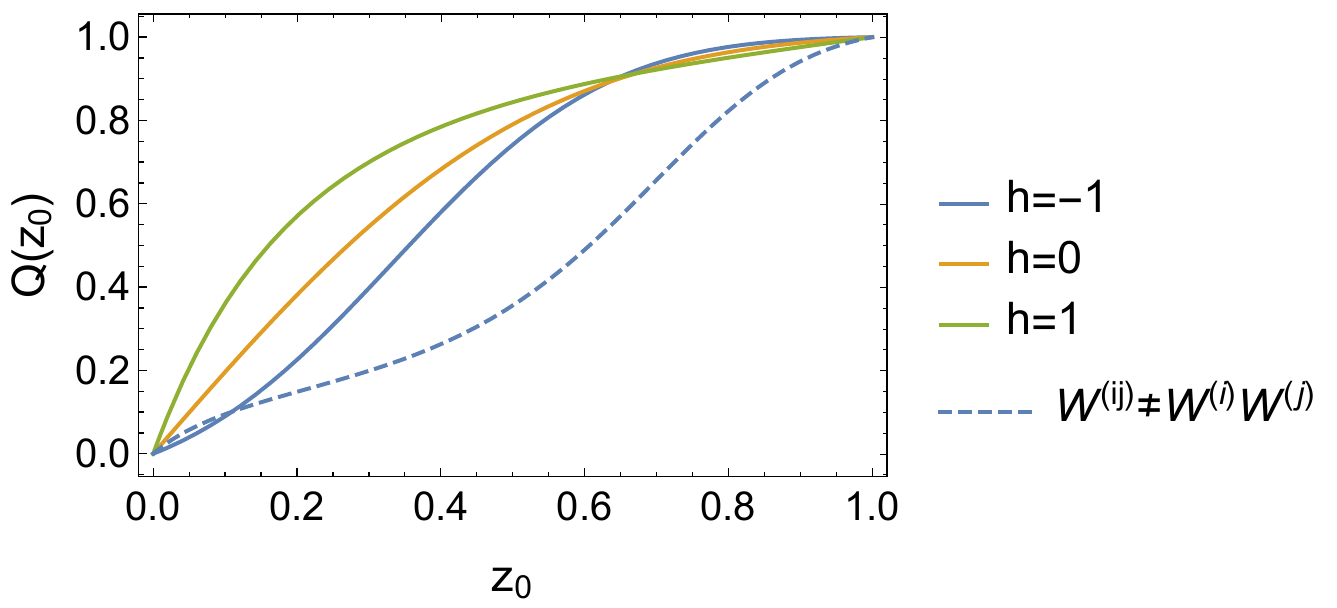} 
\end{center}
 \caption{Left panel: phase diagram of the dynamics for the mechanistic diploid Moran model described in \sref{sec:HW}. Non-neutral dynamics are depicted by grey arrows. The blue dashed line shows the form of the neutral CM described in \eref{eq_HW_freq}. The orange arrow shows a single deterministic trajectory in the neutral system. Since the population size is fixed such that $y_1+y_2<N$, the area above the dashed black line is outside the dynamical region. Right panel: fixation probabilities as a function of $z_0$, the initial condition on the CM, for the diploid Moran model. In solid colours the fitness of genotypes is parameterised using Eq.~(\ref{W_alpha_beta}). The dashed line gives the dynamics for the parameterisation $\alpha^{(11)}=\alpha^{(13)}=\alpha^{(23)}=1$, $\alpha^{(12)}=\alpha^{(22)}=-1$ and $\alpha^{(11)}=2$. In all cases, $N=100$. }
\label{fig:HW}
\end{figure}


Having set up the model, we can proceed to the next stage of the analysis; identifying the fast and slow variables, as described in \sref{sec:first_stage}. We can obtain a CM by setting selection equal to zero ($s=0$). In this case the CM is given by \eref{eq_HW_freq}. We can now linearise the system about this CM to obtain the Jacobian, and use this matrix to obtain $u^{ \{ 1 \} }$ and $v^{ \{ 1 \} }$, the left and right eigenvectors of the Jacobian corresponding to the zero eigenvalue. Finally, using \eref{eq_general_ABar} and \eref{eq_general_BBar}, we can obtain an effective description for the system dynamics in terms of $z=y^{(1)}$ on the CM. However, in order to make a comparison between this effective theory and the classic theory, we must express the effective theory in terms of $x^{(1)}$, the frequency of A$^{(1)}$ alleles (see \eref{eq_HW_classic}). Since $y^{(1)}=(x^{(1)})^{2}$ on the CM, we must therefore make the transformation $x^{(1)}=\sqrt{z}$. The full calculation is given in \aref{sec:app_HW}. We note that while there are some subtle mathematical points that need to be attended to, these should not distract from the key points of the method. Our final result is that we can approximate the dynamics of the mechanistic model by an SDE of type \eref{SDE_simple} with terms
\begin{eqnarray}
& & \tilde{A}( x^{(1)} ) = s x^{(1)} ( 1 - x^{(1)} ) \left[ \alpha^{(23)} - \alpha^{(33)} +  \left( \alpha^{(13)} + 2 \alpha^{(22)} - 6 \alpha^{(23)} \right. \right. 
\nonumber \\
&+& \left. 3 \alpha^{(33)} \right) x^{(1)} + 3 \left( \alpha^{(12)} - \alpha^{(13)} - 2 \alpha^{(22)} + 3 \alpha^{(23)} - \alpha^{(33)} \right) (x^{(1)})^{2} 
\label{eq_HW_effective_A} \nonumber \\
& & \left. + \left(  \alpha^{(11)} - 4 \alpha^{(12)} + 2 \alpha^{(13)} + 4 \alpha^{(22)} - 4 \alpha^{(23)} + \alpha^{(33)} \right) (x^{(1)})^{3} \right] \,, \\
& & \tilde{B}(z) = z ( 1 - z ) \,. \label{eq_HW_effective_B}
\end{eqnarray}

While the form of \eref{eq_HW_effective_A} appears a little complicated, we can in fact show that this becomes identical to \eref{eq_HW_classic} under the assumption that each term $\alpha^{( \alpha \beta )}$ can be decomposed into the sum of contributions from each allele;
\begin{eqnarray}
\alpha^{( \alpha \beta )} = \alpha^{(\alpha)} + \alpha^{(\beta)} \,,
\end{eqnarray}
and that $\alpha^{(\alpha)}$ take the precise values
\begin{eqnarray}
\alpha^{(1)} = 1 \,, \qquad \alpha^{(2)} = h \,, \qquad \alpha^{(3)} = 0 \,. 
\end{eqnarray}
Since these values give a precise mapping from Eqs.~(\ref{eq_HW_effective_A}) and (\ref{eq_HW_effective_B}) to \eref{eq_HW_classic}, we can now ask what consequences this has for $W^{( \alpha \beta )}$. This matrix now takes the form
\begin{eqnarray}
W = \left(
\begin{array}{ccc}
 1 + 2 s & 1 + s + s h & 1 + s \\
1 + s + h s & 1 + 2 s h  & 1 + s h \\
 1+ s s & 1 + h s & 1 \\
\end{array}
\right) \,.
\end{eqnarray}\label{eq_HW_param}
This is in fact \emph{exactly} the form of $W$ that we would expect at leading order in $s$ if 
\begin{eqnarray}
W^{( \alpha \beta )} = W^{( \alpha )}W^{( \beta )} \,, \qquad W^{(1)} = 1 + s\,, \quad W^{(2)} = 1 + s h \,, \quad W^{(3)} = 1 \,,
\end{eqnarray}
i.e. if the fitness of each genotype is the same as that postulated in \eref{eq_HW_classic} and the success of genotype pairings is a multiplicative function of the individual genotype fitness. A similar mapping exists if we assume an additive interaction of the genotypes fitness on their pair sexual success or an averaging effect (see \aref{sec:app_HW}). It is testament to the great physical intuition of the founders of population genetics that this captures their assumptions. While we have essentially recovered here a known result, it is worth noting that of course the approach we have taken is not without merits. In particular, it allows us to explore a far richer fitness landscape than the original model (see \fref{fig:HW}).

\subsection{Stochastic epidemics on networks}
\label{sec:epidemics}
The networks which we have discussed so far in this paper have described the way in which islands interact with each other through migration. The islands have had populations of individuals which are already themselves large (in the sense that they are of order $N$, the large parameter in the system). These are metapopulation models, since the whole system is a population of populations~\cite{hanski_1999}. However, in many network models the nodes are composed of a single individual, rather than a population. It is not immediately obvious how to apply the diffusion approximation in this case, and hence the reduction method of Sec.~\ref{sec:example}. However there are situations when the programme we have outlined so far can be pursued, and we now describe such a case.

The example is taken from stochastic models of infectious diseases, rather than population genetics, but it illustrates the points we wish to make clearly. We will follow Ref.~\cite{parrarojas_2016}, where more details can be found. The model is an SIR model, which means that the individuals are either susceptible to the disease (in which case they are in class $S$), infected with the disease (in which case they are in class $I$), or recovered from having the disease (in which case they are in class $R$). We assume that the disease is such that individuals recover, and do not die as a consequence of having the disease. Our interest will be in the properties of an epidemic which might occur, and we assume that this takes place over a much shorter timeframe than the demographic processes of birth and death. So the only processes which occur in the model are (i) infection, and (ii) recovery.

The network structure enters because we assume that each individual has a fixed number of contacts from which the infection is acquired, even though the contacts themselves will change. Therefore we may imagine all individuals located at the nodes of a network, where the degree of that node is equal to the number of contacts characteristic of that individual. The network is considered in the so-called dynamic limit, in which the network structure is assumed to evolve much more quickly than the epidemic, so that the only role of the network is to encode the number of connections a given individual has to other individuals. The variables at the microscale are therefore $S_k, I_k$ and $R_k$ where $k$ labels the degree of the node on which individuals of the type $S, I$ and $R$ are located. As is often done in SIR models, we assume that the population is closed, so that at time $t$ $S_k(t)+I_k(t)+R_k(t)=N_k$, where $N_k$ is independent of time. This implies that one set of individuals --- for example those which have recovered --- can be removed: $R_k(t)=N_k-S_k(t)-I_k(t)$. The large parameter in the model is the total number of individuals: $N=\sum^K_{k=1} N_k$, where $K$ is the maximum degree of the network. The specific network of interest in Ref.~\cite{parrarojas_2016} was a truncated Zipf distribution where the probability of an individual having degree $k$ is given by $d_k \propto k^\alpha$, with $-3 < \alpha < -2$ and $k=1,\ldots,K$, but the method we describe is also applicable to other distributions.

We have stressed throughout this article that one needs to start with the microscopic description, at the level of single individuals, to avoid ambiguities. However here, to avoid too much formalism, we will move directly to the mesoscopic model which can be derived from the microscopic description~\cite{parrarojas_2016}:
\begin{align}
\frac{ds_k}{d\tau} &= -\beta ks_k\sum^K_{l=1} li_l + \frac{\eta_1^{(k)}(\tau)}{\sqrt{N}},\label{eqn:sk}\\
\frac{di_k}{d\tau} &= \beta ks_k\sum^K_{l=1} li_l-\gamma i_k + \frac{\eta_2^{(k)}(\tau)}{\sqrt{N}},\label{eqn:ik}
\end{align}
where 
\begin{align}
\left\langle \eta_\mu^{(k)}(\tau) \eta_\nu^{(l)}(\tau ')\right\rangle &= B^{(k)}_{\mu \nu}(\bm{x}) \delta_{k l} \delta(\tau - \tau '),
\label{eqn:noise_corr_SIR_1}
\end{align}
and where $k=1,\ldots,K$ and $\bm{x}$ is a vector of all the $2K$ variables. Here $\beta$ and $\gamma$ are the infection and recovery rates respectively, and $N$ is assumed to be sufficiently large that $s_k = S_k/N$ and $i_k = I_k/N$ can be assumed to be continuous. We will not give the precise form of $B^{(k)}_{\mu \nu}(\bm{x})$, $\mu,\nu=1,2$, here, nor the form of the rescaled time $\tau$.

The problem we face is clear from Eqs.~(\ref{eqn:sk}) and (\ref{eqn:ik}): this is a stochastic system with $2K$ variables, where in many cases of interest $K$ will be not be small. This is then another case where we wish to reduce the number of variables, and where the reduction method described above may be useful. In fact, one can effect a partial reduction before applying the technique of Sec.~\ref{sec:example} by making the ansatz $s_k(\tau)=d_k\theta(\tau)^k$, which replaces the $K$ equations (\ref{eqn:sk}) by the single equation
\begin{equation}
\frac{d\theta}{d\tau} = - \beta \theta \sum^{K}_{l=1} l i_l + \frac{1}{\sqrt{N}}\xi(\tau),
\label{theta_eqn}
\end{equation}
where $\xi(\tau)$ is a Gaussian noise with zero mean which is related to $\eta_1^{(k)}(\tau)$. 

After this initial manoeuvre we attempt to further reduce this $(K+1)$-dimensional system by searching for fast and slow modes. Examination of the Jacobian~\cite{parrarojas_2016} reveals that there is one $K$-fold degenerate eigenvalue which may be significantly greater in magnitude than the remaining eigenvalue. If we denote the ratio of the magnitude of the former to the magnitude of the latter by $\epsilon$, then as long as $\epsilon$ is small there will be effectively $K$ fast modes and $1$ slow mode, so we would expect to be able to reduce the motion to a two-dimensional SS where one of the variables is $\theta$ and the other is the slow one just mentioned. It is found that this additional variable is $\lambda(\tau) = \sum^{K}_{l=1} l i_l(\tau)$~\cite{parrarojas_2016}, so that Eq.~(\ref{theta_eqn}) simply becomes $d\theta/d\tau = - \beta \theta \lambda + \xi(\tau)/\sqrt{N}$. The equation for $\lambda(\tau)$ is more complicated, involving as it does three different noise terms, and we do not give it here.

Although the system has been reduced to a manageable two dimensional model, it is always worthwhile to perform numerical investigations to see if it is possible to make further reductions, since it is still the case that two dimensional stochastic systems are difficult to study analytically. In this case, it is found that typically the noise term in the theta equation is much smaller in magnitude to the noise terms in the lambda equation. Therefore we can try to omit the noise term in Eq.~(\ref{theta_eqn}) and combine the three noises in the $\lambda$ equation together to obtain the further simplification:
\begin{eqnarray}
\frac{d\theta}{d\tau} &=& - \beta \theta \lambda, \nonumber \\
\frac{d\lambda}{d\tau} &=& \lambda\left( \beta \phi(\theta) - \gamma \right) + \frac{1}{\sqrt{N}} \bar{\sigma} \zeta(\tau),
\label{ultimate_reduction}
\end{eqnarray}
where $\phi(\theta)=\sum^K_{k=1} k^2 d_k \theta^k$, $\zeta(\tau)$ is a Gaussian noise with zero mean and $\langle \zeta(\tau) \zeta(\tau') \rangle = \delta(\tau - \tau')$, and $\bar{\sigma}$ is a function of $\theta$ and $\lambda$ which is not given here~\cite{parrarojas_2016}. The model defined by Eq.~(\ref{ultimate_reduction}) is a semi-deterministic mesoscopic model, in the sense that one of the dynamical equations is deterministic, having no noise term. In fact it can be identified as the Cox-Ingersoll-Ross (CIR) model~\cite{CIR}, for which some analytic results are known. 


\begin{figure}[t]
\begin{center}
 \includegraphics[width=0.45\textwidth]{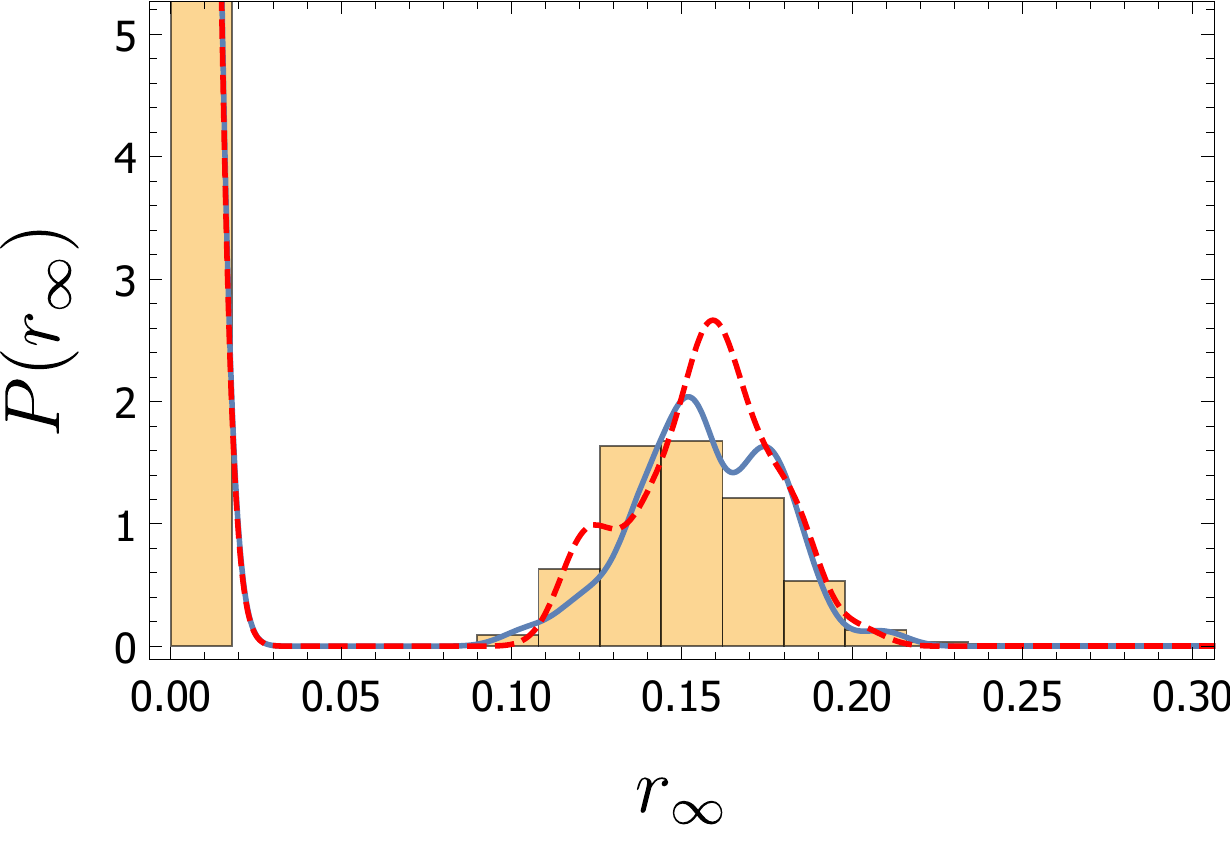}
\end{center}
 \caption{The distribution of epidemic sizes for $N=20 000, K=1000$ and $\alpha=-2.5$. The histogram gives the result for the full model, the solid blue line is the result for the reduced mesoscopic model and the red dashed line the result for the semi-deterministic mesoscopic model given by Eq.~(\ref{ultimate_reduction}).}
\label{fig:r_infinity}
\end{figure}


As a test of the various reductions that have been made on this model we will investigate how well they capture the distribution of epidemic sizes as given by $r_\infty$, the number of recovered individuals at the end of the epidemic~\cite{parrarojas_2016}. The results are shown for a particular case in Fig.~\ref{fig:r_infinity}, and indicate that even the very much simplified model given by Eq.~(\ref{ultimate_reduction}) gives a good representation of the result.

\section{ Revealing noise-induced selection via fast variable elimination }
\label{sec:noise_induced}
In \sref{sec:example}, we laid the groundwork for conducting fast-variable elimination in stochastic systems, while in \sref{sec:applications} the utility of this approach was demonstrated. We have essentially shown that, given a separation of timescales exists, the dynamics of a high dimensional system can be approximated by the projection of these dynamics onto an SS. Thus far, this is true for both the deterministic and stochastic component of the dynamics (see Eqs.(\ref{eq_general_ABar})~and~(\ref{eq_general_BBar})). However, there is a further consideration that we have until now omitted; noise-induced dynamics in the slow subspace. While accounting for such dynamics is more technically involved than the method outlined in \sref{sec:example}, the resultant dynamical behaviour can be very interesting, and so it is worth describing the key aspects here.

Noise-induced dynamics are the phenomena whereby the projection of the deterministic dynamics onto the slow-subspace (see \eref{eq_general_ABar}) does not entirely describe the time-evolution of the average dynamics when demographic noise is accounted for. In the context of a system that features a CM, this means that rather than there being no average dynamics along the CM, a bias in fact emerges in some direction. Although this is a second order effect (it scales inversely with the population size and disappears in the infinite population size limit), it can in fact completely govern the dynamics along the CM, where the first order terms that control the collapse of the system to the CM disappear. If the system under consideration is one featuring competing organisms, this bias can be interpreted as noise-induced selection (or drift-induced selection in a population-genetics context). The origin of these bias terms can be interpreted in various ways.

First, and perhaps most intuitively, noise-induced dynamics can be graphically understood as resulting from a bias in how fluctuations taking the system off the CM (or SS) return to the CM. Under certain scenarios, such as when the CM is strongly curved or the trajectories to the CM are divergent, fluctuations off the CM do not return, on average, to the point on the CM from which they originated (see \fref{fig:PG_projection}). This introduces a bias that stochastically `ratchets' the dynamics in a preferred direction.

Second, these noise-induced dynamics can be understood as a mathematical consequence of making a non-linear change of variables into the slow-subspace of the stochastic system. As we have mentioned, the SDEs that we work with should be interpreted in the It\={o} sense. In this context, different rules of stochastic calculus apply, and accounting for the additional terms that arise in the analysis gives rise to additional terms in the reduced description on the CM. For readers not familiar with the analysis of SDEs, It\={o} calculus can in some sense be loosely understood as arising in the same way as Jensen's inequality; the average of a function of a random variable is not necessarily the same as the function evaluated at the average of that random variable.

Finally, in a more biological context, noise-induced dynamics in systems of competing organisms can be understood resulting from a selective pressure to reduce variance in reproductive output (see Gillespie's Criterion~\cite{gillespie_1974,hansen_2017}). However, in contrast to Gillespie's original study, variance between the reproductive output of organisms can arise as a result of the dynamics of the organisms, rather than being assumed \textit{a priori}.

Of course, the existence of these noise induced dynamics is not in itself dependent on a system exhibiting a separation of timescales. We have already mentioned Gillespie's Criterion, which demonstrates that selection for a genotype with a lower variance in reproductive output takes place in a model with only a single variable (measuring the frequency of one of two genotypes). Further, in a similar single-variable population genetics context that assumes a well-mixed population of finite size, it has been shown that noise-induced selection favours genotypes that increase the population size~\cite{houch_2012,houch_2014}. However, while fast-variable elimination does not generate noise-induced dynamics itself, it does often give us a means of quantifying and analytically understanding the effect. 

As in the earlier sections, rather than transforming into the fast-slow basis of the problem, removing the fast variables and then transforming back into the original variables, we take a shortcut by using a non-linear projection of the dynamics onto the CM. The reason for this is two-fold. Firstly, it is more straightforward practically; it is in the original, biologically relevant variables that we can better understand the behaviour of the system. Secondly, as has long been recognised~\cite{coullet_1983,roberts_2015}, the non-linear transform that yields the slow-fast basis is not always obvious. In order to obtain the non-linear projection, second order perturbation techniques can be used to deduce the effective dynamics. The full calculation is too long to reproduce here, however a clear and coherent explanation is given in Ref.~\cite{parsons_rogers_2015}. There it is shown that if a system has $M$ variables and a one-dimensional slow-subspace, then the equation for the reduced/effective dynamics can be expressed by
\begin{eqnarray}
\frac{ \mathrm{d} z }{ \mathrm{d} \tau} = \bar{ A }( z ) + \bar{ A }^{ \rm S }( z )  + \sqrt{ \frac{1}{N} } \zeta(t) \,. \label{eq_SDE_reduced_induced}
\end{eqnarray}
The term $\bar{ A }( z )$ and the white noise term $\zeta(t)$ retain the forms given in Eqs.~(\ref{eq_general_ABar})-(\ref{eq_general_BBar}) (that is, they are respectively components of the deterministic dynamics and noise projected onto the SS). Most interesting is the term $ \bar{ A }^{ \rm S }( z )$ as it is this that controls the noise induced dynamics. It is of order $1/N$ (induced as it is by demographic noise) and takes the explicit form
\begin{eqnarray}
\bar{ A }^{ \rm S }( z )= \frac{1}{N} \sum_{i,j=1}^M & & \left[ \frac{ \mathrm{d} u^{\{1\}}_i }{ \mathrm{d} z } u^{\{1\}}_j + u^{\{1\}}_i  \frac{ \mathrm{d} u^{\{1\}}_j }{ \mathrm{d} z } - 2 u^{\{1\}}_i u^{\{1\}}_j \sum_{k=1}^{M} \left( v^{\{1\}}_k  \frac{ \mathrm{d} u^{\{1\}}_k }{ \mathrm{d} z } \right) \right. \nonumber \\ 
& & \left. - u^{\{1\}}_i u^{\{1\}}_j \sum_{k=1}^{M} \left( \frac{ \mathrm{d} v^{\{1\}}_k }{ \mathrm{d} z } u^{\{1\}}_k \right) \right] \left. B_{ij}(\bm{x}) \right|_{\mathrm{CM}} \,,
\label{eq_general_ABar_S}
\end{eqnarray}
where we recall that $\bm{v}^{\{1\} }$ and $\bm{u}^{\{ 1\} }$ are the right and left eigenvectors of the neutral Jacobian on the CM corresponding to the zero eigenvalue (see \sref{sec:example}). The terms which feature $\mathrm{d} u^{\{ 1\} }_i / \mathrm{d} z $ capture how quickly the fast directions change as a function of $z$, and can thus be understood as the component of the noise-induced dynamics that arises from the non-linearity of trajectories to the CM. Meanwhile the term $\mathrm{d} v^{\{ 1\} }_k / \mathrm{d} z $ is the component of the noise-induced dynamics that arises from curvature of the CM itself~\cite{parsons_rogers_2015}. Note that if both $v^{\{ 1\} }_k$ and $u^{\{ 1\} }_k$ are independent of $z$ (that is, the CM is linear and the direction of fast trajectories to the SS do not vary along the SS to first order in the selection strength), then there can be no noise-induced dynamics at this order. This is the case in the models discussed in Sections \ref{sec:example} and \ref{sec:D_island_Moran}. Finally, there are also cases where, despite the CM being curved or the trajectories to the CM being divergent, there is still no noise induced selection as the terms in \eref{eq_general_ABar_S} all cancel. This is the case in both Sections \ref{sec:SLVC} and \ref{sec:HW} for the parameterisations given in those sections.

In what follows we will illustrate how the effective description provided by \eref{eq_SDE_reduced_induced} can be used to analytically tackle some particular problems of interest. In \sref{sec:LV_PG} we will address a two-species Lotka-Volterra model. Unlike in \sref{sec:SLVC} however, we will allow the species to have distinct birth, death and competition rates at leading order. Calculation of the effective system will reveal both that slow-living species and species that increase the global carrying capacity are stochastically selected for. In \sref{sec:heterogamety} we will describe a population genetic model of transitions between modes of sex determination. Although this population genetic model is much more complicated than the haploid Moran model, the presence of a CM in the neutral limit allows us to analytically characterise a noise-induced bias favouring the substitution of dominant neutral mutations.

\subsection{Two-species Lotka-Volterra type models}\label{sec:LV_PG}

\begin{figure}[H]
\begin{center}
 \includegraphics[width=0.45\textwidth]{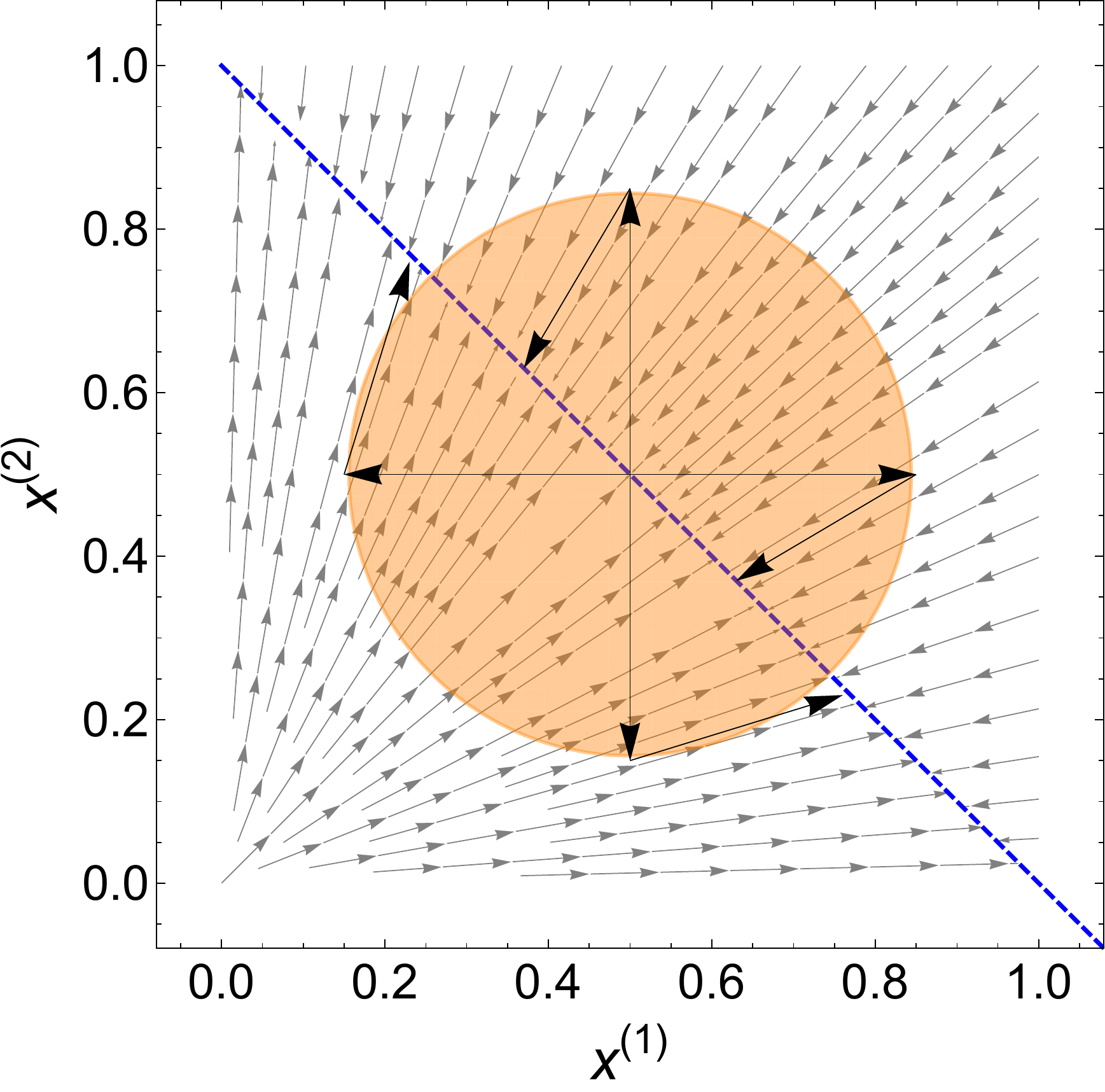} \includegraphics[width=0.45\textwidth]{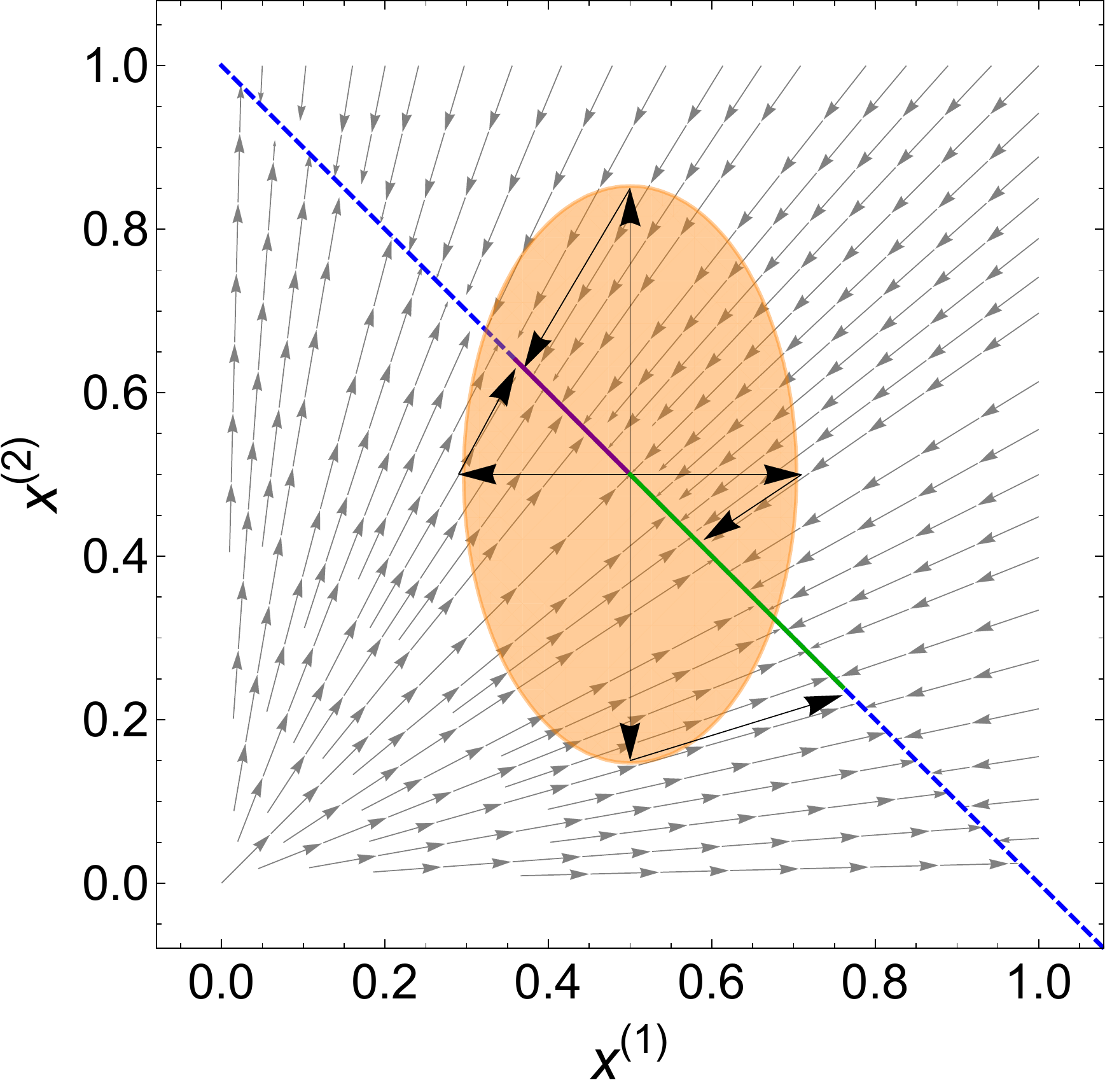}
 \includegraphics[width=0.9\textwidth]{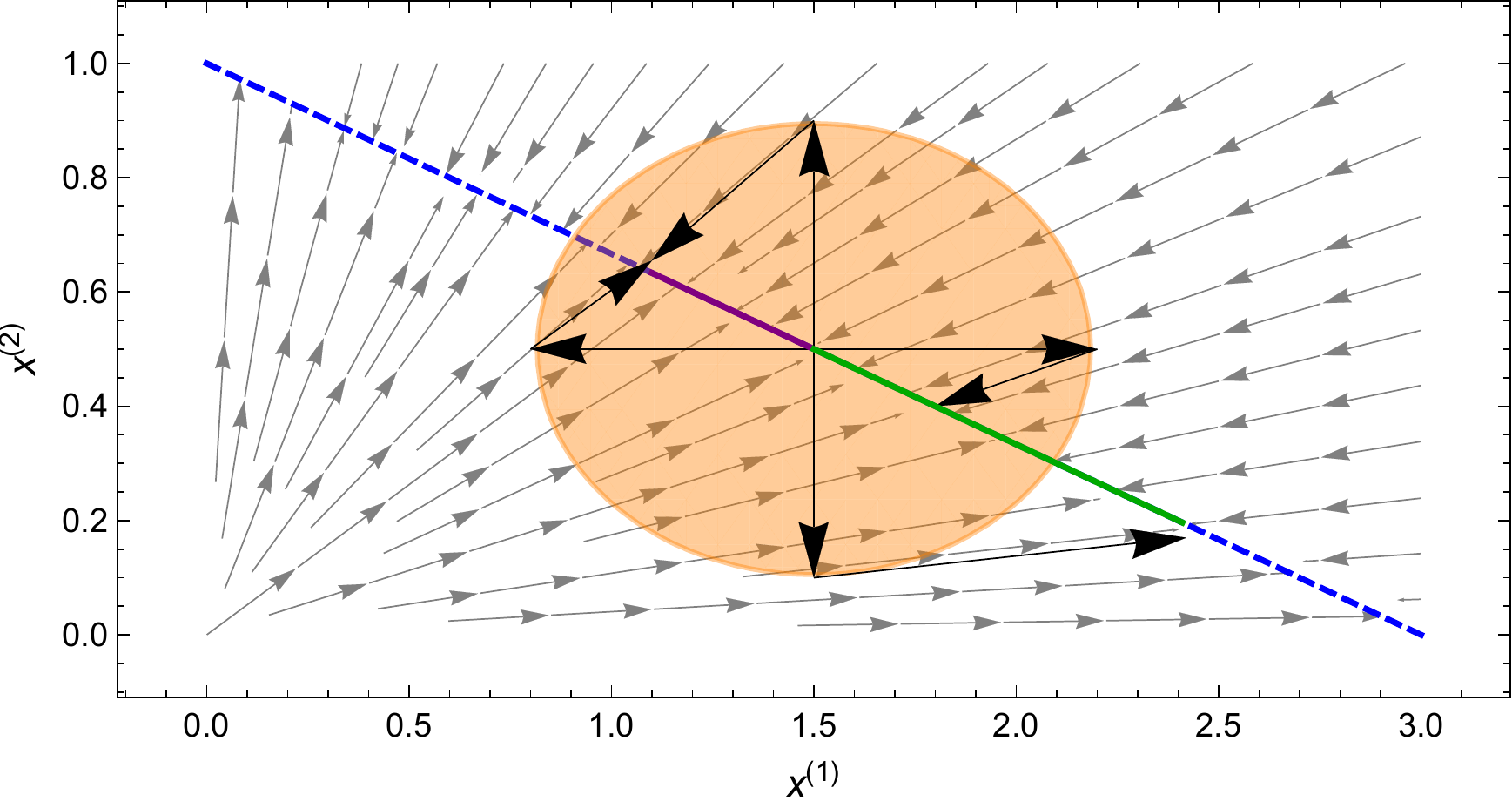}
 \caption{Phase plots for the deterministic dynamics of the neutral Lotka-Volterra model described in \sref{sec:LV_PG} under three different parameter scenarios. In each the CM is plotted as a blue dashed line. The surface of the orange ellipses is indicative of the distribution of fluctuations arising from the centre of the ellipse. Arrows from the centre of the ellipse (going up, down, left and right) show possible fluctuations away from the CM that occur with equal probability. Arrows travelling from the end of these fluctuations back to the CM show the trajectories along which these fluctuations are quenched. Bias in the direction to which fluctuations are mapped are illustrated by green and purple lines. \textbf{Top left:} System with two species identical ($b^{(1)}=b^{(2)}$, $d^{(1)}=d^{(2)}$, $c^{(1)}=c^{(2)}$).  Fluctuations down and right are projected back to the CM in the direction of $x^{(1)}$, but this is perfectly countered by fluctuations up and left which are projected back to the CM with in the direction of $x^{(2)}$. \textbf{Top right:} System with two species, the second reproducing and dying at a faster rate ($b^{(1)}-d^{(1)}=b^{(2)}-d^{(2)}$ $b^{(2)}>b^{(1)}$, $d^{(2)}>d^{(1)}$, $c^{(1)}=c^{(2)}$). As a result of this asymmetry, species $2$ experiences greater demographic fluctuations (the ellipse is larger in direction $x^{(2)}$). Consequently, fluctuations off the CM are projected back to the CM with a bias that favours $x^{(1)}$. \textbf{Bottom:} System with two species, the first increasing the carrying capacity of the system ($b^{(1)}=b^{(2)}$, $d^{(1)}=d^{(2)}$, $c^{(2)}>c^{(1)}$). The asymmetry in the angle of the CM now induces a bias in the projection of fluctuations back to the CM that favours $x^{(1)}$. 
}
\label{fig:PG_projection}
\end{center}
\end{figure}

We begin with a two-species Lotka-Volterra model of a similar type to that described in \sref{sec:M_allele_SLVC} (see \eref{parameters_non_neut}), except we now make the restriction
\begin{eqnarray}
b^{(1)} - d^{(1)} = \tilde{b}( 1 + \epsilon ) \,, \quad c^{(11)}&=&c^{(21)}\equiv c^{(1)} \,, \nonumber \\
b^{(2)} - d^{(2)} = \tilde{b} \,, 	\quad	\qquad	\quad c^{(21)}&=&c^{(22)}\equiv c^{(2)} \,.
\end{eqnarray}
By taking the limit of large system size, we can apply the diffusion approximation, as described in \sref{sec:setup}. The system is now approximated by an SDE of the same form as \eref{SDE_simple}. However, since the system is in two variables, it is difficult to analyse. We next follow the approach taken in \sref{sec:first_stage} and identify a CM. A CM exists under the above parameterisation if $\epsilon $ is equal to zero. It now takes the form
\begin{eqnarray}
 x^{(2)} = \frac{1}{c^{(2)}} \left( \tilde{b} - c^{(1)} x^{(1)} \right)  \, . \label{eq_PG_CM}
\end{eqnarray}
Notice that in isolation, species $1$ exists at $x^{(1)}=\tilde{b} / c^{(1)}$ and species $2$ at $x^{(2)}=\tilde{b} / c^{(2)}$. Further, if we assume that $c^{(2)}>c^{(1)}$, then increasing the frequency of species $1$ in the population increases the joint carrying capacity of the species, as can be seen in \fref{fig:PG_projection}. If additionally $\epsilon>0$, such that species $1$ reproduces at a lower rate that species $2$, then this can be interpreted as a model of public good production. Species $1$ pays a cost $\epsilon$ to its reproductive rate to increase the carrying capacity of both species. Species $2$ pays no cost, but still enjoys a reduced death rate in the presence of species $1$ as a result of the lower competition parameter of species $1$ (interpreted here as resulting from the production of a public good). However, in isolation, species $2$ exists at lower numbers than species $1$, interpreted here as resulting from the non-production of the public good.

We can now briefly describe the deterministic dynamics. In the neutral limit, where $\epsilon=0$, the population grows to a point on the CM described by \eref{eq_PG_CM}. We now move away from the neutral limit, such that $1\gg \epsilon>0$; the system grows to a point on the SS (which is equal to the CM at leading order in $\epsilon$) after which the system moves along the SS until species $1$ is driven to extinction. We now use fast-variable elimination to characterise the dynamics when demographic noise is accounted for.

As discussed in \sref{sec:first_stage}, our first task is to identify the left and right eigenvectors of the system evaluated on the CM, \eref{eq_PG_CM}. Note that while the right-eigenvectors are not so important in \sref{sec:second_stage} (where there was no noise-induced selection), they become very important here, featuring as they do in \eref{eq_general_ABar_S}. We find
\begin{align}\label{eigenvectors_pg_simple_left}
\bm{u}^{\{ 1\} } = \frac{ 1 }{ \tilde{b} }\,\left( \begin{array}{c} 
\tilde{b} - c^{(1)} z \\
-c^{(2)} z
\end{array} \right), 
\ \ \bm{u}^{\{ 2\} } =  -\frac{ 1 }{ \tilde{b} }\,\left( \begin{array}{c} 
\tilde{b} - c^{(1)} z \\
\tilde{b} c^{(2)}/c^{(1)} + c^{(2)} z
\end{array} \right).
\end{align}
\begin{align}\label{eigenvectors_pg_simple_right}
\bm{v}^{\{ 1\} } =\,\left( \begin{array}{c} 
1 \\
- c^{(1)}/c^{(2)}
\end{array} \right), 
\ \ \bm{v}^{\{ 2\} } = -\frac{ c^{(1)} }{ \tilde{b} - c^{(1)} z }\,\left( \begin{array}{c}
z \\
(\tilde{b} - c^{(1)} z)/c^{(2)}
\end{array} \right).
\end{align}

We can use these expressions for the left and right eigenvectors to obtain an approximate description of the dynamics in the SS using \eref{eq_SDE_reduced_induced} with expressions for $\bar{A}( z  )$, $\bar{A}^{ \rm S }( z ) $ and $ \bar{B}(z)$ directly from Eqs.~(\ref{eq_general_ABar}), (\ref{eq_general_ABar_S}) and (\ref{eq_general_BBar}), where $z$ is the value of $x^{(1)}$ on the SS;
\begin{eqnarray}
 \bar{A}( z ) &=&  -  \epsilon z \left( \tilde{b} - c^{(1)} z\right)  \,,\\
 \bar{A}^{ \rm S }( z ) &=& \frac{1}{N} \frac{2}{ \tilde{b}^{2} }  z \left( \tilde{b} - c^{(1)} z\right) \left( c^{(2)} (\tilde{b} + d^{(2)} ) - c^{(1)} (\tilde{b} + d^{(1)} ) \right) \,, \\
 \bar{B}(z) &=&   \frac{2}{ \tilde{b}^{2} }  z \left( \tilde{b} - c^{(1)} z\right) \left[ z  \left( c^{(2)} (\tilde{b} + d^{(2)} ) - c^{(1)} (\tilde{b} + d^{(1)} ) \right) + \tilde{b} ( d^{(1)} + \beta)\right] \,. \nonumber \\ 
\end{eqnarray}


An analysis of these terms reveals a number of points. Firstly, and as we might expect, in the limit $c^{(1)}=c^{(2)}\equiv c^{(0)}$, $b^{(1)}=b^{(2)}\equiv b^{(0)}$ and $d^{(1)}=d^{(2)}\equiv d^{(0)}$ the noise induced term $\bar{A}^{ \rm S }( z )$ vanishes and we are left with a system that takes the same form as that in \sref{sec:SLVC}. Secondly, we examine the limit in which $\epsilon=0$. Now $\bar{A}( z )=0$ as there are no dynamics along the CM in the infinite population size limit. However, if the population has a finite size the noise induced term $ \bar{A}^{ \rm S }( z ) $ is not zero. Thirdly, continuing with the system when $\epsilon=0$, and now additionally asking that $c^{(1)}=c^{(2)}\equiv c^{(0)}$ (i.e. that the carrying capacity is unchanged by the composition of the population), we find that $ \bar{A}^{ \rm S }( z ) $ is positive for all $z$ along the CM if $b_1 < b_2$. The species with the lower birth rate (and death rate, since $\tilde{b}$ is fixed) is therefore selected for. This insight, made in Refs.~\cite{parsons_quince_2007_1,doering_2012,chotibut_2017}, is a result of the fact that phenotypes which are reproducing and dying more quickly are subject to greater population fluctuations (they have a larger rate of population turnover). Consequently, it is easier for the longer lived phenotype (lower birth/death rates), to invade and fixate. This phenomena can be viewed as analogous to Gillespie's Criterion.

Turning instead to the limit $\epsilon=0$, $b^{(1)}=b^{(2)}\equiv b^{(0)}$ and $d^{(1)}=d^{(2)}\equiv d^{(0)}$ reveals a similar, but biologically distinct insight. In this case though the rate of population turnover of both species is identical, one species exists at greater numbers in isolation than the other (this species has a lower value of $c^{(\alpha)}$). Again while $\bar{A}( z )=0$, the noise-induced term $ \bar{A}^{ \rm S }( z ) $ is non-zero and drives the dynamics in the finite system. We now find that $\bar{A}( z )>0$ for all $z$ on the CM if $c^{(2)}>c^{(1)}$. That is, the species that increases the \emph{joint} carrying capacity of the species is selected for. One interpretation of this noise-induced term is that it is easier for a novel mutant species to invade a small population than a large one; thus species that increase the carrying capacity of the population receive a benefit by being more stochastically robust to invasion attempts~\cite{constable2016}. 

Note that in the above context, if $\epsilon>0$ but $N$ is finite, it is possible that $\bar{A}^{ \rm S }( z )> \bar{A}( z )$ along the length of the SS. Biologically, this provides a mechanism that allows for the evolution of public good producing behaviour despite the evolution of such behaviours being forbidden in the deterministic limit. This has been noted in more typical population genetic models~\cite{houch_2012,houch_2014} as well as models of the form described here. If the population is not well mixed but exists in space, it has been shown that for weak migration the noise induced selection for public good production is amplified both in metapopulation~\cite{constable2016,chotibut_2017} and continuous space models~\cite{hallatschek_2011}. In \cite{constable2016} it was also shown that this behaviour (whereby a species that increases the joint population carrying capacity is stochastically selected for) is generic and robust to the inclusion of a suite of environmental variables in the model that can modify population size. 

\subsection{Models of transitions between male and female heterogamety }\label{sec:heterogamety}


\begin{figure}[th]
\begin{center}
 \includegraphics[width=0.35\textwidth]{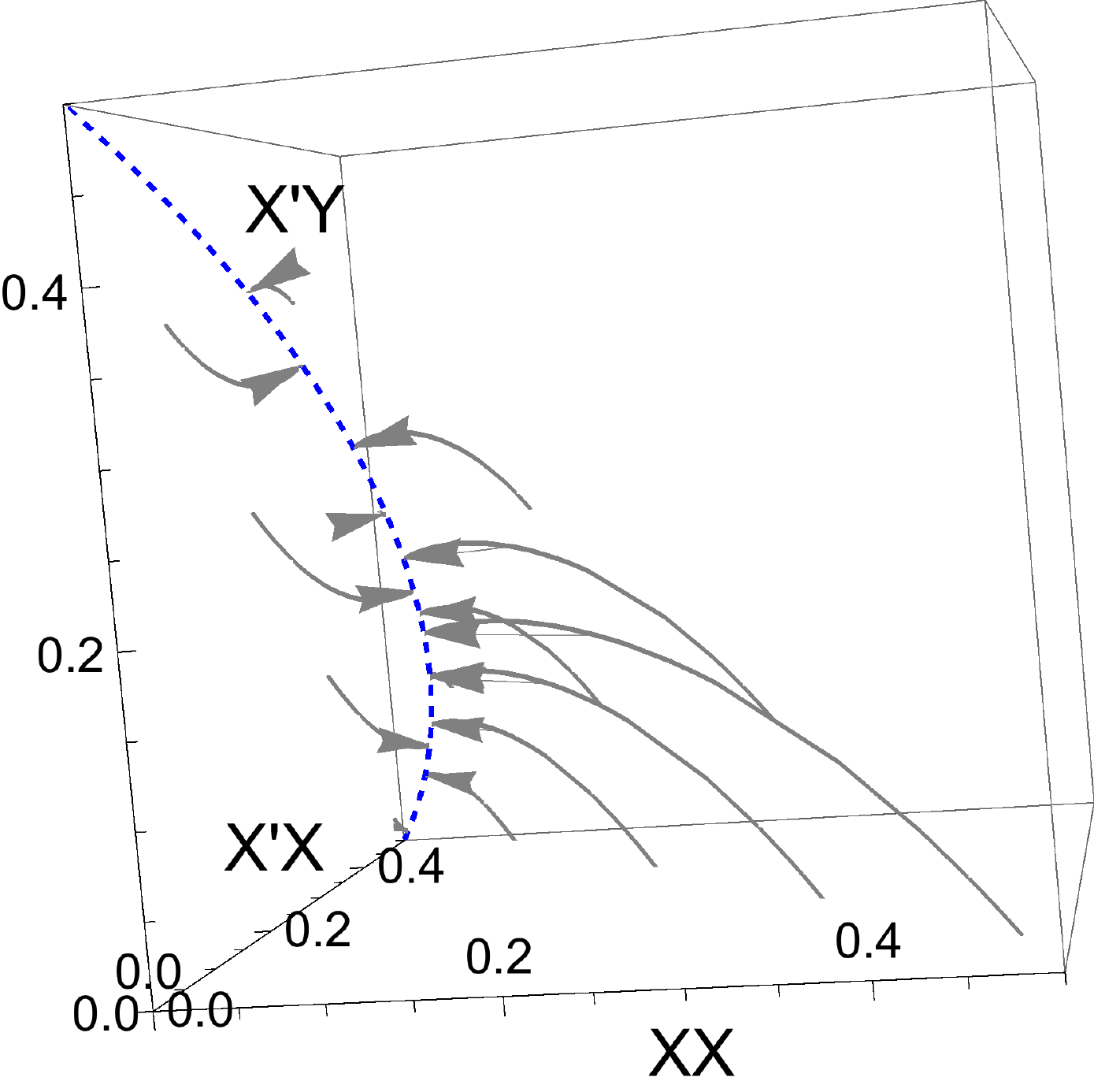} \includegraphics[width=0.45\textwidth]{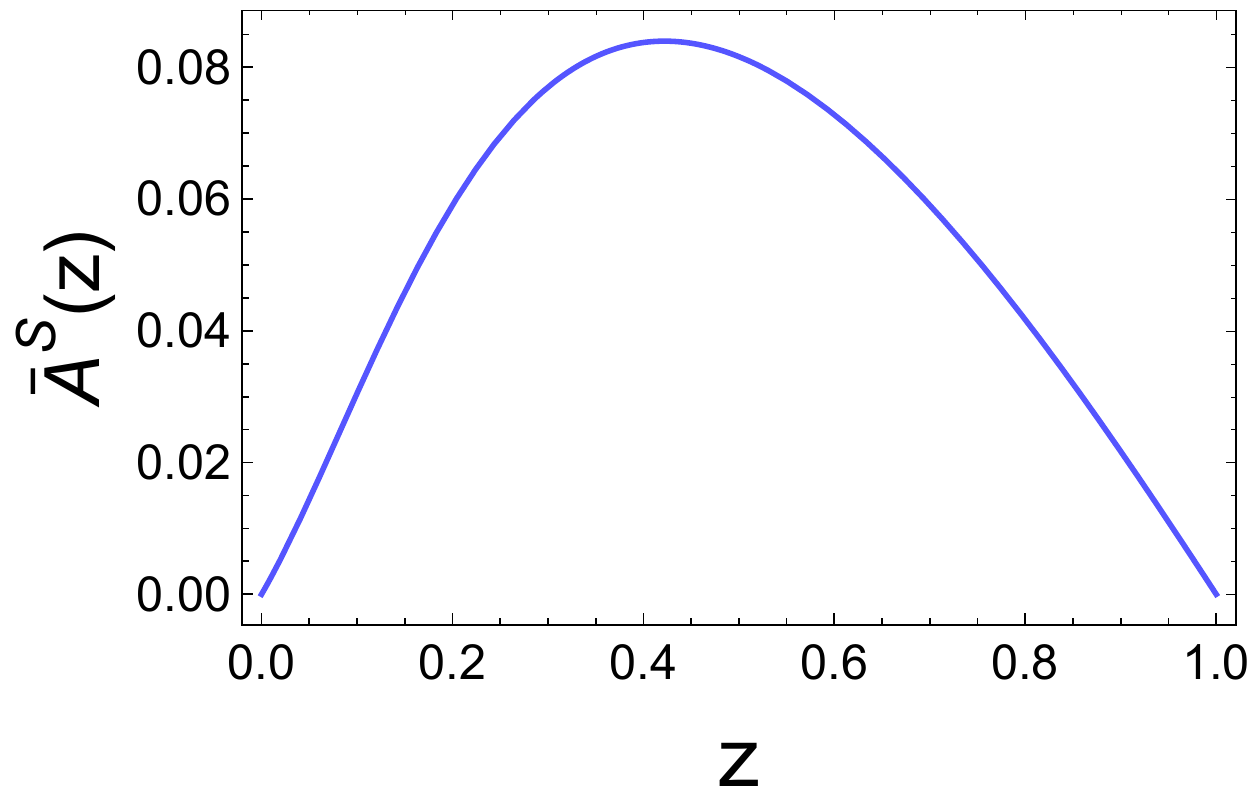}
\end{center}
 \caption{Left: Figure illustrating the neutral dynamics of the model of transitions from male to female heterogamety. Only the frequency of female genotypes is shown. The blue dashed line is the CM, defined by \eref{eq_CM_sex_det}, with male genotypes assumed here to be fixed at their value on the CM. Grey arrows indicate deterministic trajectories starting at various points and eventually reaching the CM. Right: Form of noise-induced selection on the SS (see \eref{eq_heterogamety_drift}).}
\label{fig:heterogamety_traj}
\end{figure}


In this model, a diploid population is considered in which sex is genetically determined by a dominant mutation at a single locus. In mammals, sex is determined by the dominant Y chromosome, so that XY individuals are male while XX individuals are female. Such a system is termed male heterogamety. In birds the situation is somewhat reversed. Their sex determination system features a dominant feminising W chromosome, such that ZW individuals are female, while ZZ individuals are male. This scenario is termed female heterogamety. 

Intriguingly, while these systems are relatively static in both mammals and birds, transitions between male and female heterogamety can occur in reptiles and amphibians. In this section, we will discuss how fast-variable elimination can be exploited to understand the impact of genetic drift on these transitions.

We consider a system of male heterogamety comprised of XX females and XY males. A mutation arises on the X chromosome, changing it to an X$^{\prime}$ and rendering it dominant to the Y, such that X$^{\prime}$Y individuals are female. Genotypes X$^{\prime}$X can also be produced, which are female, as can YY genotypes, which are male. This renders a system of five genotypes. Along with the absorbing state in which the system is entirely XX and XY (male heterogamety), there is also and absorbing state when the system is entirely X$^{\prime}$Y and YY (female heterogamety). Note that this state is analogous to that found in birds up to some relabelling of the chromosomes. We wish to understand the transitions between these states.

We can construct a population genetic model of this process in a very similar way to the diploid Moran model (see \aref{sec:app_HW}), except with matings restricted to being between males and females. We assume a fixed population size $N$. Males and females of each genotype encounter each other proportional to their frequency in the population. They produce a progeny, that inherits its chromosomes (alternatively, alleles at a single locus) in a Mendellian fashion from its parents. Simultaneously, in order that the population size is fixed to $N$, another individual is picked to die. In order to account for selection against certain genotypes, the probability that each genotype is selected to die can be a weighted probability. The detailed model set-up is given in \cite{veller2017}, however here we will discuss only the main results.

Let the frequency of XX, X$^{\prime}$X and X$^{\prime}$Y (female) and XY and YY genotypes (male) be respectively given by $\bm{x} = (x^{(1)},x^{(2)},x^{(3)},x^{(4)},x^{(5)})$. Due to the condition of fixed population size, we can then express $x^{(5)}$ as $x^{(5)}=1-x^{(1)}-x^{(2)}-x^{(3)}-x^{(4)}$. If all genotypes are equally fit, then this four-dimensional system exhibits a one-dimensional CM that connects the absorbing states, characterised in \cite{bull1977}. It is given by 
\begin{eqnarray}
& & x^{(1)} = \frac{(1-z)^2}{2(1+z)^2},\quad x^{(2)} = \frac{z(1-z)}{(1+z)^2},\quad x^{(3)} = \frac{z}{1+z},\nonumber \\
& &  x^{(4)} = \frac{1}{2}(1-z),\quad x^{(5)} = \frac{z}{2} \,,
\label{eq_CM_sex_det}
\end{eqnarray}
and illustrated in \fref{fig:heterogamety_traj}. By definition there are no deterministic dynamics along this line and so one might expect that transitions between the absorbing states occur with equal probability. However employing the fast-variable elimination described in the introduction to this section, we find that in finite populations, there is noise-induced (equivalently, drift-induced) selection along the line:
\begin{eqnarray}
 \bar{A}( z ) &=&   0 \,,\nonumber\\
 \bar{A}^{ \rm S }( z ) &=& \frac{ 1 }{ 4 N } \frac{ z (1 - z) (1 + z)^3 (1 + z^2) \{1 + (2 - z) z [4 + z (6 + z)]\} }{ [1 + z (2 + 3 z)]^3 } \,, \label{eq_heterogamety_drift} \\
 \bar{B}(z) &=&  \frac{1}{4} \frac{ z (1 + z )^{3} \{ 1 + z \left[ 1 + z (6 - z (6 + (3 - z) z)) \right] \} }{  \left[1 + z (2 + 3 z)\right]^2} \,,\nonumber
\end{eqnarray}
where $z$ is the value of $ 1 - 2 x^{(4)}$ on the CM (see \eref{eq_CM_sex_det}). We find that $ \bar{A}^{ \rm S }( z ) $ is positive along the entire length of the CM (see \fref{fig:heterogamety_traj}). Since $z=0$ corresponds to the absorbing state in which the population is entirely comprised of XX and XY genotypes, and $z=1$ corresponds to the absorbing state in which the population is entirely comprised of X$^{\prime}$Y and YY genotypes, we can conclude that the dominant mutation X$^{\prime}$ is selected for along the entire CM. Therefore, if a dominant X$^{\prime}$ mutation occurs in a resident XX|XY population, it is more likely to invade and fixate than a recessive X mutation in a resident X$^{\prime}$Y|YY population.

The above picture, whereby dominant sex-determining mutations are more likely to invade and fixate, is recapitulated for successive invasions. If the resident population is X$^{\prime}$Y|YY, a dominant $Y$ mutation can occur, Y$^{\prime}$, yielding X$^{\prime}$Y$^{\prime}$ males. Again five genotypes can emerge following random mating including X$^{\prime}$X$^{\prime}$ (female) and Y$^{\prime}$Y (male). This dominant mutation is once again more likely to invade, creating an emergent directionality to evolution with substitution rates of neutral dominant mutations being five times higher than those of recessive mutations~\cite{veller2017}. 

Intriguingly, the behaviour described here recapitulates the empirical observation that sex chromosomes typically evolve to include a cascade of inhibitory mutations~\cite{wilkins1995}, with more recent mutations at the top of the cascade. The ``bottom up'' hypothesis~\cite{vandoorn2014b} therefore pictures the sequential invasion of sex-determining genes, each one of which represses the action of the previous mutation. While it can be argued that the drift-induced phenomenon described here is not solely responsible for this pattern, and other deterministic explanations have been suggested~\cite{vandoorn2010}, it is nevertheless interesting to see the emergence of the pattern in a model with minimal assumptions.

In \cite{veller2017}, the above analysis is extended to include biologically relevant selection. In addition more complicated models of transitions between sex chromosomes are also explored via the same timescale-separation arguments. 

\section{Discussion}
\label{sec:discussion}

Our aim in this paper has been to describe and review an approach to the systematic modelling and analysis of a class of models in population genetics. Readers with a background in theoretical physics, and statistical physics in particular, will have found many of the ideas and techniques familiar. These include: the care taken in distinguishing between the form of the models at the microscale, the mesoscale and the macroscale; the idea that the model should be formulated and unambiguously defined at the microscale and that the corresponding mesoscopic and macroscopic models should be derived from the former by some form of systematic approximation procedure; the nature of these approximation procedures (the diffusion limit and fast mode elimination); the formalism of non-equilibrium statistical physics including master equations, FPEs and SDEs.   

While the techniques and philosophy that we use come from statistical physics, the motivation and the models are taken from population biology, and especially from population genetics. Due to the difficulty of performing analytical calculations with models that have many variables and parameters, researchers in these fields frequently create models for rather specific situations and to answer one very limited and definite question. These models may be restricted to one particular situation, but they at least have the possibility of being analysed. On the other hand the danger with this way of doing things is that the subject becomes characterised by a series of models with conflicting assumptions and little overlap. We have taken a different path: we have not held back from setting-up general models, even though they typically contain very many variables and parameters, but have instead tried to systematically approximate these models to obtain ones which are more tractable. In this way the variables that we focus on are those which are likely to be the most important in the medium to long term (the slow modes) and the parameters which are chosen as being important at late times are combinations of the initial parameters of the original model which would have been impossible to guess \textit{a priori}.

There are also some modellers who would argue that to study more complex models, and especially those which feature individuals with more than one or two attributes and which have many parameters, one should turn to agent based models. Here there is no attempt at mathematical analysis and a computer simulation is set up in which the agents (\textit{i.e.}~the individuals) are simply allowed to interact with each other according to a set of rules. These rules will have many parameters associated with them, which will not in general easily map onto the parameters of the type we defined here when formulating the microscopic models. While agent based models might be useful in some situations, the difficulty with the whole approach is that one may have no idea what aspect of the model is responsible for a particular outcome that one is interested in. In fact, this is frequently put forward as a strength of the agent-based method: behaviour \textit{emerges} without having to have been programed-in. We believe that our approach straddles those where the models are simple and tailored for a particular situation and those agent based models which allow for complex systems, but which may be unable to provide insight into the underlying reasons for particular outcomes.

Of course, fast-variable elimination has a long history in theoretical biology and ecology. Most prominently, the form of the Michaelis-Menten function~\cite{briggs_1925} and the Hollings type II functional response~\cite{dawes_2013} are derived based on arguments that assume a separation of timescales. However, these techniques tend to be employed most frequently in deterministic settings. The field of population genetics provides a notable exception to this rule, with the crucial role of genetic drift providing motivation for studies that employed fast-variable elimination in a stochastic setting. We have explicitly discussed the classical assumption of diploid populations at Hardy-Weinberg frequencies in \sref{sec:HW}, where a separation of timescales was assumed \textit{a priori}. However, there also exists a range of studies which apply more sophisticated techniques (see, for example \cite{nagylaki1980SM,ethier_1980,ethier_1988,stephan_1999,hossjer_2016,newberry_2016}). 

The relatively frequent occurrence of stochastic fast-variable elimination in population genetics is the result of two factors. The first, as we have already mentioned, is that population genetics models are often inherently stochastic. The second is that a common assumption in many population genetic models (even when fast-variable elimination techniques are not required) is that selection is weak relative to the other processes. If then the question we wish to ask is which of two alleles (or genotypes) fixes in a population, the stage is set for methods based on the elimination of fast-variables to be useful: by definition, if there is no selection, there must be a CM connecting states in which either allele is fixed; if selective forces are weak, the CM is simply replaced by an SS; if there are only two alleles competing, regardless of the other variables, the SS is one-dimensional, allowing an analytic treatment of the resulting effective equation.

Given the ubiquity of the above listed features in models of population genetics, it is therefore perhaps surprising that there are in fact not many more studies employing fast-variable elimination. Perhaps some of the reasons for this stem from the apparent difficulty of dealing with these types of problem, with previous studies~\cite{stephan_1999} relying on the perturbative techniques described by Gardiner~\cite{gardiner_2009}. While certainly thorough, this approach is perhaps lacking in intuition, taking place as it does within the context of the FPE and involving operators. In contrast, we believe that there is much to say for the approaches that we have outlined in this paper. As they are described within the context of SDEs (entirely equivalent to the FPE), and as the effective equation is constructed from physically meaningful quantities (such as the left and right eigenvectors), it is relatively straightforward to apply the methodology. We note that while other approaches to fast-variable elimination in the SDE setting are well practiced (see~\cite{constable_2013}, which gives a more complete review of this area of the literature), these too have not cemented themselves as methods within the population genetics literature. The fact that, as described in \sref{sec:heterogamety}, the techniques we have presented are yielding novel insights to models first developed in the 1970s is testament to the untapped potential of the approach.

In terms of applications of the method, we have tried to some extent to move away from using the Moran model, which has been a very popular starting point among some researchers over the last decade or two. Instead we have introduced competition between individuals to control population growth. There are many other types of interaction between individuals that could be included, and this is an interesting question for the future. But even when only competition between individuals is included, some interesting points emerge. For example, as discussed in Sec.~\ref{sec:M_allele_SLVC}, the model with competition can be mapped into a Moran model, but only if the selection is frequency dependent, which suggests that the traditional use of frequency independent selection may be too restrictive. In addition, the mapping gives a direct connection between the competition rates and the elements of the payoff matrix. Only relative payoffs of a certain type appear in the mapping between the two models, which again could be used to constrain the nature of the payoffs. This would be useful, since the lack of prescriptions available to guide the choice of payoffs is a significant weakness in the application of game theory to the subject. Of course, our methods are only valid when selection is weak, and moving beyond this regime, such as when strong mutualisms are accounted for, is known to invalidate this mapping between the SLVC and models of game theory~\cite{chotibut_2016}. Extending such a study using the methods we have presented is also an interesting avenue for future studies. 

Perhaps one of the most surprising insights that fast-variable elimination has provided is the occurrence of noise-induced dynamics in otherwise deterministically neutral systems. These dynamics, described in \sref{sec:noise_induced}, can be interpreted as drift-induced selection and can give rise to surprising behaviours, such as selection reversal. While historically such behaviour has been observed and characterised in simpler settings~\cite{gillespie_1974}, it has been often dismissed as a small second order effect of little biological significance~\cite{hansen_2017}. This is because noise induced selection is inversely proportional to the size of the system, and therefore, it is argued, likely to be swamped by other processes. However, this misses the fact that when selection is weak (a frequent assumption in population genetics) or when a population has a small effective population size (such as when a system is spatially distributed) noise-induced terms can in fact dominate the dynamical behaviour. 

Moreover, utilisation of fast-variable elimination techniques has led to an increasing awareness that noise-induced selection appears in many distinct evolutionary systems in ecology, epidemiology and population genetics. We have already discussed some of these in detail in \sref{sec:noise_induced}. However more examples are arising with increasing frequency. In \cite{lin_mig_1}, stochastic Lotka-Volterra models for multiple islands (similar to the models described in \sref{sec:D_island_SLVC}) were used to show that demographic stochasticity induces a selective bias for species that disperse at a higher rate between identical islands, where there is no selection deterministically. Intriguingly, this bias persists even when the islands are inhomogeneous and a deterministic selective pressure exists for slower dispersal~\cite{lin_mig_2}. In \cite{kogan_2014} meanwhile, a model of viral evolution was investigated, showing that noise-induced dynamics selected for viral strains with a fast infection and recovery time in SIS and SIR models.

It is clear that the methods we have described in this paper hold a great degree of utility, whether one is seeking to consolidate existing disparate models, add more biological detail to well understood systems or uncover and characterise entirely new dynamics. Beyond drawing attention to this fact, we hope that as presented, this work will prove of value to readers with both biology and statistical physics backgrounds. For the biologists, this paper has aimed to show that these techniques are not as conceptually formidable as other studies may have suggested. For the statistical physicists, this paper has aimed to show that many interesting and relevant problems in biology still exist, as yet untapped, and amenable to analysis with their approaches. 

\section*{Acknowledgments}

GWAC thanks the Finnish Center for Excellence in Biological Interactions and the Leverhulme Early Career Fellowship provided by the Leverhulme Trust for funding.

\appendix

\section{Calculational details and background for \sref{sec:HW} }
\label{sec:app_HW}

In this appendix we give analytical support for the system discussed in Sec.~\ref{sec:HW}. We begin by briefly reviewing the classic result (which assumes a separation of timescales \textit{a priori}) before moving on to the fast-variable elimination itself.

\noindent \textbf{Classic approach.} The classic treatment begins by considering the dynamics of alleles, rather than the genotypes that carry them. The probability of an allele reproducing is proportional to its frequency in the population (which is assumed to be given by Hardy-Weinberg frequencies) weighted by the fitness of the genotypes carrying the allele. So, if there is a frequency $n^{\rm (A^{(1)}) }/(2N)$ of allele A$^{(1)}$, then at Hardy-Weinberg frequencies the frequency of A$^{(1)}$A$^{(1)}$ genotypes is approximately $[n^{\rm (A^{(1)}) }/(2N)]^{2}$ and the frequency of A$^{(1)}$A$^{(2)}$ genotypes is $2 n^{\rm (A^{(1)}) }(N-n^{\rm (A^{(1)}) })/(2N)^{2}$ (see \eref{eq_HW_freq}). Assuming that the A$^{(1)}$ allele has a fitness $1+s$ when it occurs in an A$^{(1)}$A$^{(1)}$ genotype, and a fitness $1+sh$ when it occurs in a A$^{(1)}$A$^{(2)}$ genotype, the transitions rates can be written
\begin{eqnarray}
& & T( n^{\rm (A^{(1)}) } + 1 | n^{\rm (A^{(1)}) } ) = \nonumber \\
& & \overbrace{\left[ (1 + s) \left( \frac{ n^{\rm (A^{(1)}) } }{ 2 N } \right)^{2} +  (1 + s h )  \frac{ n^{\rm (A^{(1)}) } }{2N} \left(1 - \frac{ n^{\rm (A^{(1)}) } }{2 N} \right)  \right] }^{\rm{Weighted\,prob.\,A^{(1)}\,birth}} \times \overbrace{\left(1 - \frac{ n^{\rm (A^{(1)}) } }{2 N} \right)}^{neutral\,prob.\,A^{(2)}\,death}\,, 
\nonumber\\ \nonumber \\
& & T( n^{\rm (A^{(1)}) } - 1 | n^{\rm (A^{(1)}) } ) = \nonumber \\
& & \underbrace{\left[ (1 + s h ) \frac{ n^{\rm (A^{(1)}) } }{2N}\left(1 - \frac{ n^{\rm (A^{(1)}) } }{2 N} \right) +  \left(1 - \frac{ n^{\rm (A^{(1)}) } }{2 N} \right)^{2} \right]}_{\rm{Weighted\,prob.\,A^{(2)}\,birth}} 
\times \underbrace{ \frac{ n^{\rm (A^{(1)}) } }{2 N} }_{neutral\,prob.\,A^{(1)}\,death} \,.\nonumber
\end{eqnarray}
Note that since only ``half'' of the A$^{(1)}$A$^{(2)}$ genotype carries the A$^{(1)}$ allele, an extra factor of a half arises in terms that involve both $n^{\rm (A^{(1)}) }$ and $n^{\rm (A^{(2)}) } = 2 N - n^{\rm (A^{(1)}) }$. Applying the diffusion approximation with $x^{(1)}=n^{\rm (A^{(1)}) }/2N$ and assuming $s$ is small one obtains \eref{eq_HW_classic}. Note that since the diffusion approximation is here conducted with a large parameter $2N$, an additional factor $1/2$ occurs before the diffusion term in the FPE formulation (see \eref{FPE_simple}) and the noise term in the SDE formulation (see \eref{SDE_simple}).

\noindent \textbf{Mechanistic approach.} We want to build a similar model, but more mechanistically using the genotype frequencies. We will denote their numbers $n^{(1)}$ (A$^{(1)}$A$^{(1)}$), $n^{(2)}$ (A$^{(1)}$A$^{(2)}$) and $n^{(3)}$ (A$^{(2)}$A$^{(2)}$), so that the fixed size of the population is $n^{(1)} + n^{(2)} + n^{(3)} = N$. (Note the slight difference in notation with the main text; $n^{\rm ( A^{(1)} A^{(1)} )} \equiv n^{(1)}$, $n^{\rm ( A^{(1)} A^{(2)} )} \equiv n^{(2)}$ and $n^{\rm ( A^{(2)} A^{(2)} )} \equiv n^{(3)}$). We now denote by $W^{( \alpha \beta )}$ the fitness of genotype $ \alpha $ reproducing with genotype $ \beta $, and assume $W^{( \alpha \beta )}$ is symmetric. The transition rates may now be constructed by accounting for the probability of each different pairing and death. For instance, the probability that $n^{(1)}$ increases by $1$ only is the probability that any pairing occurs (given by the product of frequencies of the genotypes in the pairing) multiplied by the probability that pairing gives rise to an A$^{(1)}$A$^{(1)}$ genotype by Mendelian inheritance, multiplied by the probability that an A$^{(2)}$A$^{(2)}$ genotype dies (given by the frequency of A$^{(2)}$A$^{(2)}$ in the population). Accounting for these terms for each possible transition gives us the following transition rates:
\begin{eqnarray*}
& & T( n^{(1)} + 1 ,n^{(2)}|n^{(1)} ,n^{(2)}) = \frac{1}{N^{3}}\left[ W^{(11)} (n^{(1)})^2  + 2 \frac{1}{2} W^{(12)} n^{(1)} n^{(2)}  \right. \nonumber \\
& & \left. + \frac{1}{4} W^{(22)} (n^{(2)})^2 \right] (N - n^{(1)} - n^{(2)}), \\
& & T( n^{(1)} - 1 ,n^{(2)}|n^{(1)} ,n^{(2)}) = \frac{1}{N^{3}} \left[ \frac{1}{4} W^{(22)} (n^{(2)})^{2}  + 2 \frac{1}{2} W^{(23)}  n^{(2)} (N - n^{(1)} - n^{(2)}) \right. \nonumber \\
& & \left. +   W^{(33)} (N - n^{(1)} - n^{(2)})^2 \right] n^{(1)}, \\
& & T( n^{(1)} ,n^{(2)} + 1|n^{(1)} ,n^{(2)}) = \frac{1}{N^{3}} \left[ 2 \frac{1}{2} W^{(12)} n^{(1)} n^{(2)}  + 2 W^{(13)} n^{(1)} (N - n^{(1)} - n^{(2)})   \right. \nonumber \\
& & \left. +  \frac{1}{2} W^{(22)} (n^{(2)})^{2}  + 2 \frac{1}{2} W^{(23)} n^{(2)} ( N - n^{(1)} - n^{(2)}) \right] (N - n^{(1)} - n^{(2)} ), \\
& & T( n^{(1)} ,n^{(2)} - 1|n^{(1)} ,n^{(2)}) = \frac{1}{N^{3}} \left[ \frac{1}{4} W^{(22)} (n^{(2)})^2 + 2 \frac{1}{2} W^{(23)} n^{(2)}  (N - n^{(1)} - n^{(2)}) \right. \nonumber \\
& & \left. + W^{(33)} (N - n^{(1)} - n^{(2)} )^2\right] n^{(2)}, \\
& & T( n^{(1)} +1 , n^{(2)} -1|n^{(1)} ,n^{(2)}) = \frac{1}{N^{3}} \left[ W^{(11)} (n^{(1)})^2  + 2 \frac{1}{2} W^{(12)} n^{(1)} n^{(2)} \right. \nonumber \\
& & \left. + \frac{1}{4} W^{(22)} (n^{(2)})^2 \right] n^{(2)}, \\
& & T( n^{(1)} -1 , n^{(2)} + 1|n^{(1)} ,n^{(2)}) = \frac{1}{N^{3}} \left[ 2 \frac{1}{2} W^{(12)} n^{(1)} n^{(2)} + 2 W^{(13)} n^{(1)}\times \right. \nonumber \\
& & \left. (N - n^{(1)} - n^{(2)}) + \frac{1}{2} W^{(22)}  (n^{(2)})^2 + 2 \frac{1}{2} W^{(23)} n^{(2)} (N - n^{(1)} - n^{(2)} ) \right] n^{(1)}.
\end{eqnarray*}
We now set $W^{( \alpha \beta )}=1 + s \alpha^{( \alpha \beta )}$ and assume $s$ is small to obtain a diffusion approximation (see \eref{FPE_simple} and \eref{SDE_simple}) with $y^{(1)}=n^{(1)}/N$ and $y^{(2)}=n^{(2)}/N$;
\begin{eqnarray}
& & A^{(1)}(\bm{y}) = -y^{(1)}(1 - y^{(1)}) + y^{(2)} ( y^{(1)} + \frac{y^{(2)}}{4} ) + s \left[ -\alpha^{(33)} y^{(1)} \right. \nonumber \\
& & + (\alpha^{(11)} - 2 \alpha^{(13)} + 2 \alpha^{(33)}) (y^{(1)})^2 + (\alpha^{(12)} - 2 \alpha^{(23)} + 2 \alpha^{(33)}) y^{(1)} y^{(2)} \nonumber \\  
& & + \frac{ 1 }{ 4 } \alpha^{(22)} (y^{(2)})^2  -   (y^{(1)})^3 (\alpha^{(11)} - 2 \alpha^{(13)} + \alpha^{(33)} ) -  (y^{(1)})^2 y^{(2)}\times \nonumber \\
& & \left. (2 \alpha^{(12)} - 2 \alpha^{(13)} - 2 \alpha^{(23)} + 2 \alpha^{(33)}) - y^{(1)} (y^{(2)})^2 ( \alpha^{(22)} - 2 \alpha^{(23)} + \alpha^{(33)} ) \right],
\nonumber \\
& & A^{(2)}(\bm{y}) = 2 \left[ y^{(1)} (1 - y^{(1)} ) - y^{(2)} ( y^{(1)} + \frac{ y^{(2)} }{4} )  \right] + s \left[ 2 \alpha^{(13)} y^{(1)} \right. \nonumber \\
& & + y^{(2)} (\alpha^{(23)} - \alpha^{(33)}) - 2 \alpha^{(13)} (y^{(1)})^2 + 
 y^{(1)} y^{(2)} (\alpha^{(12)} - 4 \alpha^{(13)} - \alpha^{(23)} + 2 \alpha^{(33)}) \nonumber \\
& & +  \frac{1}{2} (y^{(2)})^2 (\alpha^{(22)} - 6 \alpha^{(23)} + 4 \alpha^{(33)}) -  (y^{(1)})^2 y^{(2)} (\alpha^{(11)} - 2 \alpha^{(13)} + \alpha^{(33)}) \nonumber \\
& & \left. -  2 y^{(1)}(y^{(2)})^2 (\alpha^{(12)} - \alpha^{(13)} - \alpha^{(23)} + \alpha^{(33)}) -  (y^{(2)})^3 (\alpha^{(22)} - 2 \alpha^{(23)} + \alpha^{(33)}) \right], \nonumber \\ 
\label{eq_HW_A}
\end{eqnarray}
and 
\begin{eqnarray}
B^{(11)}(\bm{y}) &=& y^{(1)} (1 + y^{(1)}) + \frac{1}{4} (y^{(2)})^2 - 2 (y^{(1)})^3 - 2 (y^{(1)})^2 y^{(2)} \nonumber \\
&+& y^{(1)} y^{(2)} \left( 1 - \frac{1}{2} y^{(2)} \right)\,, \nonumber \\
B^{(22)}(\bm{y}) &=& 2 (y^{(1)} + y^{(2)} ) - 2 (y^{(1)})^2 - 6 y^{(1)} y^{(2)} - \frac{ 5 }{ 2 } (y^{(2)})^2 + (y^{(2)})^3 \nonumber \\
&+& 4 (y^{(1)})^2 y^{(2)} + 4 y^{(1)} (y^{(2)})^2\,, \nonumber \\
B^{(12)}(\bm{y}) &=& - 2 (y^{(1)})^2 - y^{(1)} y^{(2)} - \frac{1}{4} (y^{(2)})^3 + 2 (y^{(1)})^3 + (y^{(1)})^2 y^{(2)} - \frac{1}{2} y^{(1)} (y^{(2)})^2\,, \nonumber \\
B^{(21)}(\bm{y}) &=& B^{(12)}(\bm{y}) \,. \label{eq_HW_B}
\end{eqnarray}
We note however that unlike in the classic approach (where the noise correlation matrix has a prefactor $N^{-1}/2$) this noise term has the standard pre-factor of $N^{-1}$.

Our end goal is to compare the mechanistic description in terms of genotypes given by Eqs.~(\ref{eq_HW_A}) and (\ref{eq_HW_B}) to that proposed by \eref{eq_HW_classic}. To this end the following calculation will involve two steps. The first will be removing the fast degrees of freedom in the model (see \fref{fig:HW}) as described in the main text. The second will be transforming from the genotype frequencies of the mechanistic model ($y^{(1)}$ and $y^{(2)}$) into the allele frequencies ($x^{(1)}$) in which the classic approach is couched. We note that in both these steps there are some additional mathematical complications that arise that, while formally important, are not crucial to understanding the basic principles of the method or insights developed.

In the first step, removing the fast-variables, a noise-induced selection term arises. Such terms are discussed in \sref{sec:noise_induced}. In the second step, the non-linear transformation we employ makes use of the It\={o} calculus necessary, giving rise to extra terms in the transformed equation~\cite{risken_1989}. However in the problem at hand we will see that these two effects cancel precisely. Readers who only wish to understand the principles of the problem and physical intuitions developed are therefore free in this section to bypass references to noise-induced dynamics and It\={o} calculus.

In order to employ the fast-variable elimination procedure, we begin with \eref{eq_HW_A}. From this we can determine that a CM exists, given by \eref{eq_HW_freq}. We wish to remove the fast variables as described in the \sref{sec:example} and \sref{sec:noise_induced}. To begin, we must identify the left and right eigenvectors of the Jacobian evaluated on the CM that correspond to the zero eigenvalue, $\bm{u}^{\{1\}}$ and $\bm{v}^{\{1\}}$. These are given by
\begin{align}\label{eigenvectors_HW}
\bm{u}^{\{ 1\} } = \sqrt{ z } \,\left( \begin{array}{c} 
2 
\\
1
\end{array} \right), 
\ \ \bm{v}^{\{ 1\} } =  \frac{ 1 }{ \sqrt{z} }\,\left( \begin{array}{c} 
\sqrt{z}
\\
1 - 2 \sqrt{z}
\end{array} \right),
\end{align}
where $z$ is $y_1$ on the CM. We can now use these, along with Eqs.~(\ref{eq_general_ABar})-(\ref{eq_general_BBar}) and \eref{eq_general_ABar_S} in the main text to calculate the effective SDE. We find 
\begin{eqnarray}
\bar{A}(z) &=&  2 s z (1 - \sqrt{z})  (\alpha^{(23)} - \alpha^{(33)}+ 3 z (\alpha^{(12)} - \alpha^{(13)} - 2 \alpha^{(22)} + 
      3 \alpha^{(23)} - \alpha^{(33)})  \nonumber \\
& &+ \sqrt{z} (\alpha^{(13)} + 2 \alpha^{(22)} - 6 \alpha^{(23)} + 3 \alpha^{(33)}) \nonumber  \\
& &+ z^{3/2} (\alpha^{(11)} - 4 \alpha^{(12)} + 2 \alpha^{(13)} + 4 \alpha^{(22)} - 
      4 \alpha^{(23)} + \alpha^{(33)}) 
)\,, \\
 \bar{B}(z) &=& 4 z^{3/2} - 4 z^2 \,,
\end{eqnarray}
from our standard analysis (see \sref{sec:example}). From our more involved analysis given in \sref{sec:noise_induced} we also find a noise-induced term
\begin{eqnarray}
\bar{A}^{S}(z) =  \frac{1}{N} \left( \sqrt{z} - z \right)\,.
\end{eqnarray}
We withhold attaching any biological meaning to this for the time being, however we can note that it arises entirely as a result of the curvature of the CM since the trajectories to the CM are not divergent (see \sref{sec:noise_induced}). 

We now want to make a change of variables from $z$ into $x^{(1)}=\sqrt{z}$, where $x^{(1)}$ is the frequency of A$^{(1)}$ alleles in the population (recall from Eq.~(\ref{eq_HW_freq}) that $y^{(1)}=(x^{(1)})^{2}$ on the CM and that $z$ is our variable $y^{(1)}$ on the CM). This is a non-linear transformation, and we must therefore employ It\={o} calculus~\cite{risken_1989}. This yields a reduced SDE with
\begin{eqnarray}
\tilde{A}( x^{(1)} ) &=& \left( \frac{ \mathrm{d} x^{(1)} }{ \mathrm{d} z } \right) \left\{ \bar{A}[ (x^{(1)})^{2} ] + \bar{A}^{S}[ (x^{(1)})^{2} ] \right\} + \frac{1}{2N} \left( \frac{ \mathrm{d}^{2} x^{(1)} }{ \mathrm{d} z^{2} } \right) \left\{ \bar{B}[ (x^{(1)})^{2} ] \right\} \,, \\
\tilde{B}( x^{(1)} ) &=& \left( \frac{ \mathrm{d} x^{(1)} }{ \mathrm{d} z }  \right)^{2} \bar{B}[ (x^{(1)})^{2} ] \,.
\end{eqnarray}
Here the term in $\tilde{A}( x^{(1)} )$ involving a second derivative of $x^{(1)}$ is that which arises from a proper implementation of It\={o} calculus. However, after some algebra we find that the term $\bar{A}^{S}[ (x^{(1)})^{2} ]$ cancels this It\={o} term, and so we find the above expression for $\tilde{A}( x^{(1)} )$ simplifies to 
\begin{eqnarray}
\tilde{A}( x^{(1)} ) &=& \frac{ \mathrm{d} x^{(1)} }{ \mathrm{d} z } \bar{A}[ (x^{(1)})^{2} ] \,,
\end{eqnarray}
which is written in full in \eref{eq_HW_effective_A}, while $\tilde{B}( x^{(1)} )$ is given by \eref{eq_HW_effective_B}. Note that in this particular case, these expressions are those that we would have obtained if we had naively ignored both noise-induced selection and It\={o} calculus. Finally, we draw attention to the fact that once we account for the extra factor $1/2$ that occurs in front of the noise correlation term $B(x^{(1)})$ in the classic approach (which we recall comes from the diffusion approximation resulting in a coefficient of $N^{-1}/2$ in the FPE), the noise terms are identical in both the effective and classic equations, \eref{eq_HW_effective_B} and \eref{eq_HW_classic}.

With the effective SDE now in hand, we are free to choose which values of $W^{( \alpha \beta )} = 1 + s \alpha^{( \alpha \beta )}$ we wish. We will look at some very simple parameterisations. First we consider a multiplicative fitness function where the fitness $W^{( \alpha \beta )}$ of any pairing of genotypes $ \alpha $ and $ \beta $ is product of the fitness of the genotypes;
\begin{eqnarray*}
W^{( \alpha \beta )} &=& W^{( \alpha )}W^{( \beta )} \\
&=& ( 1 + s \alpha^{( \alpha )} ) ( 1 + s \alpha^{( \beta )} ) \\
&=&  1 + s (\alpha^{( \alpha )} + \alpha^{( \beta )} ) + \mathcal{O}(s^{2}) \,.
\end{eqnarray*}
Substituting this into \eref{eq_HW_effective_A}, we find that this is exactly the same form as that posited in classic population genetics (see \eref{eq_HW_classic}) if $\alpha^{(1)} = 1$, $\alpha^{(2)}=h$ and $\alpha^{(3)}=0$.

We next consider a system, similarly parameterised, but where the genotype pairings have the average fitness of the two genotypes;
\begin{eqnarray*}
W^{( \alpha \beta )} &=& \frac{ W^{(\alpha)} + W^{(\beta)} }{ 2 } \\
&=& 1 + s \frac{ (\alpha^{(\alpha)} + \alpha^{(\beta)} ) }{ 2 } \,.
\end{eqnarray*}
Again we get a form similar to that in \eref{eq_HW_classic} if $\alpha^{(1)} = 1$, $\alpha^{(2)}=h$ and $\alpha^{(3)}=0$, except now selection acts half as quickly. The case of additive fitness can be tackled in precisely the same fashion by setting $W^{(\alpha \beta)} =  W^{(\alpha)} + W^{(\beta)}  = 2 + s( \alpha^{(\alpha)} + \alpha^{(\beta)} )$, and then rescaling by a factor two.

%


\end{document}